%% file: class_astroph.tex
\documentclass{aa}
\usepackage{graphicx}
\usepackage{txfonts}
\usepackage{rotating}
\usepackage{natbib}
\usepackage{colortbl}
\usepackage{longtable,lscape}
\bibpunct{(}{)}{;}{a}{}{,} % to follow the A&A style

\newenvironment{jh}{}{}
\newenvironment{tjh}{}{}

\newcommand{\micron}{\hbox{$\mu{\rm m}$}}                      % um

\newcommand{\Msun}{\mbox{$M_{\odot}$}}
\newcommand{\Lsun}{\mbox{$L_{\odot}$}}

\newcommand{\Menv}{\mbox{$M_\mathrm{env}$}}

\newcommand{\Tbol}{\mbox{$T_\mathrm{bol}$}}
\newcommand{\Lbol}{\mbox{$L_\mathrm{bol}$}}
\newcommand{\Lsmm}{\mbox{$L_\mathrm{smm}$}}
\newcommand{\prob}{\hbox{P}}

\defcitealias{paperI}{Paper~I}
\defcitealias{class}{Paper~II}
\defcitealias{outflows}{Paper~III}

\begin{document}
   \title{Star formation in Perseus}

   \subtitle{II. SEDs, classification and lifetimes}

   \author{J. Hatchell\inst{1},  G. A. Fuller\inst{2}, J. S. Richer\inst{3}, T.J.Harries\inst{1}, E. F. Ladd\inst{4}}
   \authorrunning{Hatchell et al.}
   \titlerunning{SCUBA Perseus survey --- II. Classification}
   
   \offprints{hatchell@astro.ex.ac.uk}

   \institute{School of Physics, University of Exeter, Stocker Road, Exeter EX4 4QL, U.K.
              \and School of Physics \& Astronomy, University of Manchester, P.O. Box 88, Manchester M60 1QD, U.K.
              \and Cavendish Laboratory, Cambridge CB3 0HE, U.K.
              \and Department of Physics and Astronomy, Bucknell University, Lewisburg, PA 17837, U.S.A.
}

   \date{}

   \abstract{Hatchell et al.~(2005) (Paper~I) published a submillimetre continuum map of the Perseus molecular cloud, detecting the starless and protostellar cores within it.}
{To determine the evolutionary stage of each submm core in Perseus, and investigate the lifetimes of these phases.}
{We compile spectral energy distributions (SEDs) from 2MASS (1--2~\micron), Spitzer IRAC (3.6, 4.5, 5.8, 8.0~\micron),  Michelle (11 and 18~\micron), IRAS (12, 25, 60, 100~\micron), SCUBA (450 and 850~\micron) and Bolocam(1100~\micron) data.  Sources are classified starless/protostellar on the basis of infrared and/or outflow detections and Class~I/Class~0 on the basis of \Tbol, $\Lbol/\Lsmm$ and $F_{3.6}/F_{850}$.  In order to investigate the dependence of these evolutionary indicators on mass, we construct radiative transfer models of Class~0 sources.}
{Of the submm cores, 56/103 (54\%) are confirmed protostars on the basis of infrared emission or molecular outflows.  Of these, 22 are classified Class~1 on the basis of three evolutionary indicators, 34 are Class~0, and the remaining 47 are assumed starless.  Perseus contains a much greater fraction of Class~0 sources than either Taurus or Rho Oph.  \begin{jh}We derive estimates for the correlation between bolometric luminosity and envelope mass for Class~I and Class~0 sources.\end{jh} }
{\begin{jh}Comparing the protostellar with the T~Tauri population, the lifetime of the protostellar phase in Perseus is $0.25\hbox{--}0.67$~Myr\end{jh} (95\% confidence limits).  The relative lifetime of the Class~0 and Class~1 phases are similar, confirming the results of Visser et al.~(2002) in isolated cores.   We find that for the same source geometry but different masses, evolutionary indicators such as \Tbol\ vary their value.  It is therefore not always appropriate to use a fixed threshold to separate Class~0 and Class~I sources.  More modelling is required to determine the observational characteristics of the Class~0/Class~I boundary over a range of masses.}


\maketitle

%%%%%%%%%%%%%%%%%%%%%%%%%%%%%%%%%%%%%%%%%%%%%%%%%%%%%%%%%%%%
\section{Introduction}
\label{sect:introduction}

The submillimetre (submm) surveys of the Perseus molecular cloud with SCUBA
\citep[hereafter Paper~I]{paperI} and Bolocam \citep{enoch06} have for the first time
produced a fairly complete census of the star formation activity in
this cloud, identifying all the submm cores above their flux detection
limits.  \begin{jh}Previously, \end{jh}these submm cores have been discussed in terms of
their masses, clustering and their relationship to the molecular cloud
\citep{paperI,enoch06,kirk06} but so far there has been little attempt
to identify their evolutionary status.  With an identification of the
protostars amongst the $\sim100$ submm cores, questions of the star
formation rate, the timescales of the prestellar and protostellar
phases, and the detailed evolution of protostars can be addressed.

The timescales involved in protostellar formation are particularly
important as constraints on the underlying physics controlling the
protostellar collapse.  If cores form at the point where turbulence
\begin{jh}rapidly\end{jh} completely dissipates, then collapse should occur on roughly a
free-fall time.  If, on the other hand, \begin{jh}magnetic fields and
  ambipolar diffusion play an important role in the evolution of the
  cores, then the time for a core to collapse can be greatly increased
  \citep{tassismouschovias04} \end{jh}.  Current stimates for the ratio of the
  prestellar to the protostellar lifetime vary from 2--3
  \citep{visser02,leemyers99} to 10 \citep{wt94,jessopwt00}.

Estimates for the lifetime of an embedded low-mass
  protostar are $1\hbox{--}2\times 10^5$~Myr
  \citep{greene94,kenyonhartmann95,wilking89,kenyon90}.  These early
  estimates were based on infrared surveys (eg. IRAS) with very
  incomplete counts of protostars.  Now that we have a submm-based
  census of the protostellar counts in Perseus, combined with much
  more complete estimates of the T~Tauri population down to below the
  brown dwarf limit \citep{muench03,wilking04,jorgensen06a}, and new
  age estimates for the T~Tauri population using updated
  pre-main-sequence (PMS) evolutionary tracks \citep{wilking04,mayne06}, it
  is time to revisit these estimates.

A catalogue of protostellar sources also provides a valuable resource for
followup study (eg. with interferometers such as ALMA) as subsamples
with particular properties (starless/protostellar, mass, luminosity)
can be identified in order to investigate specific aspects of protostellar or starless core evolution.

An established method of differentiating between protostars and
starless cores is on the basis of their spectral energy distributions
(SEDs).  Perseus has now been surveyed not only in the mm/submm by
SCUBA and Bolocam but also across the infrared by 2MASS, Spitzer and
IRAS, so fluxes are available across a wide range of wavelengths.
Using these publicly available data we can compile a set of spectral
energy distributions for the submm sources in Perseus which is based
on a homogenous dataset.  These consistent spectral energy
distributions are an important tool in determining evolutionary stage.

In this paper we collate the spectral data on the submm sources in
Perseus and discuss the evolution and lifetimes of protostars within
the cloud.  In Sect.~\ref{sect:seds} we discuss the derivation of
fluxes for the SEDs, including the new mid-infrared observations with
Michelle on UKIRT, and in Sect.~\ref{sect:results} the resulting SEDs.
Sect~\ref{sect:classification} discusses the resulting classification
and its implications.  In Sect~\ref{sect:radtrans} we present models
of early Class~0 sources and their implications for the
classification.  In Sect.~\ref{sect:lifetimes} we consider the
lifetime of the embedded protostellar phase and our conclusions are
summarised in Sect.~\ref{sect:summary}.

\begin{jh}Throughout, we assume a distance of 320~pc for the Perseus molecular cloud based on the
  Hipparcos distance of IC348 \citep{dezeeuw99}, consistent with our
  earlier paper \citepalias{paperI}, though other recent studies
  \citep{kirk06,enoch06} have assumed a closer distance of 250~pc
  based on extinction studies \citep[][and references
  therein]{cernisstraizys03}.
\end{jh}

% \section{New observations}
% \label{sect:obs}
% 
% 
% \subsection{Michelle 11 and 18~\micron\ imaging}
% \label{michelle}
% 
% 11 and 18~\micron\ data were taken in 1999 using the Michelle imaging camera on UKIRT.
% Details...(Gary?)
% 
\section{Spectral energy distributions}
\label{sect:seds}

We have compiled a spectral energy distribution (SED) for each submm
source in Perseus, based on an extended version
of the 850~\micron\ catalogue given in Paper~I. With a mass limit of
of 0.1~\Msun\ for a 30~K protostellar envelope or 0.5~\Msun\ for a 10~K
starless core, the sources we expect to detect in the submm are those at the
earliest stages of protostellar evolution -- starless, Class~0 and
Class~I.

\subsection{Contributions to the SED}

Spectral energy distributions were compiled from fluxes either from
images or catalogues from the following datasets: 2MASS (1-2\micron),
Spitzer IRAC (3.6, 4.5, 5.8, 8.0~\micron), Michelle (11 and
18~\micron), IRAS (12, 25, 60, 100~\micron), SCUBA (450 and
850~\micron, \citetalias{paperI}) and Bolocam (1100~\micron, \citealt{enoch06}).

\subsubsection{SCUBA 850 \micron\ and master catalogue}

SCUBA 850\micron\ fluxes are taken from our SCUBA map
\citep{paperI}.  The JCMT beam size at this wavelength is $14''$.

\begin{jh} The SCUBA chop results in poor sensitivity and noise artefacts in reconstructed maps on angular scales $\geq 100''$ \citep[see][Fig.
  1]{johnstone00a,visser02}.  To avoid false detections and large
  uncertainties on measurements of the extended flux we take out the
  large spatial scales by applying an unsharp mask (USM) filter.  This
  filter removes structure on scales of $>2'$ ($40,000$~AU at
  320~pc).  This was implemented by convolving the map with a $2'$ FWHM
  Gaussian and subtracting the resulting smoothed map from the
  original.  This filtering also selects appropriate spatial scales
  for candidate protostars \citep[eg.][]{motte98}.\end{jh}

To find integrated source fluxes for the 850\micron\ sources we ran
Clumpfind (2D version) \citep{williams94} and Sextractor
\citep{sextractor} on the SCUBA data from Paper~I.  Clumpfind
allocates pixels (and therefore flux) below the highest closed contour
to sources using a friend-of-friends method.  We also used Sextractor
to extract fluxes within a fixed $90''$ aperture.  We ran Clumpfind
with a base level of 3$\sigma$ and increments of 1$\sigma$ ($\sigma =
35\hbox{ mJy/beam}$ at 850\micron\ and 200 mJy/beam at 450\micron) and
a minimum pixel count of 5 ($3''$ pixels).

The catalogue for which we calculate SEDs (the master catalogue)
consists of {\em (a)} all sources with a peak flux above $5\sigma$ at
850\micron\ and a Clumpfind detection from the SCUBA 850\micron\ USM
map; {\em (b)} Clumpfind detections with a peak flux above $3\sigma$
at 850\micron\ and either an identification in Paper~I, or a Bolocam
1.1mm detection with Clumpfind (Enoch et al.~2005).  We additionally
included the well known Class~I source IRAS~03410+3152 which is a
$3\sigma$ 850\micron\ detection with Clumpfind in this category.  The
requirement for a Clumpfind detection means that we can allocate
850\micron\ source integrated fluxes to every source presented here in
a consistent way.  The $5\sigma$ detection limit provides a peak flux
limited catalogue.  Our aim in the wider compilation including the
$3\sigma$ sources with multiple detections is to include as many real
sources as possible while minimising the number of artefacts.

The requirement of 5 pixels above $3\sigma$ with a $5\sigma$ peak
should include any true point source with a peak above $5\sigma$ while
ruling out noise spikes.  The corresponding point source mass limit at
850\micron\ is 0.5~\Msun\ at 10~K.  The maximum
mass within one beam of a starless core which escapes our detection
limit is therefore 0.5~\Msun (see Sect.~\ref{sect:masses} for
the details of the mass calculation).  The masses calculated from flux
within the Clumpfind area do not take into account low level flux $<
3\sigma$ outside this area.  Where the integrated flux is less than
the peak flux per beam, the peak flux per beam is used to calculate
masses.  This affects 12 low mass sources, mainly Bolocam detections
with an 850\micron\ peak $< 5\sigma$.  The lower requirement of 5
pixels above $3\sigma$ with a peak of at least $3\sigma$ corresponds
to a mass of 0.3~\Msun\ at 10K.

Several further sources which we identified by eye in
\citetalias{paperI} were also not identified by Clumpfind as separate
sources but merged into one clump.  Sextractor $90''$ aperture fluxes
are also included in the SEDs for comparison with the Clumpfind fluxes
to give an idea visually of the uncertainties due to the differing
clump extraction methods.  The \citetalias{paperI} source numbers (in
some cases multiple) corresponding to each Clumpfind source are listed
in Table~\ref{tbl:crossrefs} along with the cross-identification from
the 2MASS and Bolocam catalogues and other known names for each
source.  The source positions listed in Table~\ref{tbl:crossrefs} and
used for cross-identification with sources at other wavelengths are
the positions of peak 850\micron\ flux.  Pointing with SCUBA is
typically good to within $3''$.

% no, these now separated
% These included the well known sources NGC1333
% IRAS~4A/B (41/42) and L1448~N/NW (27/28) were not separated by
% Clumpfind at 850\micron.  We modified the catalogue to separate these
% pairs of sources and allocated each half of the integrated 850\micron\
% flux to each. 

\subsubsection{SCUBA 450\micron }

SCUBA 450\micron\ integrated fluxes (beam size $8''$) were also taken from our SCUBA maps \citepalias{paperI}.  As for the 850\micron\ data, 2D Clumpfind was run on the
unsharp masked map.  Contours were set to $3\sigma$, $3\sigma +
2^n\times 1\sigma$ with $n=0\hbox{--}6$ and $\sigma = 200 \hbox{mJy/beam}$.  For many of the 850\micron\ sources, no corresponding 450\micron\ peak was identified largely due to the higher relative noise level on the 450\micron\ map.  The 450\micron\ fluxes are generally integrated over a smaller area than the 850\micron\ fluxes due to the higher noise level at this shorter wavelength.

\subsubsection{Bolocam}

Bolocam 1.1mm fluxes were taken from \citet{enoch06} where Bolocam
peak positions were within $20''$ of the 850\micron\ peak.  The large
offset allowed takes account of the comparatively large beam of
Bolocam ($30''$ compared to SCUBA's $14''$).  The median offset
between SCUBA and Bolocam peaks was $9''$.

\subsubsection{IRAS}

Upper limits on the IRAS fluxes were taken from HIRES processed maps.
The HIRES processing was carried out with 120 iterations at 60 and
100\micron\ and 40 at 12 and 25\micron, following the method used
successfully on the L1448 cluster \citep{wolfchase00}.  The resulting
beams are typically 50--$90''$ at 100\micron\ and 25--$60''$ at
12\micron.  Despite the HIRES processing many of the sources are
still confused.  As the 100\micron\ beam sizes are usually large compared to the
850\micron\ sources, the FIR flux for each submm core falls within one
HIRES beam, and therefore the HIRES peak flux per beam for each source
can be used as an estimate of the total flux.  There may also be a
contribution from nearby neighbours where sources are tightly packed,
and for that reason the IRAS HIRES fluxes are marked as upper limits
on the SEDs.

\subsubsection{Michelle on UKIRT}

Observations were carried out using MICHELLE on UKIRT\footnote{The
  United Kingdom Infrared Telescope (UKIRT) is operated by the Joint
  Astronomy Centre on behalf of the U.K. Particle Physics and
  Astronomy Research Council.} on the nights of 14 September to 21
September 2002.  The instrument has a field of view of $67''$ by
$50''$ with $0.21''$ pixels.  Using 11.6$\mu$m\ and 18.5$\mu$m
narrowband filters contiguous regions were imaged to cover the
brighter SCUBA submillimetre peaks, concentrating on the regions known
to contain clusters: the IC348(HH211), L1448, L1455 and NGC1333
regions.  During the observations the seeing had a FWHM of
$\sim0.7''$ at 11.6$\mu$m and $\sim1''$ at 18.5$\mu$m.  Source fluxes
were measured in $2''$ or $2.5''$ aperatures and the typical 1-sigma noise
levels reached in the observations were 30~mJy at 11.6$\mu$m and
between 90~mJy and 200~mJy at 18.5$\mu$m.

\subsubsection{Spitzer IRAC}

Spitzer IRAC maps at 3.6, 4.5, 5.8, 8.0~\micron\ observed as part of
the Cores to Disks legacy project \citep{c2d} were extracted from the
archive using Leopard.  Source positions and fluxes were extracted
using Sextractor \citep{sextractor} and source detection was carried out
on the median map.  Fluxes were measured within isophotal areas down
to a threshold of twice the local background, which was typically
0.1--0.2~MJy~Sr$^{-1}$ at 3.6~\micron\ and 6--15~MJy~Sr$^{-1}$ at
8~\micron.  Spitzer sources were considered to be associated with
SCUBA sources where they lay within $12''$ or the Clumpfind source FWHM,
whichever was the larger, of the peak.  No preselection was made on Spitzer colours, unlike \citet{jorgensen06a}.
%(Where apertures were used the apertures were $5\times 1.2'' = 6''$).

\subsubsection{2MASS}

JHK fluxes were taken from the 2MASS point source catalogue for
sources with rising (red) spectral indices $H-K>0.8$ which lay within
$12''$ or the submm source FWHM (as determined by Clumpfind),
whichever was the larger, of the SCUBA 850\micron\ peak.

\subsection{Results}
\label{sect:results}

SEDs for each source are shown in Fig.~\ref{fig:seds}.  Every SED
includes the 850\micron\ flux from Clumpfind, a Bolocam 1100\micron\
flux or limit, IRAS HIRES upper limits, and Spitzer IRAC and 2MASS
fluxes or upper limits.  Additionally, an 850\micron\ flux in a $90''$
aperture is shown for the majority of sources (those identified by
SEXTRACTOR), a 450\micron\ integrated flux for sources identified by
Clumpfind on the 450\micron\ map, and Michelle 11.6 and 18.5\micron\
fluxes where available.  For a few well-known sources further flux
constraints are available in the literature but we have not included
these as our aim was to compile as homogeneous a dataset as possible.
The differing flux extraction methods, in particular the necessarily
different areas for flux measurement, result in some mismatches in the
SEDs.  \begin{jh}In a very few cases there are apparent mismatches between a
submm source and an infrared source with falling spectrum (sources 12, 13, 58, 62, 65, 70).\end{jh}

\subsubsection{\Tbol\ and \Lbol}

For each source we calculated the bolometric temperature \Tbol\ and
bolometric luminosity \Lbol\ using a piecewise integration between the
flux measurements following \citet{myersladd93}.  Where there is no
Spitzer, Michelle or 2MASS detection, both \Tbol\ and \Lbol\ are taken
as upper limits to reflect the fact that the HIRES fluxes are upper
limits.  The bolometric temperature \Tbol\ is robust to calibration
uncertainties or missing fluxes with even a factor 2 change between
submm and MIR wavebands changing \Tbol\ by only a few kelvin.  The
bolometric luminosities are set largely by the far-infrared limits set
by IRAS HIRES, which we estimate in uncrowded regions are good to 30\%
but in crowded regions may lead to overestimates of the luminosity due
to flux contributions from neighbouring sources.

\subsubsection{Masses}
\label{sect:masses}

We calculate envelope masses based on a constant dust temperature of
$T_\mathrm{d} = 10$~K and a 850\micron\ dust opacity of
0.012~cm$^2$~g$^{-1}$.  The temperature of 10~K was required by the
constraints of the SCUBA 850\micron\ and IRAS HIRES 100\micron\ fluxes
on model greybodies (assuming \citet{oh94} icy coagulated dust type
5).  Even at 10~K 13\% of the 100~micron upper limits fall below the
prediction of the greybody.  Greybodies at $T_\mathrm{d} = 10$~K are
plotted on the SEDs in Fig.\ref{fig:seds}; note that the 850~micron
aperture fluxes ($90''$ diameter aperture, solid triangles) are often
lower than the Clumpfind fluxes and the 450~micron fluxes, which also
often fall below the greybody, are calculated over a smaller area.
Higher temperatures, even as high as 12~K, were ruled out for the main
component of mass in most sources.  For comparison, models of Class~0
and Class~I sources \citep{shirley02,young03} suggest that the
isothermal dust temperature with which to calculate representative
masses should be $13.8\pm2.4$~K for Class~0 sources and $16\pm4~K$ for
Class~1 sources, consistent with 10~K.  Temperatures derived from
modelling the cores in Perseus as Bonnor-Ebert spheres \citep{kirk06}
indicate temperatures between 10--19K.  The 10~K to which we are
constrained by the greybodies lies at the low end of the expected
temperature range, at least for the protostellar sources if not the
starless cores.  Thus our SEDs point to detection of a larger fraction
of cold dust than is often measured for protostars, for which mean
dust temperatures of up to 30~K are often assumed when calculating
masses from submm fluxes.

A reason for the detection of more cold dust in our sample could be
that the submm fluxes are taken from the Clumpfind deconvolution
rather than more limited apertures, even though the map was
prefiltered to remove structure on scales above $2'$ before running
Clumpfind.  IRAS HIRES fluxes also provide tighter upper limits on the
source fluxes than the IRAS PSC by reducing the contribution from
nearby emission. However, the Clumpfind components are no bigger than
the IRAS HIRES beam at 100\micron\ (typically $80''$) so the SED
should give a fair representation of the clump.  Considering all the
uncertainties in calculating masses, we find that the true masses
could be lower than our estimates but are unlikely to be higher.
Firstly this is a result of the low $T_d$ compared to studies which
assume higher temperatures - a factor of 3 compared to 15~K or 6
compared to 30~K.  Secondly, we assume 320~pc whereas 250~pc is also
an arguable distance for Perseus \citep{cernisstraizys03}, which would
reduce masses by a factor 1.9.  Note, though, that our masses are
consistent with the Bolocam study of Perseus \citep{enoch06}, which
also assumed 10~K, once their lower assumed distance is taken into
account.  Our dust opacity of 0.012~cm$^2$~g$^{-1}$, taken from
\citet{oh94} assuming a gas/dust ratio of 161 by mass, is also at the
low end of the range.  Some authors prefer 0.02~cm$^2$~g$^{-1}$ at
850~\micron, which also would reduce masses by a factor 1.7.  In
summary, with a different choice of factors and fluxes measured in a
more limited aperture, masses could be up to a factor of 10 lower than
these we give.

The results for masses along with $T_\mathrm{bol}$ and
$L_\mathrm{bol}$ are given in Table~\ref{tbl:sourceproperties}.

\section{Classification}
\label{sect:classification}

\begin{figure*}[t]
\centering
\includegraphics[scale=1.20,angle=-90]{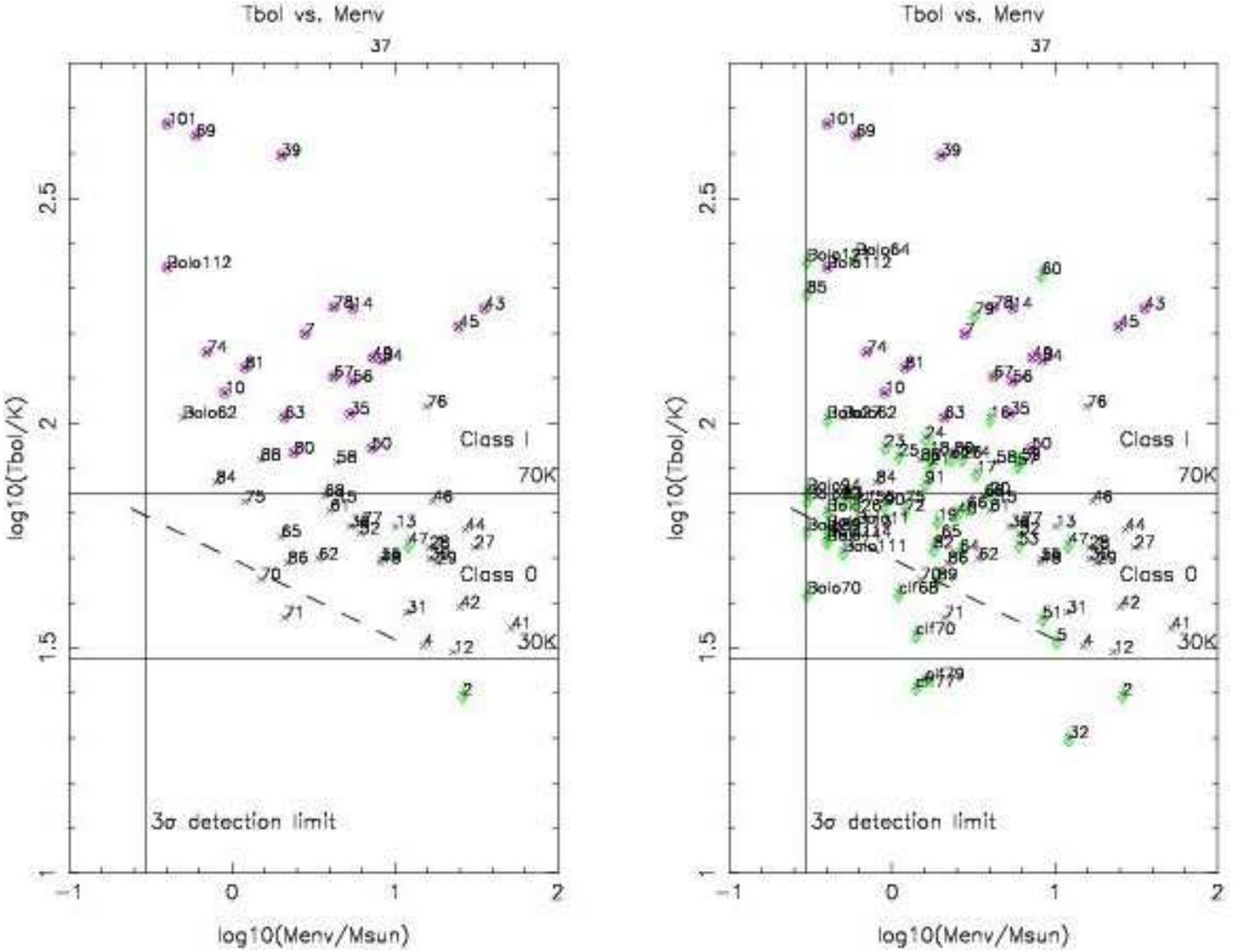}\\
\caption{Evolutionary indicators vs. envelope mass: \Tbol vs.\Menv.  {\bf Left:} confirmed protostars only.  {\bf Right:} all sources.  Upper horizontal line: Class~0/Class~I boundary at \Tbol = 70~K \citep{chen95}.  Lower horizontal line: 30~K.  Solid vertical line: SCUBA detection limit.  Dashed diagonal line: Early Class~0 models (Sect.~\ref{sect:radtrans}).  Class~I sources (see Sect.~\ref{sect:class01}) are marked by circles (magenta). }
\label{fig:menv_tbol}
\end{figure*}

The shape of the SED only depends on the dust distribution relative to
the heating source and this depends on many parameters -- envelope
mass, which determines the overall optical depth
\citep{ivezicelitzur97}; clumpiness \citep{indebetouw06}; orientation
of any outflow cavity \citep{whitney03}; and location of heating
source on the near or far side of the clump.  However, there are well
known general trends with evolutionary stage, as the overall envelope
mass reduces and the size of the outflow cavity increases
\citep{whitney03,arcesargent06}.  Bearing the caveats in mind, we
apply a number of evolutionary indicators to classify of our objects
based on their SEDs.  As we are largely interested in statistics,
individual misclassifications due to an edge-on or pole-on orientation
are less important as to some extent they cancel out across the
population as a whole.  What is important is that we use a consistent
set of measurements across our sample, and therefore can directly
compare between sources.

% For our SCUBA sources, we are unable to estimate the mass of the
% central star and, for sources unresolved at many wavelengths, rely on
% the spectral energy distribution (SED) for classification.  We have to
% bear in mind that t

We assume our sources fall into the usual classification scheme
of starless core, Class~0 or Class~I protostar.  Starless cores are
those that do not contain a hydrostatic core (though some starless
cores are protostellar cores which will go on to form one);
observationally they are cold and show no evidence for embedded
infrared, radio or outflow sources.  The boundary between the two
protostellar classes, Class~0 and I, is defined by the point in
evolution when the mass of the envelope equals the stellar mass.
However, we have no way of measuring the stellar mass and instead must
rely on criteria from modelling and/or observation to separate Class~0
and Class~I sources based on evolutionary indicators derived from observables.

The first two evolutionary indicators we use are the standard ones:
bolometric temperature \Tbol\ and the ratio of submm to bolometric
luminosity, \Lbol/\Lsmm.  \Tbol\ is plotted against envelope mass
\Menv\ in Fig.~\ref{fig:menv_tbol}.  $\Tbol = 70~K$ is used to draw a
line between Class~0 and Class~I embedded protostars \citep{chen95}.
As our \Tbol\ are often upper limits where there are no near- or
mid-infrared detections, we have plotted the identified protostars and
the sources with upper limits separately (the SCUBA 850\micron\ 
detection limits are also shown).  The highest mass sources are cold
Class~0 sources, which we would expect to evolve to less massive and
warmer objects.

\begin{figure*}[t]
\centering
\includegraphics[scale=1.0,angle=-90]{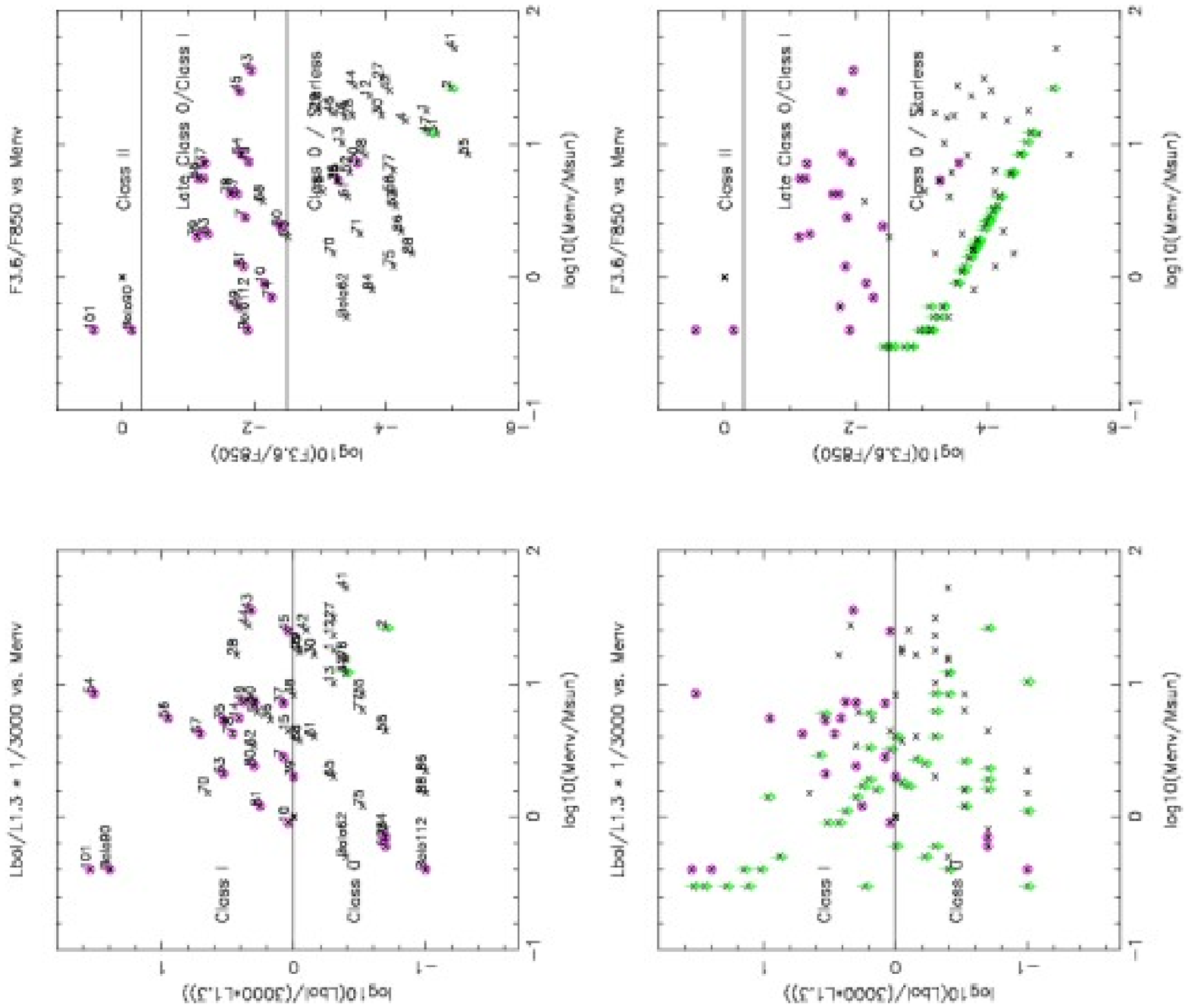}\\
\caption{Further evolutionary indicators vs. envelope mass.  {\bf Top:} confirmed protostars only.  {\bf Bottom:} all sources.  {\bf Left:} \Lbol/\Lsmm, converting $L_{850}$ to $L_{1.3}$ \citep{awb93}, {\bf Right}: $F_{3.6}/F_{850}$ with classification based on the models of \citet{whitney03}.  The boundaries between classes are shown at representative values but in fact there is overlap between classes because of the effects of orientation.  Class~I sources (on the basis of 2/3 indicators) are marked as circles (magenta) and luminosity/flux upper limits are given for sources with no Spitzer IRAC detection.}
\label{fig:evo}
\end{figure*}

The ratio of submm to bolometric luminosity is plotted in
Fig.~\ref{fig:evo}.  For the submm luminosity we estimate
$L_\mathrm{1.3}$ from the 850\micron\ integrated flux, which is
available for all our sources, measured in a bandwidth of 50~GHz and
corrected to 1.3~mm using $F_{1.3} = F_{850}\times(850/1300)^{3.5}$
assuming that dust opacity \begin{jh} is optically thin \end{jh} and
follows a power law with index $\beta = 1.5$.  The boundary between
Class~0 and Class~1 sources was calculated by \citet{awb93} to lie at
$\Lbol/L_{1.3} = 20,000$, but in order to exclude well known Class~I
sources such as L1455~IRS2 we are required to lower it to
$\Lbol/L_{1.3} = 3,000$.  The upper limits from IRAS introduce some
uncertainty in calculating \Lbol, and it may be that the differing
analysis of the IRAS fluxes here and in \citet{awb93} is responsible
for some the difference.  However,a reduction of a factor of 4 in
$\Lbol/L_{1.3}$ for the Class~0/Class~I boundary was also required by
\citet{visser02} using a similar dataset and it seems likely that the
\citet{awb93} limit was simply set too high on the basis of the
Class~I sources known at the time.  The \citet{awb93} limit was
consistent with a simple model with a single accretion rate which may
not apply in all cases: the limit should increase to account for
increased accretion rates at higher mass with models suggesting that
the boundary lies somewhere between $\Lbol \propto \Menv^{1.0}$
(alternatively, $\propto \Lsmm^{1.0}$ when \Menv\ is calculated with a
fixed $T_\mathrm{d}$ as here) and $\Lbol \propto \Menv^{2.0}\propto
\Lsmm^{2.0}$ \citep{smith00}.  Note that as envelope masses and submm
luminosities are both proportional to the submm flux, this indicator
is equivalent to plotting $\Lbol$ vs. $\Menv$ (eg.
\citet{andremontmerle94}).

A third measurement linked to the shape of the SED is the ratio of
Spitzer IRAC to submm emission.  This has the advantage of being
reliable even when the IRAS data are confused.  In Fig.\ref{fig:evo}
we also plot the 3.6/850~\micron\ flux ratio vs. the 850~\micron\ flux
together with rough ranges for this ratio based on the evolutionary
models of \citet{whitney03}, as a third evolutionary indicator.

\begin{jh}
Statistics (mean and standard deviation) for the three evolutionary indicators are given in Table~\ref{tbl:indicatorstats}.  The mean value of $\log_{10}\Tbol$ corresponds to 78~K for the total population in Perseus and 91~K for the 56 confirmed protostars (excluding the sources with upper limits on \Tbol; see Sect.\ref{sect:starlessandproto}). As a temperature of 70~K is used to divide the Class~0 and Class~I populations (see Sect.~\ref{sect:class01}) these low values indicate already that the protostellar population contains a significant fraction of deeply embedded sources.  The mean values of the other indicators ($\log_{10}(\Lbol/L_{1.3})$, $\log_{10}(F_{3.6}/F_{850})$) also support this.
\end{jh}

\begin{table}[t]
\caption{Mean and standard deviation for each of the evolutionary indicators $\log_{10}(\Tbol/K)$, $\log_{10}(\Lbol/\Lsmm)$, $\log_{10}(F_{3.6}/F_{850})$ for (top) confirmed protostars; (bottom) all sources, as plotted in Fig.~\ref{fig:evo}}   
\centering
\begin{tabular}{l c c}
\hline\hline
{\bf Confirmed protostars (56)}          &Mean      &Standard deviation\\
\hline
$\log_{10}(T_\mathrm{bol}/K)$            &1.96 (91K)    &0.33\\   
$\log_{10}(\Lbol/(3000\times \Lsmm))$ &-0.0256     &0.617   \\
\medskip
$\log_{10}(F_{3.6}/F_{850})$            &-2.92 (0.012)  &1.24  \\
{\bf All sources (103)}          &Mean      &Standard deviation\\
\hline
$\log_{10}(T_\mathrm{bol}/K)$            &1.89 (78K)  &1.04 \\  
$\log_{10}(\Lbol/(3000\times \Lsmm))$ &-0.0431    &0.691 \\  
$\log_{10}(F_{3.6}/F_{850})$            &-3.28  &1.07 \\      
\hline
\end{tabular}

\label{tbl:indicatorstats}
\end{table}

\subsection{Protostars and starless cores}
\label{sect:starlessandproto}

Near- or mid-infrared detections by Spitzer and/or 2MASS identify
54/103 sources ($52\pm7$\%) as definitely protostellar.  An additional
two sources are identified \begin{jh}as protostellar\end{jh} on the
basis of molecular outflows by our $^{12}$CO outflow survey of a
subset of this sample \citep[ Paper~III]{outflows}.  These sources are
b1-bN/S (2) and NGC1333 SK31 (47).  These were not detected by
Spitzer, in the case of b1-bN/S apparently because the driving sources
are still too deeply embedded, in the case of NGC1333~SK31 (46)
because any associated infrared source is masked by the scattered
light from the bright neighbour NGC1333~ASR114 (45).  This brings the
total of confirmed protostars in Perseus up to 56.  A number of
sources are first identified as protostellar by this survey, namely
numbers 14 and 15 (near IC348), 36 (SE of L1448), and 58, 60 and 61
(in NGC1333).

% In fact, as our outflow
% survey was biased towards higher mass cores and likely protostars, the
% true fraction of protostars undetected by IRAC is likely to be larger.
% As mass decreases, a Spitzer detection becomes less likely due to the
% overall lower luminosity -- SCUBA is much more sensitive than Spitzer
% to Class~0 sources.

The remaining $47$ sources have no positive detection of a
protostellar indicator (IR, outflow) and therefore could be starless.
Further infrared, radio or outflow observations are needed to detect
any further protostars in this remaining sample, particularly at the
low end of the mass range where any IR emission is \begin{jh}likely to
  be\end{jh} faint and outflow masses low.  Whereas the two
luminosity-based indicators rely on IRAS upper limits and are not
useful for the classification of the non-detections, $F_{3.6}/F_{850}$
on the other hand clearly places all sources with IR limits in either
the Class~0 or starless category (Fig.~\ref{fig:evo}, third panel).

\begin{jh}
  
  Table~\ref{tbl:prepro} gives the counts of starless and protostellar
  cores for each subregion of Perseus.  The percentage of starless
  cores varies between regions from 17 to 68\% compared to the overall
  mean of 46\%.  To investigate whether this variation is significant,
  we calculate the probability of these percentages occuring by chance
  assuming the sources are drawn at random from a population which is
  46\% starless.  The resulting probabilities, calculated from the
  binomial distribution, are given in the last column of
  Table~\ref{tbl:prepro}.  The small clusters (L1448, L1455, B1)
  contain small numbers each and are consistent with the cloud
  average.  NGC1333, which has a high fraction of protostars, and
  IC348 and the distributed population of isolated cores, which
  contain larger fractions of apparently starless cores, all show
  significant variation from the cloud average.  In each of these
  cases the probability of being consistent with a 46\% starless
  fraction is less than $3\%$.  This suggests that in individual
  regions, there is significant variation in the star formation rate
  on timescales of order the lifetime of a starless or protostellar
  core.  Alternatively, it is possible that the starless population in
  NGC1333 is underestimated as starless cores are hard to identify in
  richly clustered regions, and this may be biasing the cloud
  average to a low fraction of starless cores.  Excluding NGC1333, the
  percentage of starless cores rises to 55\% in the rest of the cloud,
  from which the IC348 and the isolated population are less
  significantly different.
  
  Comparing the number of submm cores in IC348 and NGC1333 to their
  PMS populations of 348 and 213 stars, respectively
  (Sect.\ref{sect:lifetimes}), shows that IC348 is three times as rich
  in submm cores as NGC1333.  Spitzer observations
  \citep{jorgensen06a} also show that the ratio of Class~II$+$III to
  Class~I sources is higher in IC348 than other parts of the cloud.
  The reason for this is that IC348 has been forming stars for longer
  ($>3$~Myr, \citealt{luhman03,muench03}) and has had time to build up
  a larger fraction of Class~II/III sources.  IC348 has certainly been
  active for longer than NGC1333 and also, possibly, than the
  remainder of the cloud, though it is harder to be definite on this
  because the Spitzer observations do not rule out a more spatially
  extended distributed population of young PMS stars, as recently
  discovered in Taurus \citep{slesnick06}.  Spitzer area selection on
  the basis of high $A_v$ biases towards younger sources still
  associated with molecular cloud material.
 
  We find 53\% of submm cores and 57\% of protostars in large clusters
  (IC348 and NGC1333).  The fraction for protostars confirms our
  pre-classification result, based on the submm cores alone, of at
  least 40\% of star formation large clusters \citepalias{paperI}.
  These fractions are also consistent with the $>50\%$ star formation
  in large clusters found by 2MASS for Perseus/Orion/Monoceros
  \citep{carpenter00} and $\sim 60\%$ by Spitzer for Orion/Ophiuchus
  \citep{allen06}.  \citet{jorgensen06a} find for the youngest
  Spitzer-selected Class~I sources in Perseus that a somewhat smaller
  fraction of 39\% lie in clusters.  The reason for the discrepancy
  with our results is that not all the Spitzer sources associated with
  submm peaks satisfy the c2d colour criteria.  \citet{jorgensen06b}
  find that positional association with MIPS sources identifies more
  sources in clusters as protostars than IRAC colours alone.  Taken
  together the results are all consistent with 40\% of current star
  formation in Perseus taking place in the distributed population of
  isolated stars and small ($< 100$ member) clusters.

\begin{table}
\label{tbl:prepro}
\caption{Counts of starless and protostellar cores by region.  The last column gives the probability (from the Binomial distribution) of the counts occuring by chance assuming the sources are drawn randomly from a population with a starless fraction of $47/103$.}
\centering
\begin{tabular}{l c c c c}
\hline\hline
Source    &Starless    &Protostellar    &\% starless  &Prob.\\    
\hline
NGC1333   &10          &26       &28\%  &0.013       \\ %36/213 = 0.17
IC348$^{1}$ &13         &6       &68\%  &0.026       \\ %19/348 = 0.05
L1448      &1          &4        &20\%  &0.20       \\ 
L1455     &2           &5        &29\%  &0.21       \\
B1        &1           &5        &17\%  &0.13       \\ 
Isolated  &20          &10       &67\%  &0.016       \\
\hline
Total     &47          &56       &46\%\\                         
\hline
\multicolumn{5}{l}{$^{1}$ Including the filament to the S of IC348 which contains HH211.}
\end{tabular}
\end{table}

\end{jh}

% On the
% basis of the discussion of minimum \Tbol\ given \Menv\, we estimate
% about a third of sources are too cold to contain protostars given
% their mass (see Sect.~\ref{sect:radtrans}) and must be starless, but this remains to be confirmed by further observations.
% 

\begin{jh}From the definite detections, the ratio of protostars to starless
cores is at least $47/56=0.8$.  Previous estimates for the
starless/protostellar ratio are higher, ranging from 2--3
\citet{visser02,leemyers99} to $10$ \citep{wt94,jessopwt00}.  This
would appear to suggest that our survey is incomplete to prestellar
cores, although it is difficult to believe that we underestimate the
number of starless cores by the factor of 12 necessary to reach the
ratio measured by \citet{jessopwt00}.  

Assuming that the star formation rate in Perseus is constant, we
estimate that the average lifetime of starless cores above our
detection limit is $0.8$ times the lifetime of the embedded
protostars.  If more protostars are later identified among the
apparently starless population, the relative lifetime of detectable
starless cores will be reduced.  On the other hand, prestellar cores
begin their life below our detection limit so the full timescale for
prestellar development will be longer than the time over which
\begin{jh}we could detect them in these observations\end{jh}.
Starless cores with a flattened central density distribution which
peaks below our mass per beam sensitivity limit could still contain
considerable (distributed) mass and remain undetected by our survey.
We conclude that prestellar timescales are roughly equal to or longer
than protostellar lifetimes.

\end{jh}

% and 31 (L1448 filament) but I don't believe it.

\subsection{Class~0 and Class~I sources}
\label{sect:class01}

Based on the three evolutionary indicators in Fig.~\ref{fig:evo}
(\Tbol, \Lbol/$L_{1.3}$, $F_{3.6}/F_{850}$), 16/103 sources are
certain Class~I objects (see Table~\ref{tbl:sourceproperties}).  A
further six are likely Class~Is based on two out of three evolutionary
indicators bringing the total number of Class~Is to 22.  More
spectral/spatial information is required in order to determine the
true nature of a further five sources (58,76,84,88, and Bolo62), which
are classified Class~I on \Tbol\ but are classified as Class~0 on the
other two indicators.  For these sources, the temperature is high but
the estimated luminosity is too low to be consistent with a Class~I
identification.  We assume these are Class~0.  The classifications
based on each evolutionary indicator, and the majority vote, are given
in Table~\ref{tbl:sourceproperties} and the spatial distribution of the
sources is shown in Fig.\ref{fig:class01}.

%(Source 88, which has high \Tbol\ but low
%\Lbol\, is an interloper -- a Herbig-Haro object where the IRAC
%luminosity comes from shocked H$_2$ emission.)

% Definite Class Is: 7,10,14,37,39,43,45,49,54,56,63,67,78,80,81,101,Bolo90
% Possible Class Is: 35, 50, 69, 74
% Unlikely Class Is (luminosity low): 58,76,84,88, Bolo62,Bolo112

Thus, of the 56 confirmed SCUBA-detected protostars, the number of
Class~Is is only 22 out of 56 (39\%).  The remaining 34
confirmed protostars must be Class~0.  On the basis of these counts,
SCUBA-detected Class~0s are more common than Class~Is in Perseus.

That raises the question of how many Class~I sources we fail to detect
with SCUBA because they fall below our detection limit.  We can only
detect 10~K envelope masses of 0.3~\Msun, and are complete above
0.5~\Msun.  Class~I sources begin by definition with an envelope mass
equal to the stellar mass and the envelope mass subsequently reduces.
For Class~Is forming stars below about 0.6~\Msun, therefore, we would
not expect to detect the envelope for more than half of the Class~I
lifetime.

We can, however, estimate the number of Class~Is which we fail to
detect with SCUBA as Spitzer is more sensitive to Class~I sources.
\citet{jorgensen06a} make an independent estimate of the number of
Class~I sources in Perseus from the Spitzer survey, which covers a
similar area as SCUBA, on the basis of their infrared spectral index
in the Spitzer IRAC and 2MASS bands.  As the infrared spectral index
classification \citep{greene94} does not include a separate category
for Class~0 sources (as at that time they were not detected in the
infrared) both Class~0 and Class~I sources are included in the
\citet{jorgensen06a} Class~I count of 54.  Of their sources, 14 lie
outside the SCUBA map boundary and 29 coincide with SCUBA-detected
protostars.  The remaining 11 sources in the \citet{jorgensen06a}
Class~I list lie within the SCUBA map but were not detected in the
submillimetre.  These could be Class~I sources with low envelope
masses below the SCUBA detection limit.  Including these Spitzer
detections, the total number of Class~I sources in our mapped area
could therefore be as high as $22+11=33$, bringing the total number of
protostars in Perseus up to $56+11=67$.  The 34 confirmed Class~0
sources therefore still make up more than half of the total number of
protostars.  This fraction may be even higher if more of the IR
non-detections later turn out to be protostellar through
identification of molecular outflows or radio sources, as these must
also be Class~0: the limits on the infrared emission rules out
Class~Is on the basis of $F_{3.6}/F_{850}$ (Fig.\ref{fig:evo}).

% Based on \Tbol, 68 sources are either Class 0 or starless, of which 22
% are confirmed Class 0 by the presence of outflows and/or IR sources.  20
% sources are definite Class Is.  A further 16 sources have an upper
% limit on \Tbol which could place them in the Class~I category.

% On the basis of the 54 sources observed in $^{12}$CO \citepalias{outflows},
% we would expect an overall outflow fraction of 50--80\% (55-85 outflow
% sources) and 26\% of all sources (27 sources) to be Class~Is, assuming
% the outflow sample to be a random selection.  This is consistent with
% the 20 definite and 16 possible Class~Is identified by SED.

%\subsubsection{Why are there so many Class~0 sources in Perseus?}

The ratio of Class~0s to Class~Is in Perseus is quite unlike
Ophiuchus and Taurus clouds where Class~I sources outnumber
Class~0s by 10:1 \citep{andremontmerle94,motteandre01} and exceeds
even Serpens where at least 25\% of sources are Class~0 (5 Class 0s
identified by \citet{hurtbarsony96} compared to 19 Class Is from
\citet{kaas04}).  Although it is possible that there are some further
undetected Class~I sources which were not identified by
\citet{jorgensen06a} it is hard to imagine (given the sensitivity of
the Spitzer c2d survey) that these outnumber the detections by 10:1.

% Rephrasing of what was said above.
% However, our source selection is flux limited and we become
% increasingly insensitive to Class~I sources as their envelope mass
% decreases.  We are complete for Class~0 sources above ~ 0.2~Msun
% (assuming 20~K).  As sources increase their Tbol during the Class~0
% and I phases, we are able to detect lower masses (to a limit of
% roughly $0.4\times (12~\hbox{K}/T)$~\Msun) but ultimately the decrease
% in envelope mass is sufficient to push every protostar below our
% detection threshold.  For lower mass envelopes this occurs sooner in
% the embedded lifetime than for higher mass envelopes which start
% further above the threshold.  Therefore we know we are missing a
% fraction of Class~I sources relative to Class~0 sources.  How many
% Class~I sources with low envelope masses are we missing?

One explanation for the large numbers of Class~0 sources in Perseus
could be that Perseus has undergone a recent burst of star formation
-- perhaps triggered by 40 Persei, a B0.5 star in Per~OB2
\citep{walawender05,kirk06} -- and this has produced a large number of
protostars which are still in the Class~0 phase.  Perseus has been
forming stars from at least 3--4~Myr ago IC348 until the present day
\citep{luhman03,muench03}. \begin{jh} Therefore \end{jh} multiple
triggering events would be required to explain both the earlier
formation of IC~348 and the star formation in the last
few $\times 0.1$~Myr, which would be necessary to explain a current
population of Class~0s.  The average star formation rate in
Perseus is consistent with steady star formation and a protostellar
core lifetime of a few~$\times 10^5$~years, in agreement with previous
studies (see Sect.~\ref{sect:lifetimes}, \citetalias{paperI}).

%We therefore think that
%triggering is an unlikely explanation for the large Class~0
%population.

\begin{figure}
\centering
\includegraphics[scale=1.0,angle=-90]{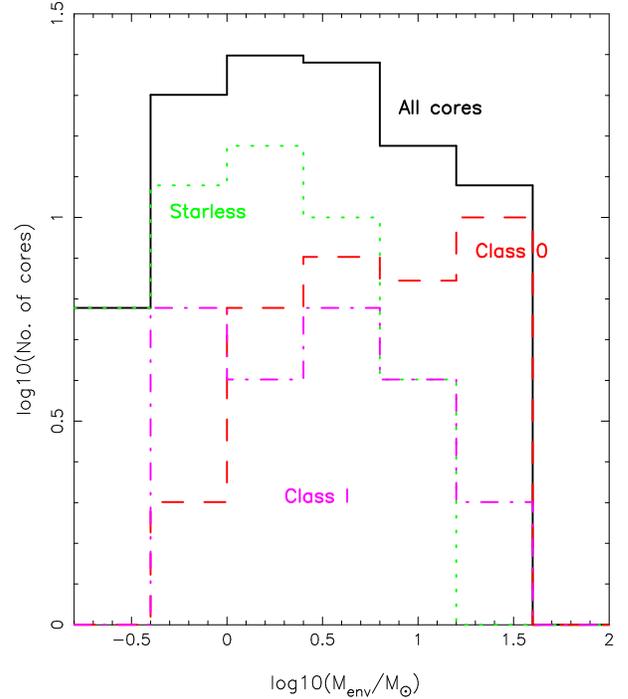}\\
\caption{Histogram of masses for all sources (solid line), Class~Is (dot-dash), Class~0s (dashed) and starless cores (dotted line).}
\label{fig:massdist}
\end{figure}

An obvious difference between Ophiuchus and Taurus on the one hand and
Perseus, Serpens and Orion on the other is that \begin{jh} the
  protostellar envelopes are more massive in Serpens, Perseus and
  Orion.  Roughly similar flux sensitivity limits mean observations
  trace the higher-mass end of the population in these more distant
  clouds (Perseus 320~pc, Serpens 259~pc, Orion 400~pc compared to
  Ophiuchus and Taurus, both at $\sim 150$~pc)\end{jh}.  The Class~0
sources in Serpens have envelope masses above 0.6~\Msun\ 
\citep{hurtbarsony96}; the Perseus~Class~0s all have masses above
0.5~\Msun (see Fig.\ref{fig:massdist}).  In fact, no Class~0 sources
with envelope masses less than $\sim 0.2~\Msun$ are found in any of
these clouds, or indeed in the recent
compilation of known Class~0 sources \citep{froebrich05}.  This is
very clearly seen in Fig.~2 of \citet{batc96}, and is partly due
selection effects: targetted submm surveys were largely based on known
infrared and radio sources and therefore only picked up Class~Is and a
limited number of high mass (luminous) Class~0s.  Our SCUBA survey, on
the other hand, is more sensitive to Class~0s than starless cores or
Class~I sources, though we may be missing the equivalents of many of
the lower-mass objects in Rho~Oph and Taurus in Perseus.  Note that the majority of Perseus Class~0 sources would
still have masses above 0.2~\Msun\ even if the masses were reduced by
the maximum uncertainty factor of 10.

% Selection effects also affect our survey: although SCUBA detects submm
% cores down to low masses, identification of protostars still relies on
% infrared or outflow detection, both of which depend on source
% luminosity.  As discussed in Sect.~\ref{sect:lvsm}, this may have an
% effect on the number of Class~0 sources detected, because more massive
% protostars can be identified relatively early during the Class~0 phase
% and are thus visible for longer.  The evidence we have suggests that
% Class~0 sources evolve quickly to near the Class~0/Class~I boundary
% and therefore this effect is small, but it needs to be tested by more
% sensitive infrared or outflow observations of the apparently starless
% sources.  

The higher mass of the Perseus population could be directly
responsible for longer Class~0 lifetimes, if the evolution of
protostars depends on mass in a way which favours long Class~0
lifetimes for higher mass objects.  Such a variation is predicted, for
example, by gravoturbulent fragmentation \citep{schmejaklessen04}.  It
is also a consequence of the competitive accretion model
\citep{bonnell01,bbb03,bonnell06}, as higher mass cores continue to
acquire material from the surrounding cloud during the Class~0 phase
and only embark on the Class~I loss of envelope mass when the
reservoir of material has run out.

% This is supported
% by smaller ratios of outflow mass loss rate to envelope mass for more
% massive envelopes, pointing to a longer envelope destruction timescale
% \citep{outflows}.  The probable decline in the IMF also contributes to
% the absence of Class~0s below $\Menv = 0.2\Msun$.

% Selection effects cannot account for the lack of
% low-mass Class~0s in our SCUBA sample, as we can detect Class~0
% sources at 0.2~Msun if they have temperatures above 15~K, at 0.1~Msun
% above 25~K, and down to 0.03~Msun (below the substellar mass limit) at
% the Class~0/Class~1 \Tbol\ boundary of 70~K.  \jh{But can we identify
%   them as protostars?}

The Class~I/Class~0 classification of protostars is also affected by
mass, because in practice the classification relies on observational
indicators such as $\Tbol$ rather than a direct comparison of the mass
in the envelope and the stellar mass, which cannot be directly
measured.  More massive sources also have greater optical depth in
their envelopes and therefore SEDs which peak at longer wavelengths
for the same dust geometry.  As a source evolves, loses envelope mass
and evacuates a cavity, it emits more infrared radiation relative to
submillimetre, and passes from Class~0 to Class~I on the basis of
\Lsmm/\Lbol\ or \Tbol.  More massive sources, with greater initial
optical depth, have to lose relatively more of their envelope mass to
reach the required \Tbol\ threshold to be classified as Class~I.  The
lower envelope mass sources have higher \Tbol\ and $\Lsmm/\Lbol$
because of lower envelope optical depth and apparently make the
transition from Class~0 to Class~I earlier in their evolution.
Fig.~\ref{fig:massdist}, which shows the mass distribution for the
cores separated by class, shows that the Class~0 sources tend to have
higher masses than the Class~Is.  In order to investigate this effect
further, we model the emission from a Class~0 geometry envelope with a
range of masses in Sect.~\ref{sect:radtrans} and discuss the
implications for the classification of Class~I and Class~0 sources,
but before we can do this we consider the relationship between mass
and luminosity.

\subsection{Relationship between \Lbol\ and \Menv.}
\label{sect:lvsm}

\begin{figure}
\centering
\includegraphics[scale=0.7,angle=-90]{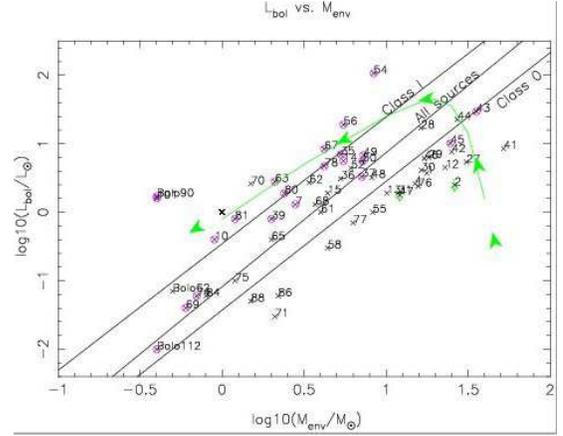}\\
\caption{Bolometric luminosity vs. envelope mass for confirmed protostars.   Class~I sources (see Sect.~\ref{sect:classification}) are marked with circles.  Power law fits to all sources, Class~0s and Class~Is are shown (see text). An evolutionary track from \citet{smith00}, scaled by a factor 30 in mass, is shown for illustration with arrows at $\log_{10}(\hbox{age}) = 3.5,4.0,4.5,5.0,5.5$~yr (in green)}
\label{fig:lvsm}
\end{figure}

The relationship between luminosity and envelope mass is shown in
Fig.~\ref{fig:lvsm}.  Mass is measured from the submm flux assuming a
constant temperature of $10$~K whereas luminosity is measured by
integration across the entire spectrum and thus is most dependent on
the IRAS fluxes.  Note that although a higher luminosity source will
have higher temperature components than a lower luminosity source of
equivalent mass, the bulk of the mass is in cold gas therefore the
constant temperature assumption in the mass calculation still holds.
There is a general trend for higher mass sources to have higher
luminosity.  But there is not a tight relationship between envelope
mass and luminosity.  This scatter is intrinsic, and due to the fact
that we are plotting not just Class~I protostars (as
\citet{reipurth93}) but also Class~0s, which have higher masses for
the same luminosity.  This scatter was also seen in the sample of
\citet{batc96}, and it is this scatter which enabled us to use
$\Lbol/\Lsmm$ which is proportional to $\Lbol/\Menv$ to estimate the
evolutionary state of a core (Sect.~\ref{sect:classification}).

Higher mass accretion rates -- and therefore higher luminosities --
are clearly driven by higher initial envelope masses.  The evolution
of luminosity with mass depends on the details of the accretion
process, and evolutionary tracks for \Lbol-\Menv have been suggested
based on various accretion models \citep[see][ and references therein]{froebrich05}.  A possible scenario is that mass accretion rates
and therefore accretion luminosity initially increase exponentially
and then reduce, resulting in \Menv--\Lbol\ tracks which rise rapidly
in luminosity (up and slightly to the left in Fig.~\ref{fig:lvsm})
during the Class~0 stage and then decrease in both luminosity and mass
(down and to the left in Fig.~\ref{fig:lvsm}).  An example of such an
evolutionary track, from \citet[][Fig.~6]{smith00}, scaled by a factor 30
in mass in order to match the high envelope masses in Perseus, is
plotted in Fig.~\ref{fig:lvsm}.  In this example the luminosity
decrease during the Class~I phase follows $\Lbol \propto \Menv^\alpha$
with power law index $\alpha$ between 1 and 2.  Of course, accretion
rates and luminosity may also vary during both the Class~0 and Class~I
evolutionary stages if accretion is variable, as is known to be the
case for mass ejection in the jet \citep[eg.][]{hartigan05}), so evolutionary tracks are probably not as simple as this model represents.

Luminosity evolution during the Class~0 stage can produce a selection
bias towards high mass Class~0s as they reach luminosities at which
infrared or outflow emission can be detected earlier in their
lifetimes.  Therefore, higher mass Class~0s may be visible as
protostars for longer.  This bias is strong if the luminosity
evolution is slow and weak if it is fast.  If mass accretion rates
rise very rapidly to a maximum during the Class~0 phase
\citep{henriksen97,schmejaklessen04,whitworthwt01,stamatellos05}, then
the Class~0s will spend most of their lifetime close to the Class~I
boundary and the number of Class~0s which remain undetected will be
small.  There is supporting evidence for this fast luminosity
evolution in Fig.~\ref{fig:lvsm} as high mass sources with low
luminosities are absent.  This suggests either that the time taken to
reach high luminosities (roughly, within two orders of magnitude of
the maximum for a given mass) is much shorter than the subsequent
evolution through the Class~0 and Class~I phases, or the proposed
evolutionary track is wrong.  For example, high mass Class~0 sources
could evolve from lower-mass Class~0 sources through competitive accretion.

% A power-law relationship for Class~I sources driving Herbig-Haro flows
% was found by \citet{reipurth93},
% $$
% \Lbol \propto S_{1300}^{1.6}$$
% .  The 1.3~mm flux $S_{1300}$ is
% proportional to envelope mass if a constant temperature is assumed.
% The range of masses increases once \Tbol\ is taken into account -- low
% flux, hot sources extend one end of the distribution to lower mass and
% high flux, cold sources at the other to higher mass and a fit to
% \Tbol-corrected data should yield a flatter gradient than with masses
% calculated at constant temperature.

We fitted a straight line to $\log_{10}(\Lbol)$ vs.
$\log_{10}(\Menv)$, assuming uncertainties in $\log_{10}(\Menv)$ of
0.5 (factor of 3) and $\log_{10}(\Lbol)$ of 0.3 (factor of 2).  For
the entire sample of known protostars, \begin{jh}
$$\log_{10}(\Lbol) =
\log_{10}(\Menv) \times (1.96\pm 0.36) - (1.09\pm 0.30).$$
Fitting just Class~0 sources, the gradient is similar but the
luminosities are lower: 
$$\log_{10}(\Lbol) = \log_{10}(\Menv) \times
(1.88\pm0.46) - (1.43\pm0.46).$$  
Fitting only Class~Is, the fit is
$$\log_{10}(\Lbol) =
\log_{10}(\Menv) \times (1.85\pm 0.52) - (0.46\pm 0.35),$$
with a gradient consistent with the index of $1.6$
found for Class~I sources driving Herbig-Haro objects
\citep{reipurth93}.\end{jh}  These fits
are plotted in Fig.~\ref{fig:lvsm}.  There are two selection effects:
firstly, low-luminosity low-mass Class~0s are selected against so the
slope of the best fit line for Class~0s and the whole sample are
probably too low; secondly, we believe some of the low mass Class~I
sources should be reclassified as Class~0s (see
Sect.~\ref{sect:radtrans}), so the slope of the best fit line for
Class~Is is probably too high.  The uncertainties on the fits are high
and the scatter is large, so they should in any case be treated with
caution.  Nonetheless, the results are consistent with Class~Is having
generally higher luminosities than Class~0s and a power law index
between 1 and 2, consistent with the \citet{smith00} model.

\section{Radiative transfer modelling of Class~0 geometry sources with different masses}
\label{sect:radtrans}

We computed radiative-transfer models of Class~0 objects using the
Monte-Carlo radiative-transfer code {\sc torus} \citep{Harries2000,
Harries2004}, The code is based on the radiative-equilibrium method
described by \citet{Lucy1999} and uses an adaptive mesh in order to
adequately resolve sub-solar-radius scales in parsec-scale
computational domains \citep{Kurosawa2004}.

\begin{tjh}

Our Class~0 geometry is based on the models described by  \cite{whitney03}.
In brief, this consists of a blackbody source of radiation surrounded by a flared
accretion disc which is in turn embedded in an infalling envelope. Low-density
bipolar cavities exist perpendicular to the disc. For simplicity we adopt ISM-type
grains, although more sophisticated models could allow for evolution of 
both the grain size distribution and the dust chemisty.  We give full details
 of the density description, along with the dust grain size distribution and
chemistry adopted, in the Appendix.

\end{tjh}

\subsection{Model parameters}

We computed models for a range of envelope masses between
0.05\,M$_\odot$ and 10$M_\odot$. The luminosity for each model was
selected by adjusting $R_*$ to follow $\log_{10}(\Lbol) \propto
1.5*\log_{10}(\Menv)$ with luminosities appropriate to masses calculated
at 30~K (consistent with \citet{batc96}).  The resulting absolute
luminosities are higher (for the same mass) than for the Class~0
sources (Sect.~\ref{sect:lvsm}) but the relative scaling (and
therefore the relative effect on \Tbol) is consistent.  A constant
$M_{\rm env}$:$M_*$ ratio of 3 \citep{whitney03} was assumed to
represent the same evolutionary stage for sources of different masses.
The basic model parameters are listed in Table~\ref{tab:model_params}.

We computed all the SEDs assuming an inclination of $90^\circ$; the
shape of the SED is essentially independent of viewing angle except
for pole-on orientations, in which case the near-to-mid IR portion of
the spectrum is enhanced by direct and scattered radiation that
escapes along the polar cavities.
 
\subsection{Model results}

% Calc other evolutionary indicators for Tim's models - done, not useful,
% 3.6 micron flux is negligible and Lsmm/Lbol roughly constant.

The dependence of \Tbol\ with \Menv\ for
Class~0 models with identical geometry is given in Table~\ref{tbl:modeltbol}.

\begin{table}[t]
\begin{tabular}{r|c c c c c c c}
$M_\mathrm{env}/\Msun$ &0.05 &0.1 &0.2 &0.5 &1.0 &3.0 &10.0\\
\hline
$T_\mathrm{bol}/K$ &67.4 &68.1 &68.1 &54.8 &48.1 &40.1 &34.9\\
\end{tabular}
\label{tbl:modeltbol}
\caption{Dependence of \Tbol\ with \Menv\ for Class~0 models with identical geometry.}
\end{table}

This dependence is plotted as a dashed line in
Fig.\ref{fig:menv_tbol}.  Low mass Class~0 sources have significantly
higher \Tbol\ than their higher-mass counterparts, for the same
geometry.  Our models of young Class~0 sources reach $\Tbol \sim 70$
for 0.2~\Msun\ and below and are on the borderline of being classified
as Class~I.  Furthermore, although we chose to model \begin{jh} a
  geometry representative of the very earliest \end{jh}Class~0 sources, a
  \Tbol--\Menv\ dependence also holds for other geometries as it is a
  fundamental property of radiative transfer, so models of more
  evolved envelopes with a higher luminosity to mass ratio and a wider
  outflow cavity would cross the Class~0/Class/I boundary at $\Tbol =
  70$~K at higher \Menv.  Low mass sources are therefore more likely
  to be classified as Class~I on the basis of \Tbol\ despite having a
  Class~0 type envelope geometry.

\subsection{Class~0 and Class~Is revisited.}

Using the models, we can revisit the classification of protostars in
Perseus and consider why we see so many more Class~0 sources compared
to lower-mass samples such as \citet{batc96}.  All the Class~0 sources
which have \Tbol\ above the predictions of the model as shown in
Fig.\ref{fig:menv_tbol} would be reclassified as Class~I sources on
the basis of \Tbol\ if their envelopes were reduced to $< 0.2~\Msun$
while retaining the Class~0 geometry.  This means that essentially all
our Class~0 sources -- excluding only the three sources b1-bN/S (2),
NGC1333~SK22(70) and HH340B(71) -- would have been classified as
Class~I had they had the same geometry but lower mass.  This
conclusion holds even if we were to calculate envelope masses assuming
a constant 30~K rather than 10~K, which reduces the masses by a factor
of 7, for consistency with \citet{batc96}: in that case 7 rather than
3 sources would retain their Class~0 classification.

Thus, if Perseus had been forming lower mass stars with envelope
masses under $0.2~\Msun$, \begin{jh} but with the same envelope
  geometry,\end{jh} the fraction of sources classified Class~0 to
Class~1 would drop to around 7/49, similar to the 20\% found by
\citet{batc96}.  This potential misclassification of
\begin{jh}Class~0-geometry sources\end{jh} with low envelope
massescould account entirely for the high number of sources classified
Class~0 in Perseus -- which we believe are truly Class~0 sources (see
below) -- compared to studies of lower mass regions where we believe
many Class~I-classified sources may in fact have Class~0 geometry.

If this is the case, and Perseus is representative, then the relative
lifetimes of Class~I and Class~0 phases are not, as earlier suggested,
10:1 \citep{andremontmerle94,motte98}, but instead roughly 1:1, as
found in Perseus and in Lynds dark clouds \citep{visser02}, assuming
that the star formation rate is constant.

How are the other SED-based evolutionary indicators shown in
Fig.\ref{fig:evo} affected by mass?  The SED-based indicator
$F_{3.6}/F_{850}$ should also decrease towards higher masses for the
same envelope geometry as the peak of the SED shifts longwards (see
Fig.~\ref{fig:model_seds}).  The remaining evolutionary indicator,
\Lbol/\Lsmm, is based on input properties of the models as \Menv\ is
derived from \Lsmm.  From Fig.~\ref{fig:lvsm}, and as also suggested
by models \citep{smith00}, the $\Lbol/\Menv$ boundary between Class~I
and Class~0 sources should rise with $\Menv$ and therefore so should
$\Lbol/\Lsmm$.  On this indicator, low mass sources should require
lower $\Lbol/\Lsmm$ ($\Lbol/\Menv$) to be classified as Class~0.  The
higher masses in our sample do not explain why we have to set a low
$\Lbol/L_{1.3}$ boundary at 3000 rather than 20,000 \citep{awb93}, as
we would expect $\Lbol/\Lsmm$ at the Class~0/Class~I boundary to rise
with mass leading to a higher threshold for Perseus.

Could many of the apparent Class~0 sources in Perseus be misclassified
and in fact be Class~Is?  As higher-mass sources have lower \Tbol\ for
the same geometry, it is the warm, high-mass sources in the Class~0
classification which might in fact be Class~Is despite \Tbol\ less
than 70~K.  This assumes that the Class~0/Class~I boundary lies
roughly parallel to the \Menv--\Tbol\ track shown on
Fig.\ref{fig:menv_tbol}, only higher in \Tbol.  Therefore sources such
as L1448~NW (27), NGC1333~IRAS2 (44), and NGC1333~IRAS7 (46), HH~211
(12), L1448~NW (27) and L1448~C (29) are the most likely sources to
have a Class~I-like geometry.  In the case of some these sources their
high \Tbol\ can be explained by their multiplicity (see
Sect.\ref{sect:multiplicity}).  However, this list contains many
classic bright Class~0 sources, which cluster in this part of the
\Menv--\Tbol\ diagram; if the number of Class~0s in Perseus is to be
reduced by reclassification then Class~0 prototypes such as HH211 and
L1448~C will be among the first sources to go.  This seems unlikely as
the collimated molecular outflows from these sources also support the
young age estimates (eg.  \citealt{hirano06,dutrey97}).  Therefore we
conclude that it is unlikely that a large number of our Class~0
sources need reclassification on the basis of the mass dependence of
the evolutionary indicators.  It is more likely that we, too, have
misclassified low-mass sources with $\Tbol$ close to 70~K as Class~I
when they should be Class~0 -- eg. sources 84 and Bolo62 which fail on
two evolutionary indicators.

It is clear the evolutionary indicators -- \Tbol, \Lbol/\Lsmm\,
$F_{3.6}/F_{850}$ -- which we use to classify protostars should all
vary with mass.  All of these evolutionary indicators are currently
set, based on models or observations, at a fixed boundary appropriate
to only a small range of masses.  How exactly they vary at the
Class~0/Class~I boundary needs further investigation.  It is clear
that more radiative transfer modelling is needed in order to define
the observable characteristics of the Class~0/Class~1 boundary at
$\Menv = M_*$ and before the numbers of sources in each class can be
determined reliably.  \begin{jh}Alternatively, it may make sense to
  consider all the embedded sources (Class~0 and I) as a single class with variation within that class.\end{jh}

% We can use the protostellar ``birthline'' to identify how many sources
% undetected in the infrared must be starless.  Sources with \Tbol upper
% limits in Fig.~\ref{fig:evo} (bottom row) can only shift further down
% and to the right in the \Menv-- \Tbol diagram.  Therefore, allowing
% for the fact that our birthline is a little too high, under half of
% the IR non-detections could possibly be protostellar; the remaining
% half must be starless cores.

\subsection{Protostellar brown dwarf candidates}
\label{sect:bd}

Our modelled Class~0 geometry, with only a narrow outflow cavity, is
intended to correspond to the earliest stage in protostellar
evolution.  If this were true then these models would define a
``birthline'' for protostars with no protostars falling below the
line.  Fig.\ref{fig:menv_tbol} shows that the majority of protostars
do indeed lie above the line.  Protostars must begin with lower
luminosities than those we have modelled, but the early phase is both
shortlived and difficult to detect in the infrared or outflows.
Protostars then are predicted to evolve steeply upwards and to the
left in the \Menv--\Tbol\ diagram \citep[eg. ]{smith00,froebrich05}.

Protostellar brown dwarfs forming in isolation, if such things exist,
will be found among the sources with envelope masses up to a few times
the substellar mass limit and fitting the Class~0 envelope model.  Our
lowest mass, lowest \Tbol\ candidate Class~0 brown dwarfs with Spitzer
detections are sources 84 (0.77~\Msun, 74~K) and Bolo62 (0.5~\Msun,
103~K), both of which are too massive and warm to be serious Class~0
brown dwarf candidates.  Earlier, more embedded candidates may be
found among the low mass sources without Spitzer detections, if any of
these can be confirmed as protostellar.  These sources include the
Bolocam sources which were only detected by SCUBA at the $3\sigma$
level and cluster around the $3\sigma$ detection limit in
Fig.\ref{fig:menv_tbol}, of which the lowest \Tbol\ candidate is
Bolo70, if this source can be demonstrated to be protostellar.  Given
the dependence of \Tbol\ on \Menv\, and that in protostars the main
source of luminosity is accretion, young brown dwarfs are likely to
rapidly evolve to $\Tbol>70$ even in the Class~0 phase.

% Given that lower-mass protostars appear warmer, we believe our SCUBA
% survey is more complete to protostars than a simple mass calculation
% based on the flux limit and 20~K would suggest.  The 850\micron\ 
% detection limit introduces a minimum \Tbol\ at which a particular
% \Menv\ can be detected.  This is plotted as a solid line in
% Fig.\ref{fig:evo}.  For Class~0 sources above 30~K the detection limit
% is $\Menv > 0.06 \Msun$, an envelope mass well below the substellar
% limit.  All our detected protostars lie above 30~K apart from the
% protostar-starless core double B1-bN/S.  For Class~I sources the limit
% is $\Menv < 0.03~\Msun$.  On the other hand, the subset of sources
% positively identified as protostars on the basis on outflow or
% infrared detections will have missed some low-luminosity sources and
% requires further investigation with more sensitive infrared or outflow
% observations.

\section{Lifetime of the protostellar phase}
\label{sect:lifetimes}

To estimate the full duration of the protostellar phase,
Class~0 plus Class~I, we can compare our census of protostars in Perseus to
the older pre-main-sequence (PMS) T~Tauri population, for which age
estimates exist, following the method of \citet{kenyon90} and
\citet{wilking89}.  Most of the Class~II and III sources near the
Perseus molecular clouds lie in the two main clusters IC348 and
NGC1333.  On larger scales, the OB association Per~OB2 contains 90\%
of the B stars in Perseus but with photometric age estimates of
6--15~Myr \citep{dezeeuwbrand85, Gimenezclausen94} we assume that
IC348 and NGC1333 contain the only star formation in the last 5~Myr.
Using the pre-main-sequence population and age estimates for these two
clusters, we can estimate the recent star formation rate and compare
to the currently forming population of embedded protostars.

The most complete estimate of the population of IC348 is
$348\pm47$~stars with $m_K \le 16.75$ in an area of $20.5\times
20.5$~arcmin$^2$ \citep{muench03}.  The mean age for this cluster is
3--4~Myr \citep[][Mayne, priv.comm.]{luhman03, muench03}.  \begin{jh}Any
star formation before this time was either at a lower level or is an
artefact due to the uncertainties in ages derived from
colour-magnitude diagrams \citep{hartmann01, burningham05}.\end{jh}

\citet{wilking04} detected a total of 213 stars in NGC1333 above a
similar magnitude limit of $m_K = 16$ (note the K-band luminosity
function is falling off rapidly at these magnitudes, so the slight
difference in sensitivity between the two PMS surveys makes little
difference to the counts).  In the centre of the northern cluster
alone, they detect 126 stars of which an estimated 15 are field stars.
Assuming the field star fraction is the same in the wider area, the
total of pre-main-sequence stars is $188\pm 25$ stars. Age estimates
for individual stars in NGC1333 vary from 0.3~Myr \citep{wilking04} to
several Myr \citep{aspin03}, though it is generally accepted to be less
than 1~Myr old.

% numbers from calc_sfr.py 4th Aug 2006, Perseus/Analysis/plife.py for
% Bayesian analysis

The total number of PMS stars in both clusters is therefore $536\pm53$
stars.  As both \citet{wilking04} and \citet{muench03} find that the
IMF is falling off towards lower masses, better sensitivity to PMS
stars within the observed cluster regions would not increase this
number significantly; this is an almost complete census of the PMS
population within these limited areas.  SCUBA, Spitzer and 2MASS
observations all have found that the fraction of stars forming in the
distributed population in Perseus, in isolated cores or small groups
outside the main IC348 and NGC1333 clusters, comprises 20--40\% of the
total \begin{jh}(ie. an additional 20--60\% over NGC1333 and IC348)
  \citep{carpenter00,paperI,jorgensen06a,megeath06,allen06}.  To get
  the total past star formation rate in Perseus we correct the star
  counts in the clusters by this factor, including the $\sim 20\%$
  uncertainties.  An additional population of 20-60\% increases the
  star count over the past 3.5~Myr to $750\pm 130$ stars.\end{jh}

\begin{jh} We compare the PMS count to the population of $56$ protostars identified in the submm
  (Sect.~\ref{sect:class01}).
  
  In order to estimate a lifetime for the protostellar population we
  have to make an assumption about the current star formation rate.
  The simplest assumption is that the star formation has been constant
  over the lifetime of star formation in the region.  There does not
  appear to be any strong evidence, from the Spitzer data or
  otherwise, that this is either a particularly active or inactive
  time for star formation in Perseus, which would justify using a
  different value than the time average for the current star formation
  rate.  Star formation in individual regions of the cloud does vary
  with time -- for example, star formation in NGC1333 only began $\sim
  1$~Myr ago.  However, the temporal variations in the star formation rate in
  individual regions are assumed to average out when spatially
  averaged over the whole cloud.  Clearly, the more regions one
  includes in such a calculation, the more likely this assumption is
  to hold, and in a single molecular cloud with just two main clusters
  and a distributed population, temporal variations in the star
  formation rate are likely to be our greatest source of error.

  The average star formation rate derived from $750 \pm 130$
  PMS stars formed over the last 3.5~Myr is 214 stars per Myr with
  95\% confidence limits of 168--306 stars per Myr, where the uncertainties include only the uncertainties on the PMS count and age.

  A lower limit of $0.26$~Myr on the protostellar lifetime is
  calculated by assuming the 56 confirmed submm-detected protostars
  are the entire protostellar population.  \end{jh} However, this
simple estimate does not take into account any corrections for either
completeness and multiplicity, both of which may increase the true
number of protostars above those detected.

\subsection{Completeness}

If we are going to compare source counts at the protostellar and PMS
stages, we need to ensure that we are comparing like with like and
selecting objects corresponding to a similar range of final stellar
masses at each stage.  The PMS studies extend well below the brown
dwarf limit.  Our SCUBA survey, on the other hand, is incomplete to
low mass protostars: the $3\sigma$ detection limit for 10~K protostars
is 0.3\Msun.  Spitzer c2d has a similar sensitivity limit to embedded
sources \citep{c2d}, though a complementary bias towards the more
evolved cores, resulting in a small number of additional MIR
detections of protostars.  It is unlikely that either of these surveys
would have detected the protostellar phase of any lower-mass PMS
objects if they form individually.  Stars of $\sim <0.1~\Msun$ may
never have had envelope masses above the detection limit of 0.3~\Msun\
and simply would not be detected by our SCUBA survey.  As envelope
mass reduces with time until it reaches zero, essentially any
protostar ultimately drops below the submm detection limit.  What
fraction of the total number of protostars do we detect?  The
variation of protostellar envelope mass with time is poorly
understood, and made more complex by disparate estimates for the
relative numbers of Class~0 and Class~I sources, so using models of
envelope evolution does not seem to be a reliable basis for an
estimate.  The detection limits are also made more complex by the
presence of protostellar disks, which can also be detected in the
submm.  Uncertainties in the envelope masses due to uncertain dust
opacities, dust temperature and cloud distance introduce further
systematic errors when comparing the two populations.  In summary, we
must take into account a fraction of undetected protostars, but
exactly how large that fraction is is hard to estimate.

% To estimate the mean fraction of the
% detectable protostellar lifetime, we then need to average this
% fraction across the entire PMS population weighted by the IMF.  In
% order to put some limits on this correction factor, we make the simple
% assumption that we detect all the protostars forming final stellar
% masses above 0.3~\Msun\ and none of those below 0.3~\Msun\.  In
% reality, the detectable fraction of the protostellar lifetime
% increases gradually with increasing mass.  From the IC348 IMF
% \citep{muench03,luhman03}, we should include 35--50\% of the PMS
% population.

% the mean mass of a star above
% 0.1~\Msun\ is 0.3\Msun, at which mass the protostar would be
% detectable for most of its Class~0 phase or roughly half its total
% lifetime (the end of the Class~0 phase being defined as when the
% envelope mass equals the stellar mass).#  when
% # comparing the PMS count with the submm count we take 80\% of the PMS
% # population (leaving out the stars which would never have been detected
% # in the submm) and then double the submm population (to take into
% # account the non-detections).

\subsection{Multiplicity}
\label{sect:multiplicity}

We also have to correct for multiplicity.  If a significant fraction
of protostellar cores form \begin{jh}multiple sources\end{jh} simultaneously, and those
sources could be resolved in the infrared studies once they reached
the PMS stage (ie. are not close binaries), then the number of protostars
is higher than the number of protostellar cores which we count and we
will underestimate the protostellar lifetime.  Note that if cores form
stars sequentially and then eject them then the number of cores will
still represent the number of stars being formed at any one time, so sequential star formation does not introduce any error in our lifetime estimate.

How much evidence is there for multiplicity in our sample?  Some of
the cores in our sample are known to be forming binaries or
higher-order systems: could these result in multiple infrared
detections?  NGC1333~IRAS4A and B are each binaries or higher order systems
\citep{lay95}; NGC1333~SVS13 is a binary \citep{anglada00}; L1448~IRS2
is a binary \citep{wolfchase00,olinger06}.  Most of these would not be
resolved at the T~Tauri stage in infrared studies equivalent to those
of IC348 or NGC1333.  The overall binary fraction in IC348 is
$19\pm5$\% \citep{duchene99} and we estimate the fraction of resolved
binaries to be $< 20\%$ of this, assuming the period distribution in
\citet{duquennoymayor91}.  On that basis 4\%$\pm$1\% of our sources will
form binaries which would be resolved in the infrared at a later
stage, which introduces only a small correction to the lifetime.

Do we see any evidence for protostar-protostar multiplicity other than
close systems which would not be resolved in the PMS studies? With a
$14''$ beam we might expect to find confusion especially in the
clustered regions, and we do.  L1448~N~A/B (27) is an obvious example.
B1-BN/S (2) is potentially a pair of protostars, though there is no
clear detection of a second infrared source or outflow
\citep{hirano99}.  Several NGC1333 sources are also multiple when
examined at higher resolution \citep{sandellknee01} or with radio
observations \citet{rac97,rr98}: IRAS2A/B (44), SSV13 (43), IRAS7
(46), and ASR114 (45) are all believed to contain multiple protostars
with separations that would be resolved in the infrared.  These
sources are among the most massive in our sample and therefore the
easiest in which to detect multiplicity.  On the above count of six
sources we estimate 10\% is a lower limit to the number of cores
containing multiple protostars, but the true number is likely to be
more as only these massive, luminous protostars have been investigated
with sufficient sensitivity and resolution.  Note that this
multiplicity is not easily identified in the infrared: many of these
sources contain a luminous protostar which swamps fainter neighbours
and in the few cases where one can resolve multiple Spitzer
detections, these appear to be due to shocked H$_2$ outflow components
rather than additional protostars.

It is also possible that some of our cores contain both a protostar
and a collapsing prestellar core.  Such sources would appear
overmassive for their luminosity and therefore classified Class~0.
Class~0 sources account for slightly more than half our total
population of cores, which places an upper limit on the number of
cores of this type. 

Increasing multiplicity also increases the protostellar lifetime, as
our submm core count again underestimates the true number of
protostars.  This effect is not entirely independent to that of
incompleteness, however, as many of the objects formed in multiple
cores will be low mass sources which would not have been detected as
protostars if isolated because of their low envelope masses.

\subsection{Bayesian probability-based estimate of protostellar lifetimes.}

\begin{figure}
\centering
\includegraphics[scale=0.7,angle=0]{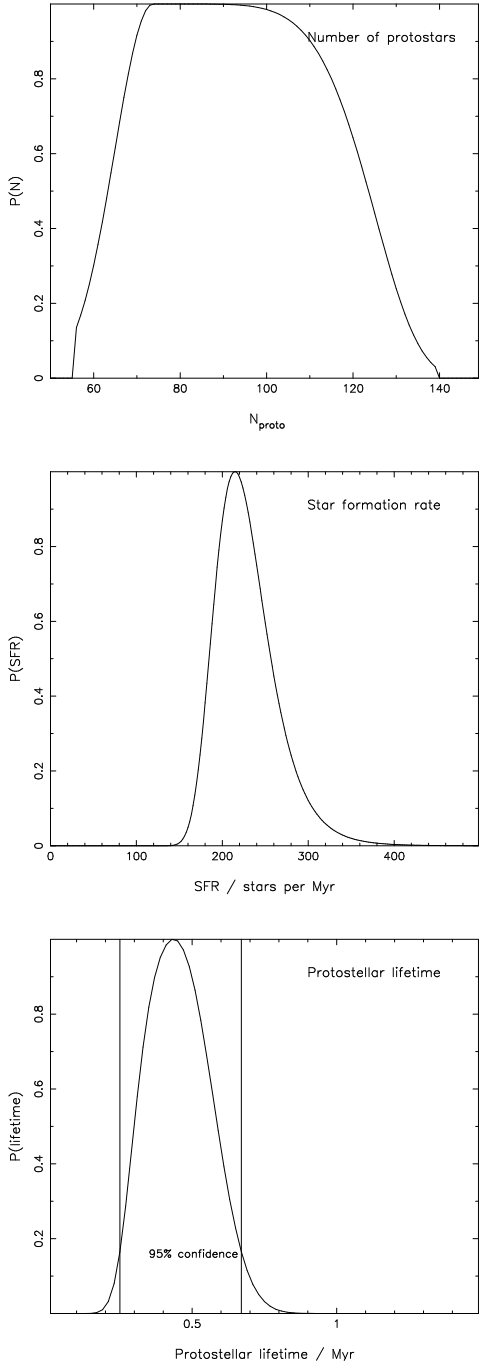}\\
\caption{Probability distribution for protostellar lifetimes.  {\bf Top:} Assumed probability distribution for the number of protostars based on SCUBA detection of 56 above $0.3\Msun$.  {\bf Centre:} Probability distribution for the star formation rate (SFR) based on the duration of star formation in IC348 and the number of pre-main-sequence stars in Perseus. {\bf Bottom:}  Resulting probability distribution for the protostellar lifetime.}
\label{fig:plife}
\end{figure}

Our current state of knowledge does not allow us to place definite
numbers on either the mean fraction of the protostellar lifetime to
which our SCUBA survey is sensitive or the multiplicity of the submm
sources.  But it does allow us to place some limits and also make some
informed estimates about the most likely scenarios.  

\begin{jh}
  We can use this additional information to improve on the simple
  lower limit on the lifetime calculate earlier and estimate the range
  of possible protostellar lifetimes.  To do this we use a Bayesian
  methodology to derive a probability distribution for the
  protostellar lifetime, based on probability distributions for the
  factors involved -- the number of protostars, number of
  pre-main-sequence stars, and duration of star formation in IC348.
  The Bayesian treatment has the advantage of allowing arbitrary
  distributions to be treated rather than making the usual implicit
  assumption that all variables follow a normal distribution.  In our
  case our state of knowledge about the number of protostars, which
  has a well-defined lower limit but a long tail to high numbers
  (allowing for high multiplicity or incompleteness), are poorly
  represented by a Gaussian, and therefore a more
  complex treatment is required.  The other parameters - the age of IC348 and the
  number of PMS stars - are represented by Gaussian variables in the
  usual way.

% We use a Bayesian methodology
%   
\end{jh}

Considering in detail the probability distribution for the number of
protostars, $\prob(N_\mathrm{proto})$, we have a lower
limit: we know that there are at least 56 protostellar cores.
Including the additional Spitzer detections, we take a count of 67
protostars to be more probable than 56.  Taking into account the known
multiplicity of at least 10\%, we would anticipate that 74 protostars
is even more probable.  We represent this by a rising probability from
56 to 74 following a partial Gaussian $\exp
-(N_\mathrm{proto}-74)^2/(2\times9^2)$.  Above this count we have to
correct for the unknown number of undetected low-mass envelopes and
the amount of multiplicity, about which our state of knowledge is very
uncertain, but again we can place some limits.  If protostellar
envelope masses mirror the IMF, then we would expect our SCUBA
detections above 0.3\Msun\ to account for only 40\% of the total
population \citep{muench03} and the total count could number as many
as 140; however, as envelopes have, on average, masses greater than
the resulting protostars and some lower-mass stars will form inside
higher-mass multiple cores we consider such a high count to be
unlikely.  To represent this, we assign a probability above the peak
probability at $N_\mathrm{proto}=74$ initially remaining high but
ultimately falling steeply to a low but finite probability to values
close to the maximum of 140, following an exponential
$\prob(N_\mathrm{proto}) \propto \exp
-((N_\mathrm{proto}-74)/54.8)^6$.  The resulting composite probability
distribution is shown in the top panel of Fig.~\ref{fig:plife}.  These
assignments of probabilities are necessarily somewhat arbitrary but
consistent with the data and our state of belief about the system.  As
we will see, the large uncertainties in $N_\mathrm{proto}$ will
manifest themselves as a broad spread in protostellar lifetimes.

The probability distribution for the age of the clusters, $\prob(A)$
we take as a simple Gaussian, mean 3.5~Myr, standard deviation
0.5~Myr.  The probability of the number of pre-main-sequence stars,
$\prob(N_\mathrm{PMS})$, is also a Gaussian given by the product of
the $536\pm 53$ stars identified in NGC1333 and IC348 and a factor
representing the $43\%\pm20\%$ additional distributed population.  The
resulting distribution is also Gaussian with a mean of 652 and a
standard deviation of 97.  From these two distributions -- age and
number of PMS stars -- we derive the corresponding distribution for
the star formation rate, shown in the middle panel of
Fig.~\ref{fig:plife}, using the formula:

$$\prob(\mathrm{SFR}) = \int_0^{\infty} \prob(A)\, \prob(N_\mathrm{PMS} = A \times \mathrm{SFR})\, dA. $$

We then assume that the star formation rate has been constant since
the onset of star formation in IC348 until the present day, and
calculate the lifetime of the protostellar phase by comparing the star
formation rate with the number of protostars, $\tau_\mathrm{proto} =
N_\mathrm{proto}/\mathrm{SFR}$.  The corresponding probability
distribution,

$$\prob(\tau_\mathrm{proto}) = \int_0^{\infty} \prob(N_\mathrm{proto})\,
\prob(SFR = N_\mathrm{proto} / \tau_\mathrm{proto}) \,dN_\mathrm{proto},
$$

is shown in the bottom panel of Fig.~\ref{fig:plife}.  

The result is that the most probable lifetime is 0.43~Myr with 95\%
confidence limits of 0.25 -- 0.67~Myr.  The distribution is not
Gaussian but skewed towards longer lifetimes, reflecting the tail of
probability towards high numbers of protostars.  \begin{jh}There is a
  large range of lifetimes with high probabilities within the 95\%
  confidence limits. The 95\% confidence range interval is much more
  robust against changes in the details of the probability
  distribution than the position of the peak probability.\end{jh}

The uncertain submm detection rate for protostars and the uncertain
multiplicity in the submm cores contribute the majority of the $\sim
40$\% uncertainties in the protostellar lifetime.  This will to a
large extent be addressed by the next generation of more sensitive
submm surveys, which have an order of magnitude better mass
sensitivity and will therefore detect envelopes down to below the
substellar mass limit.  Sensitive mid-infrared, outflow or radio
observations will be required to determine if such low-mass objects
are indeed protostars.  The uncertain age spreads in IC348 and NGC1333
and the uncertain count of recent ($< 4$~Myr) T~Tauri stars outside
these main clusters also contribute to the uncertainties.  Much recent
work has been done on age spreads in PMS clusters
\citep{hartmann01,burningham05} and these are unlikely to improve
significantly but a better count of the distributed population may
come from wide-field surveys.

% The constant star formation rate is a necessary but untested
% assumption in the calculation and is the most likely source of bias in
% the protostellar lifetime.  This is not something we can estimate
% without further data on the T~Tauri population ages in Perseus, which
% are difficult to derive precisely.  The assumption of a constant rate
% is more likely to hold when averaging over a number of star-forming
% clouds, where fluctuations in the individual star-formation rates will
% average out, than in one individual cloud.

\begin{jh} An embedded phase lifetime of $0.25\hbox{--}0.67$~Myr (95\% confidence
  limits) is consistent with the simple lower limit (0.25~Myr) we
  calculated at the beginning of this section \end{jh} but somewhat
longer than the previous estimates of $0.2\pm0.1$~Myr in Rho~Oph
\citep{wilking89,greene94} and $0.15\pm0.05$~Myr in Taurus-Auriga
\citep{kenyon90,kenyonhartmann95}.  Unlike these mid-infrared based
studies, we are comparing with both a larger sample of T~Tauri stars,
due to more sensitive photometry, and a much more complete sample of
embedded stars, which includes Class~0 objects to which the earlier
surveys were blind, so it is unsurprising that we reach a different
conclusion.
% The consistency of the
% estimates despite the wildly different sampling suggests that the
% earlier studies, though less sensitive, were selecting similar
% populations of protostars and T~Tauris.
% 
% An embedded phase lifetime of $0.23\pm0.09$~Myr is consistent
% with estimates of core destruction timescales due to outflows, which
% have a mean of 0.3~Myr \citep{outflows}, supporting the hypothesis that
% star formation is halted by the destruction of protostellar cores by
% their outflows \citep{lada85}.

% Paper I
% IC348 3-4 Myr 350 stars
% NGC1333 150 stars 1 Myr
%  => 100 stars per Myr 
% => embedded phase 0.5-1 Myr
% 
% IC348 2/5*348 = 140 stars > 0.4 Msun
% => 35 stars per Myr
% => embedded phase 3 Myr.

%\subsection{Comments on individual sources?}

\section{Summary and conclusions}
\label{sect:summary}

The Perseus molecular cloud is a well known site of star formation
\begin{jh} containing a population of pre-main-sequence stars similar to Ophiuchus and intermediate between Taurus and Orion \citep{luhmanrieke99,briceno02,luhman03}.\end{jh} We have compiled spectral energy
distributions from the near-IR to mm wavelengths (using data from
Bolocam, SCUBA, IRAS, Spitzer, Michelle (where available) and 2MASS)
for the 103 known submm cores across the molecular cloud.  These cores
lie within the boundaries of our SCUBA map of the Perseus molecular
cloud, which covers roughly the area with $^{13}$CO integrated
intensity greater than 4~K~km~s$^{-1}$ or $A_{\mathrm v} > 4$
\citepalias{paperI}.  From the presence of a near- and/or mid-infrared
source or molecular outflow, we have identified 56 cores which are
definitely protostellar.  A further 11 Class~I sources have been
identified by Spitzer, bringing the total number of definite
protostars in Perseus to 67.  There may be further protostars among
our submm detections which are not yet luminous enough to be detected
in the IR or be identified by their outflow activity
\citepalias{outflows}.

We use three evolutionary indicators to classify the
sources on the basis of their spectral energy distributions: \Tbol,
$F_{3.6}/F_{850}$ and $\Lsmm/\Lbol$.  On the basis of these
indicators, 22 protostars are Class~I (33 including additional
Spitzer-only detections) and 34 are Class~0.  This is a much higher
fraction of Class~0 sources than is found in Ophiuchus or Taurus.
On the basis of these counts, the lifetimes for the Class~0 phase are
similar to the Class~I phase, rather than much shorter as has been previously found.

\begin{jh}Protostellar envelopes in Perseus are more massive than those in
  Ophiuchus or Taurus and this is the main reason for the high number
  of Class~0-classified sources in Perseus.  It is generally true that
  higher mass envelopes have higher optical depth envelopes and
  therefore SEDs which peak at longer wavelengths, resulting in lower
  \Tbol.  \end{jh} We have demonstrated this by modelling a typical
Class~0 geometry at a range of different masses and find that the
bolometric temperature rises as the mass decreases so that a source
which would be classified Class~0 at 1~\Msun\ would be classified
Class~I below 0.2~\Msun\ on the basis of this evolutionary indicator.
Low \begin{jh} envelope \end{jh} mass sources, such as the majority in
Taurus or Ophiuchus, are therefore more likely to be classified
Class~I whereas the higher mass envelopes which are detected in
Perseus are more likely to be classified as Class~0.  This
classification issue arises because of the difficulty in defining the
Class~0/Class~I boundary in terms of observable evolutionary
indicators, all of which (\Tbol, \Lbol/\Lsmm\, $F_{3.6}/F_{850}$) vary
with envelope mass for the same geometry of source.  More
investigation with radiative transfer models is needed to determine
where the boundary should lie.  However, despite the issues in
classification we believe that the count of Class~0 sources in Perseus
is roughly correct and similar to the number of Class~Is, and that
therefore -- assuming a constant star formation rate -- the lifetimes
of these two phases, at least for stars starting from envelopes of
half a solar mass or more, are similar.

There is also a possible physical explanation for why massive clusters might
contain more Class~0 sources.  If there is competitive accretion for
material within the cloud, then more massive sources may continue to
accrete material onto their envelopes and maintain their Class~0
status for longer.  It is possible that this mechanism operates to maintain the Class~0 sources in Perseus and that this is why there are so many in the clustered environments.

\begin{jh}We find that bolometric luminosity and envelope mass are roughly correlated with $\log_{10}(\Lbol/\Lsun) \propto \log_{10}(\Menv/\Msun)^{1.96\pm 0.36}$, consistent with the results for Herbig-Haro flows \citep{reipurth93} and the protostellar models of \citet{smith00}.   Class~0 sources and Class~I sources follow approximately the same power law but the Class~0 sources have, on average, an order of magnitude lower luminosities.\end{jh}

%\jh{How to make this useful.  Where to set Lbol/Menv?  Greatest envelope mass in Rho Oph?}

\begin{jh}Our estimate for the lifetime of the entire embedded protostellar
phase is $0.25\hbox{--}0.67$~Myr.\end{jh}  This estimate, higher than
previous estimates of $\sim 0.2$~Myr, is based on our survey of
protostars with corrections for multiplicity and incompleteness, and
with counts and age estimates for T~Tauri stars from recent censuses
of the populations in IC348 and NGC1333, assuming a constant star
formation rate in Perseus over the last $3.5\pm0.5$~Myr.

The ratio of starless cores above the submm detection limit to
protostars in Perseus is at most 0.8, though this could reduce as
apparently starless cores may yet be identified as protostellar by
more sensitive IR or outflow observations.  The lifetime for a
detectable starless core ($>0.3~\Msun$ within one SCUBA beam) is
therefore at least 0.8 times the protostellar lifetime or 0.4~Myr.

% What is SF efficiency from P(core)?

\acknowledgements

The James Clerk Maxwell Telescope is operated by the Joint Astronomy
Centre on behalf of the Particle Physics and Astronomy Research
Council of the United Kingdom, the Netherlands Organisation for
Scientific Research, and the National Research Council of Canada.  JH
acknowledges support from Deutsche Forschungsgemeinschaft SFB 494 and
the PPARC Advanced Fellowship programme.  This research made use of
the Spitzer archive operated by the Spitzer Science Centre and the
SIMBAD query facility of the Centre de Donn\'ees Astronomiques de
Strasbourg.  The authors would like to thank Tim Naylor and Nathan
Mayne for useful discussions on the ages of IC348 and NGC1333, and the referee for comments which substantially improved the clarity of the discussion.

\bibliographystyle{aa}
\bibliography{perseus}

\begin{landscape}

\begin{figure}[]
\centering
\includegraphics[scale=1.0,angle=0]{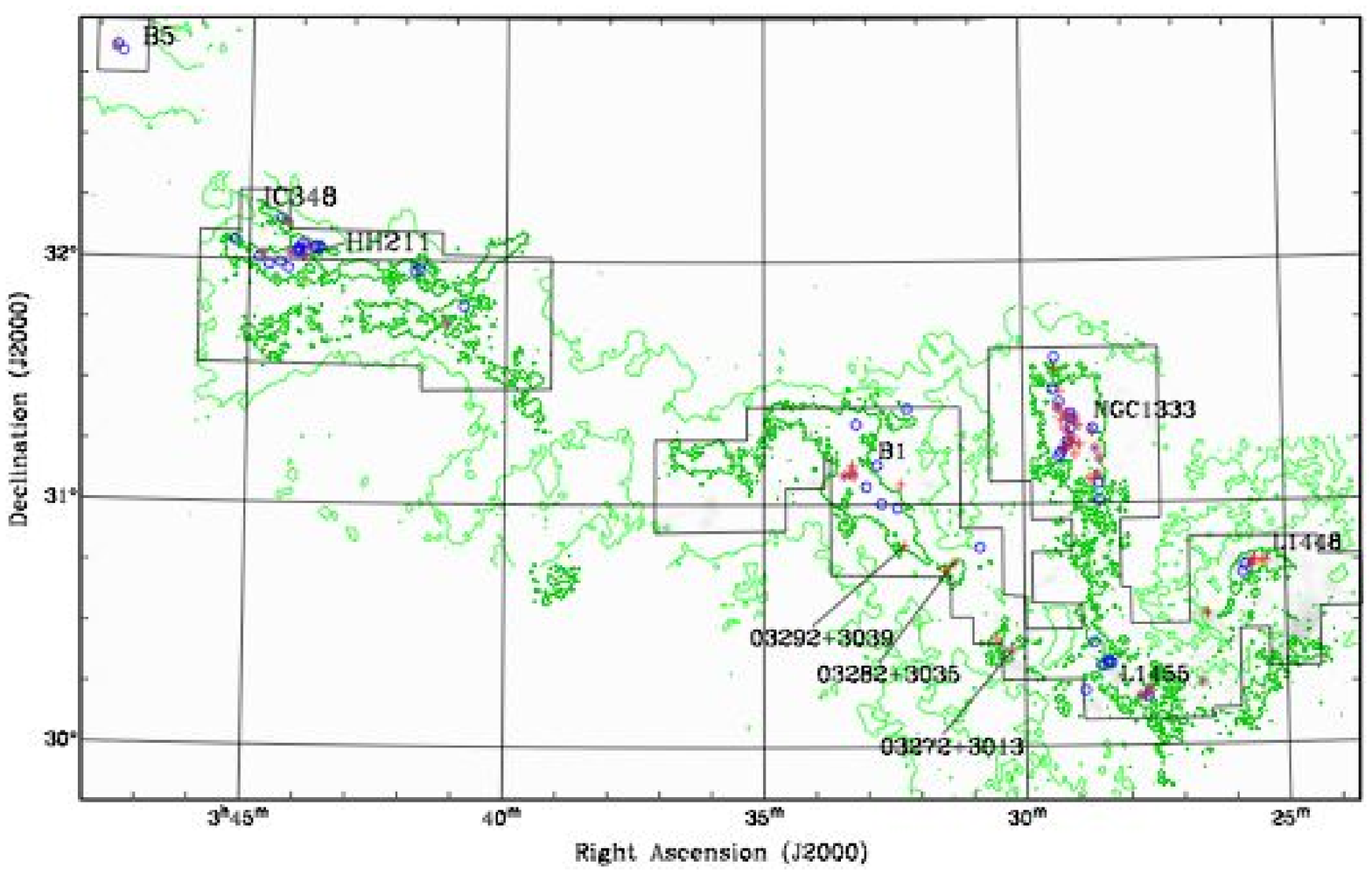}\\
% see figures.html for how to shrink kvis ps figures to manageable size

\caption{Class~0 (red crosses), Class~I (maroon stars) and starless cores (blue circles) overlaid on the Perseus molecular cloud outlined in C$^{18}$O (dark green contours 1~K~km~s$^{-1}$, \citet{paperI}) and $^{13}$CO (light green contours 3~K~km~s$^{-1}$, Bell Labs 7m survey made available by John Bally {\tt http://casa/colorado.edu/$\sim$bally}).  The area mapped by SCUBA is marked.}
\label{fig:class01}

\end{figure}

\end{landscape}

\Online

\begin{figure*}[t]
\centering
\includegraphics[scale=0.60,angle=-90]{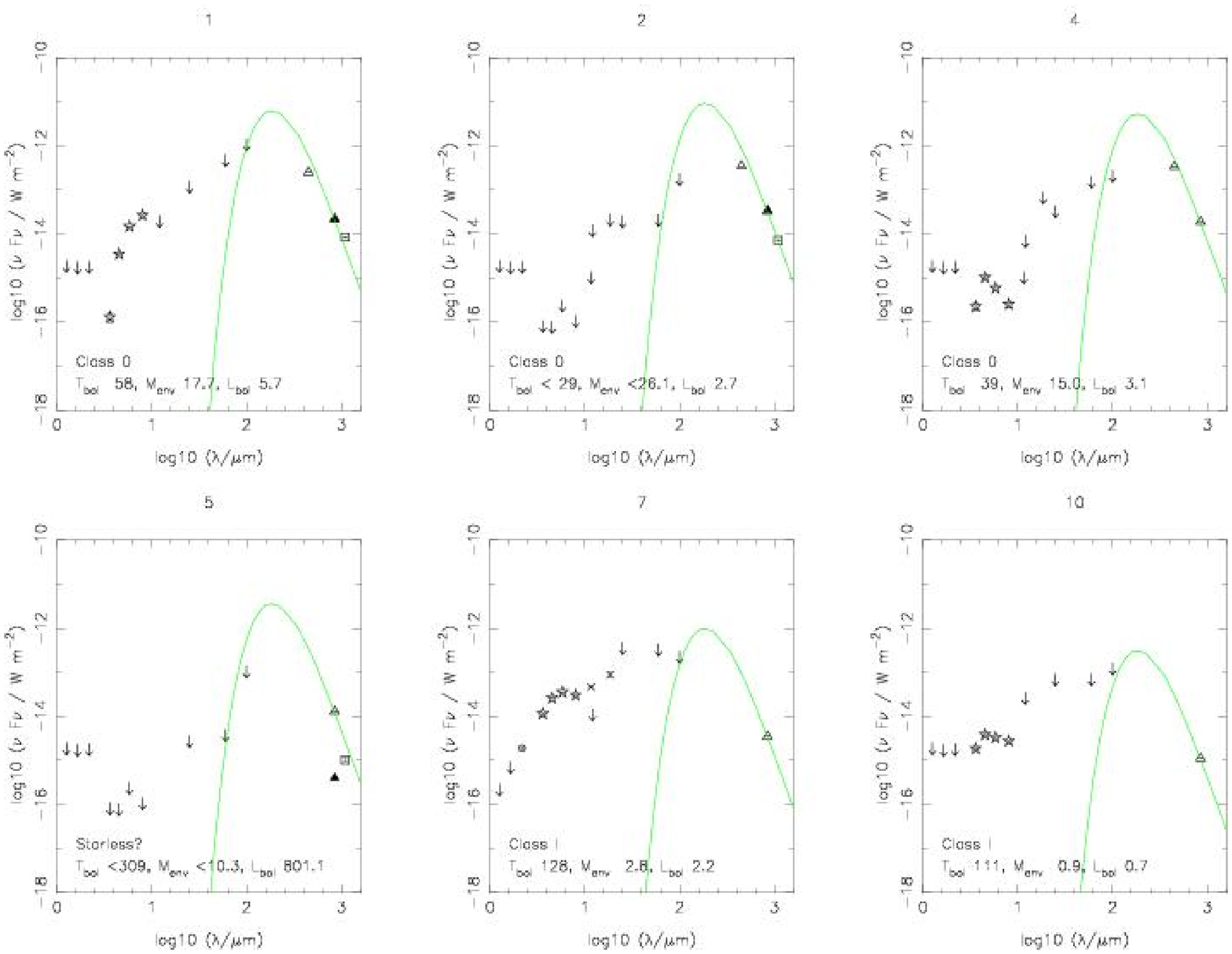}\\
\includegraphics[scale=0.60,angle=-90]{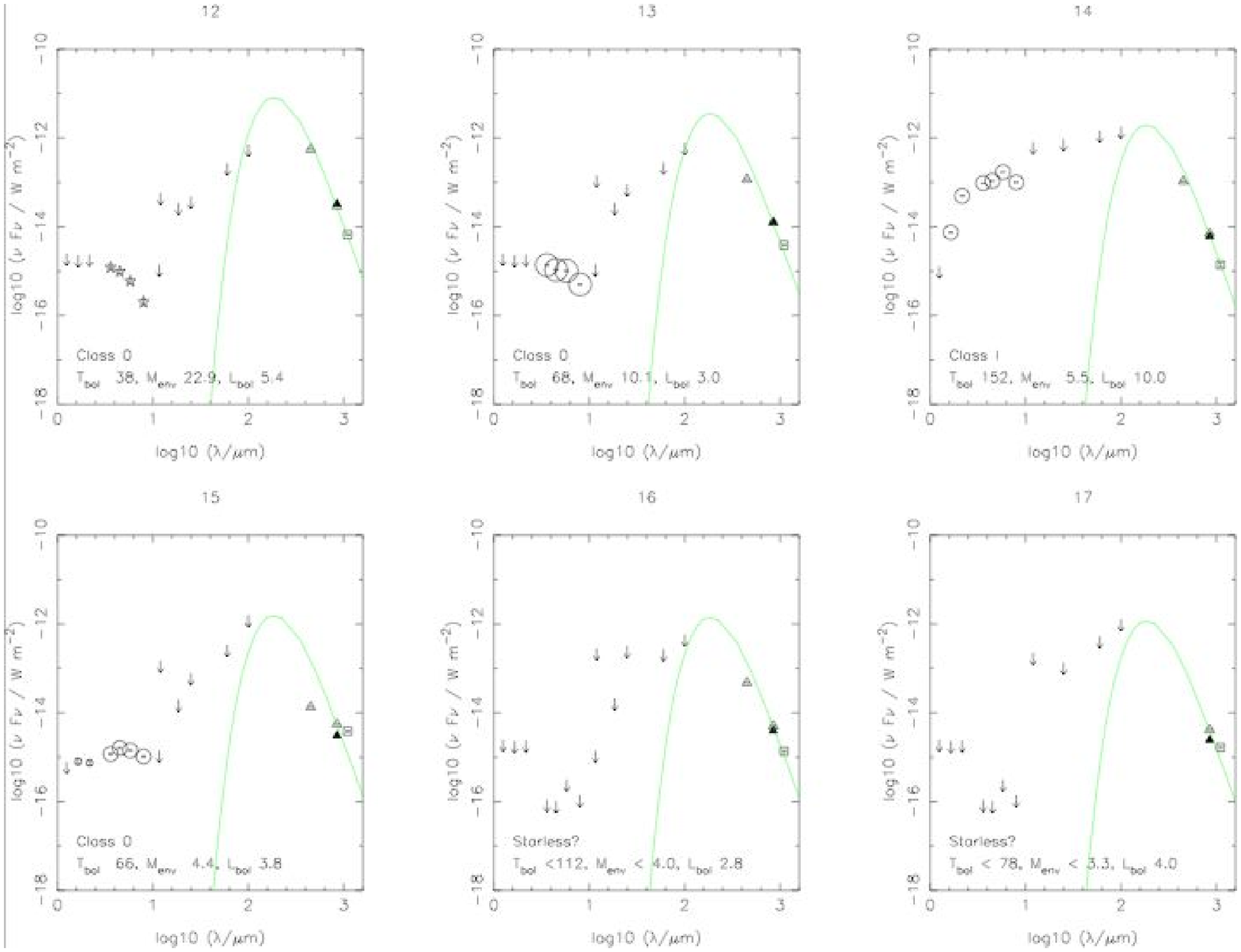}\\
\caption{Spectral energy distributions for the 103 submm detections in Perseus.  Contributions from short to long wavelength: 2MASS (circles), Spitzer IRAC (stars), Michelle ($\times$), IRAS HIRES (arrows), SCUBA (open triangles - Clumpfind; filled triangles - $90''$ diameter aperture ), Bolocam (squares). Upper limits (including IRAS HIRES) are marked as arrows and large circles denote more than one association.  A 10~K greybody passing through the 850\micron\ flux point is overplotted (green).}
\label{fig:seds}
\end{figure*}

\setcounter{figure}{6}
\begin{figure*}
\includegraphics[scale=0.60,angle=-90]{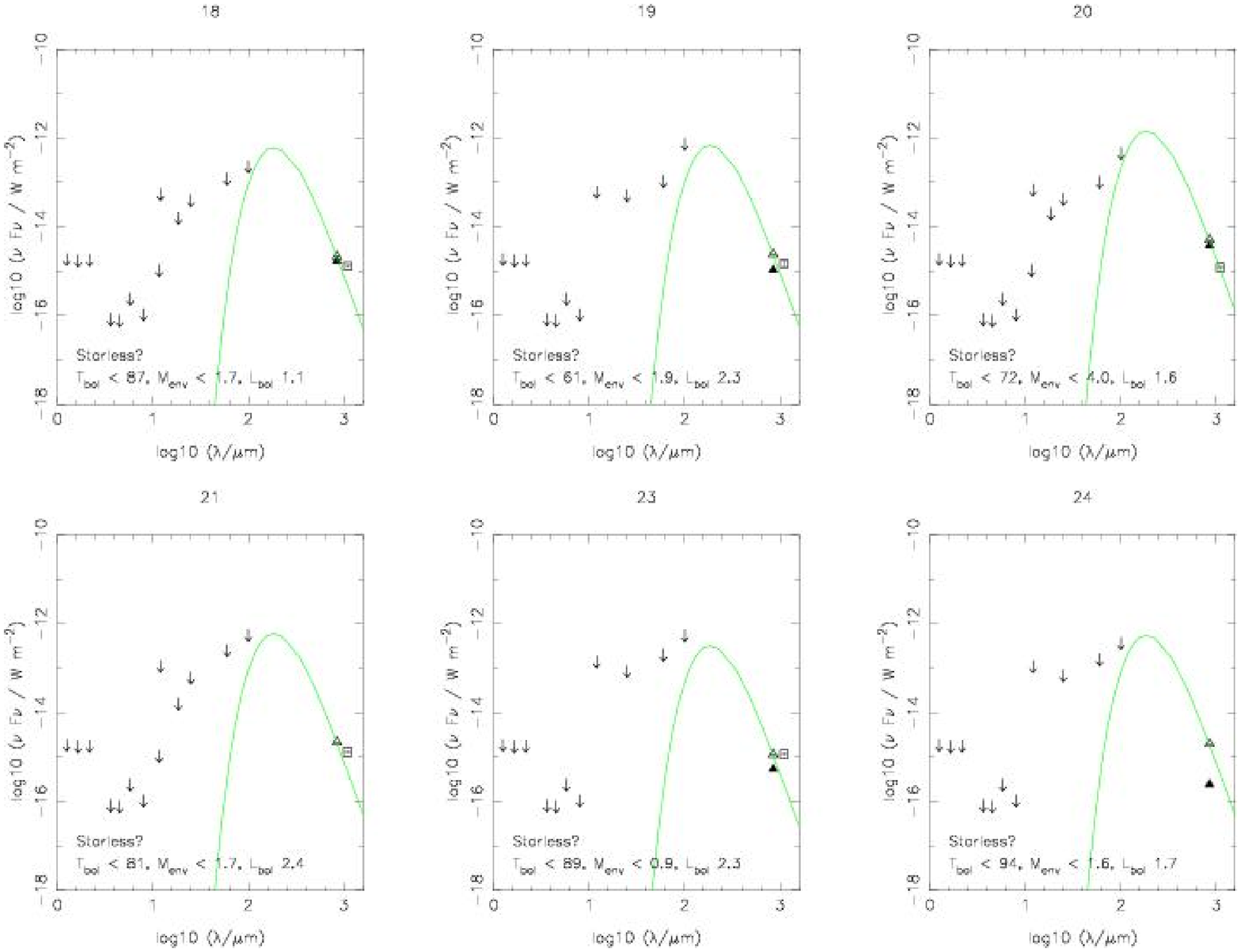}\\
\includegraphics[scale=0.60,angle=-90]{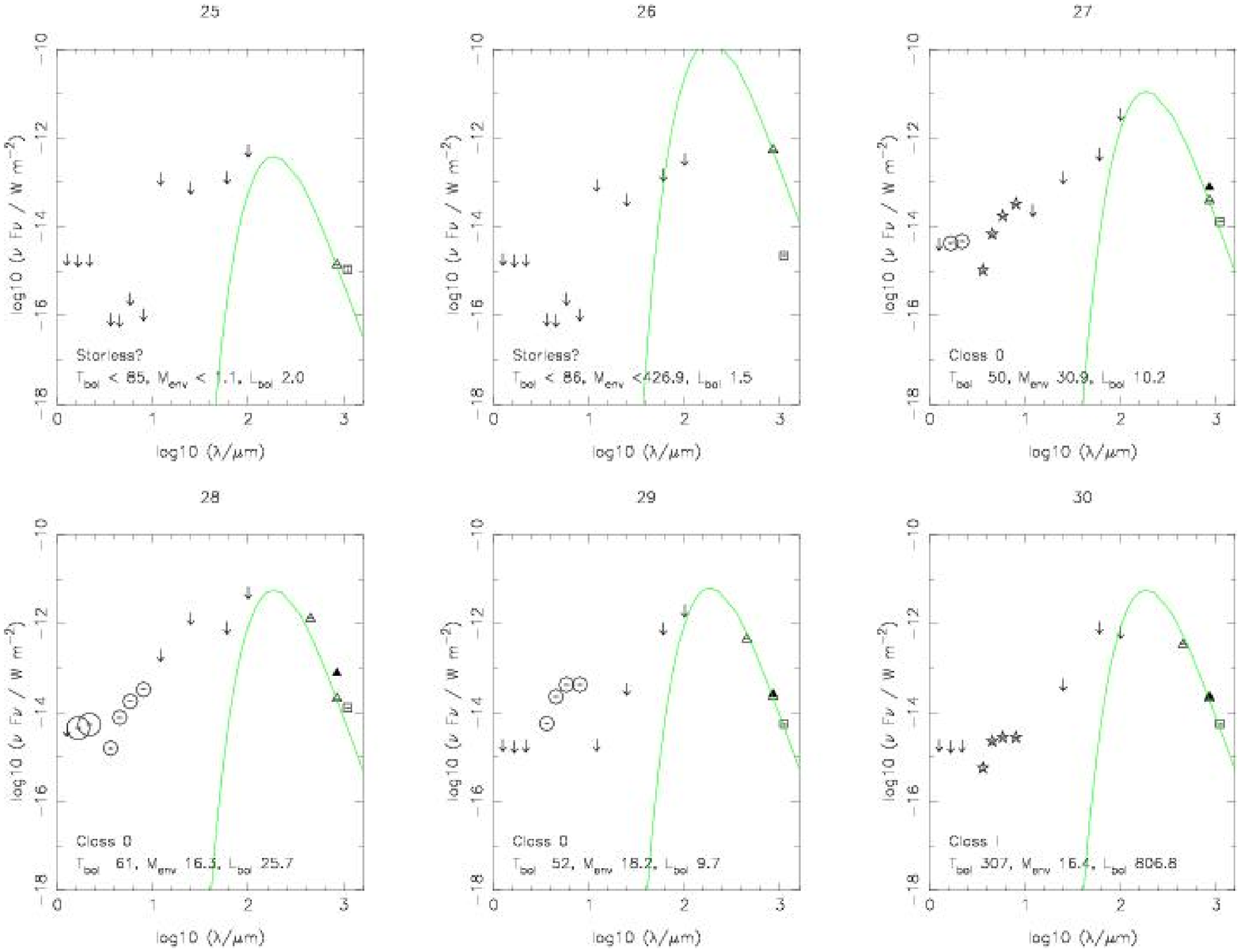}\\
\caption{Continued.}
\end{figure*}
\setcounter{figure}{6}
\begin{figure*}
\includegraphics[scale=0.60,angle=-90]{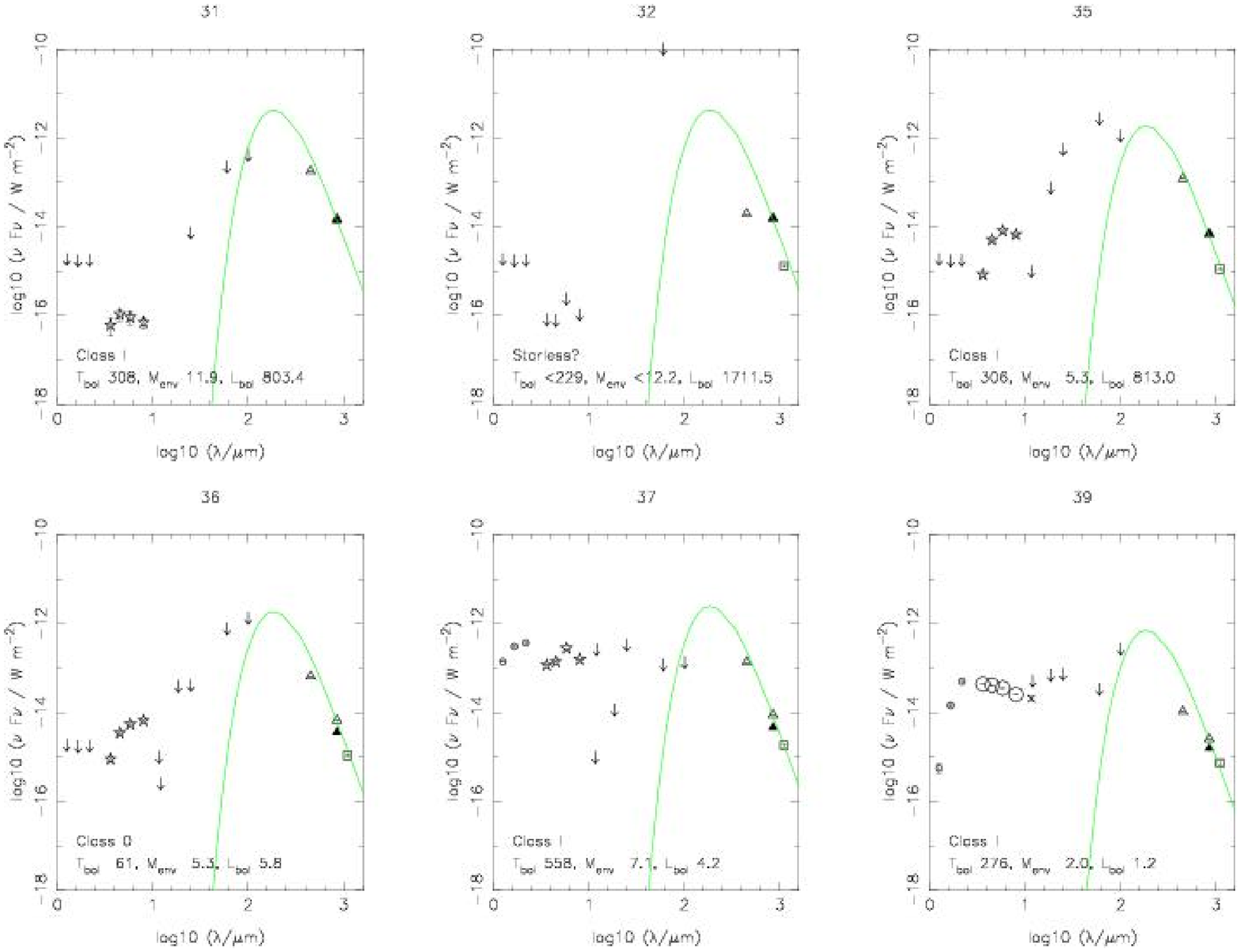}\\
\includegraphics[scale=0.60,angle=-90]{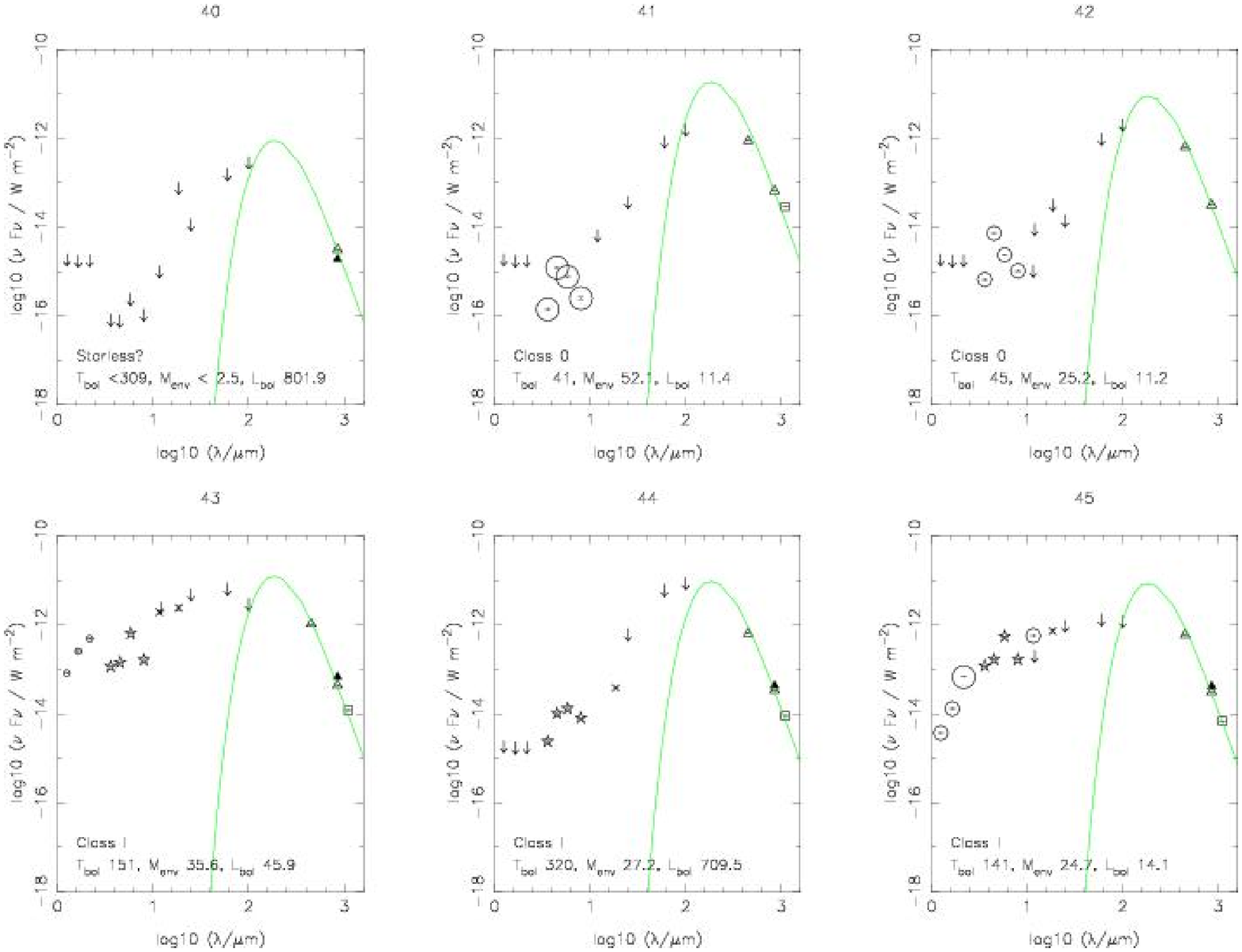}\\
\caption{Continued.}
\end{figure*}
\setcounter{figure}{6}
\begin{figure*}
\includegraphics[scale=0.60,angle=-90]{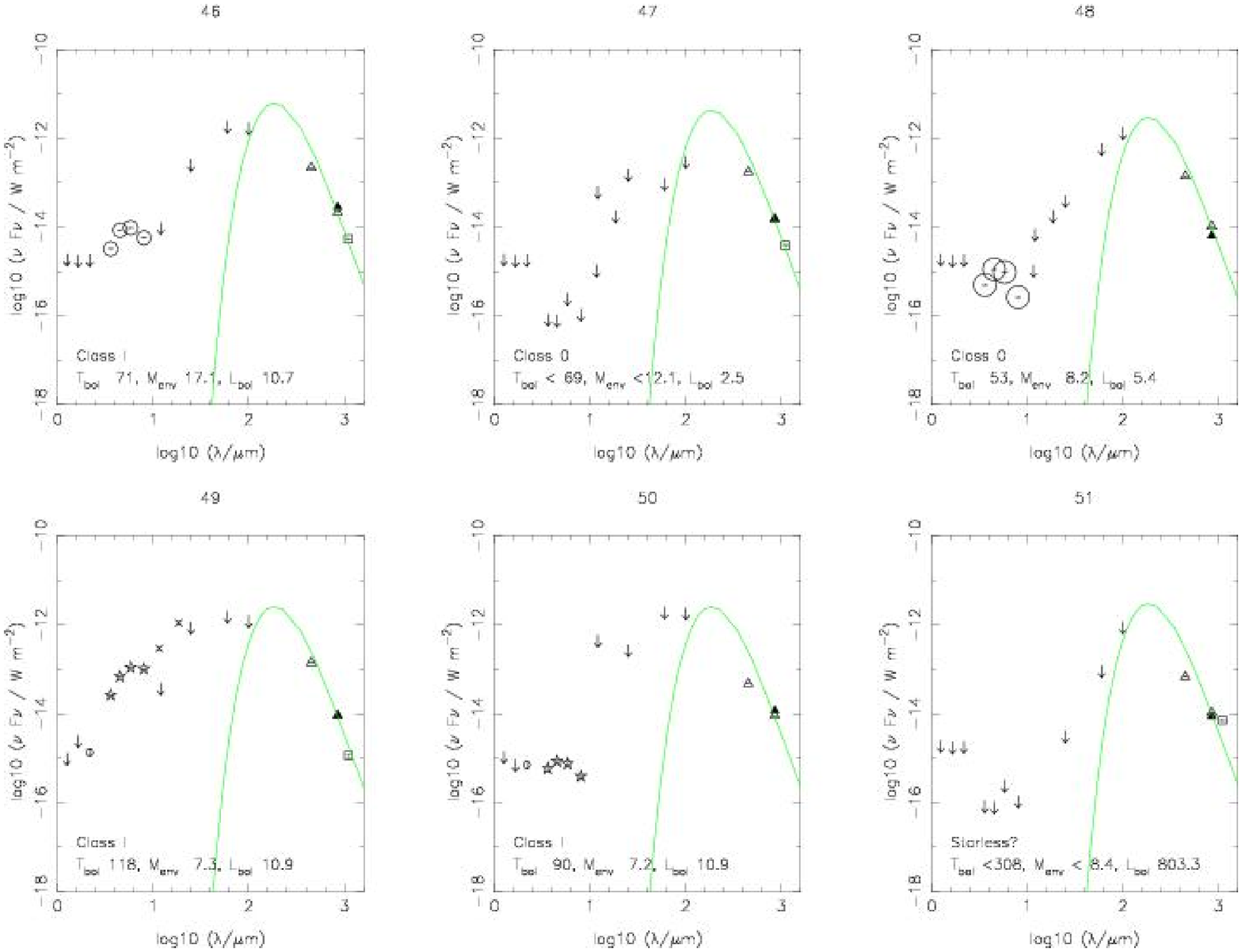}\\
\includegraphics[scale=0.60,angle=-90]{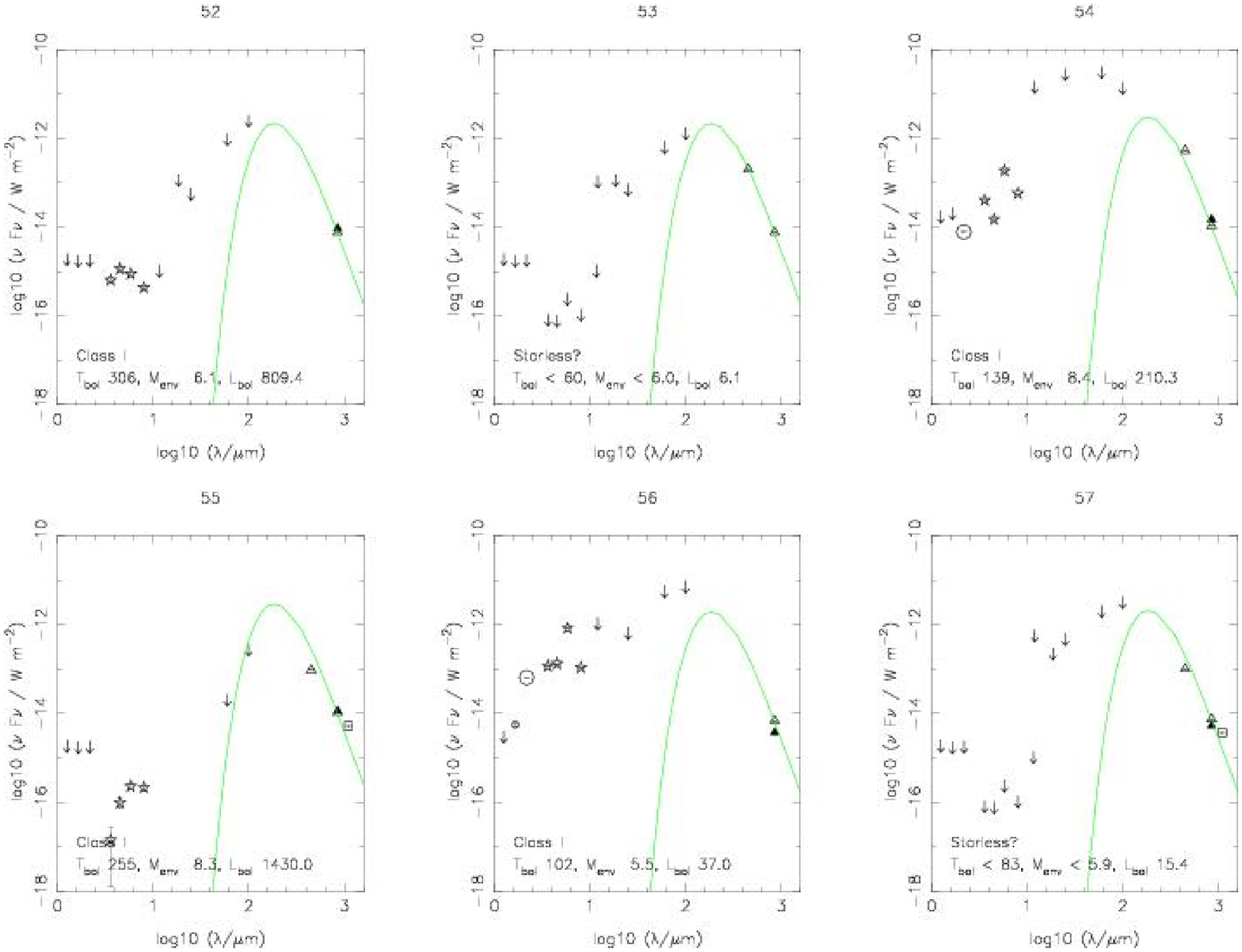}\\
\caption{Continued.}
\end{figure*}
\setcounter{figure}{6}
\begin{figure*}
\includegraphics[scale=0.60,angle=-90]{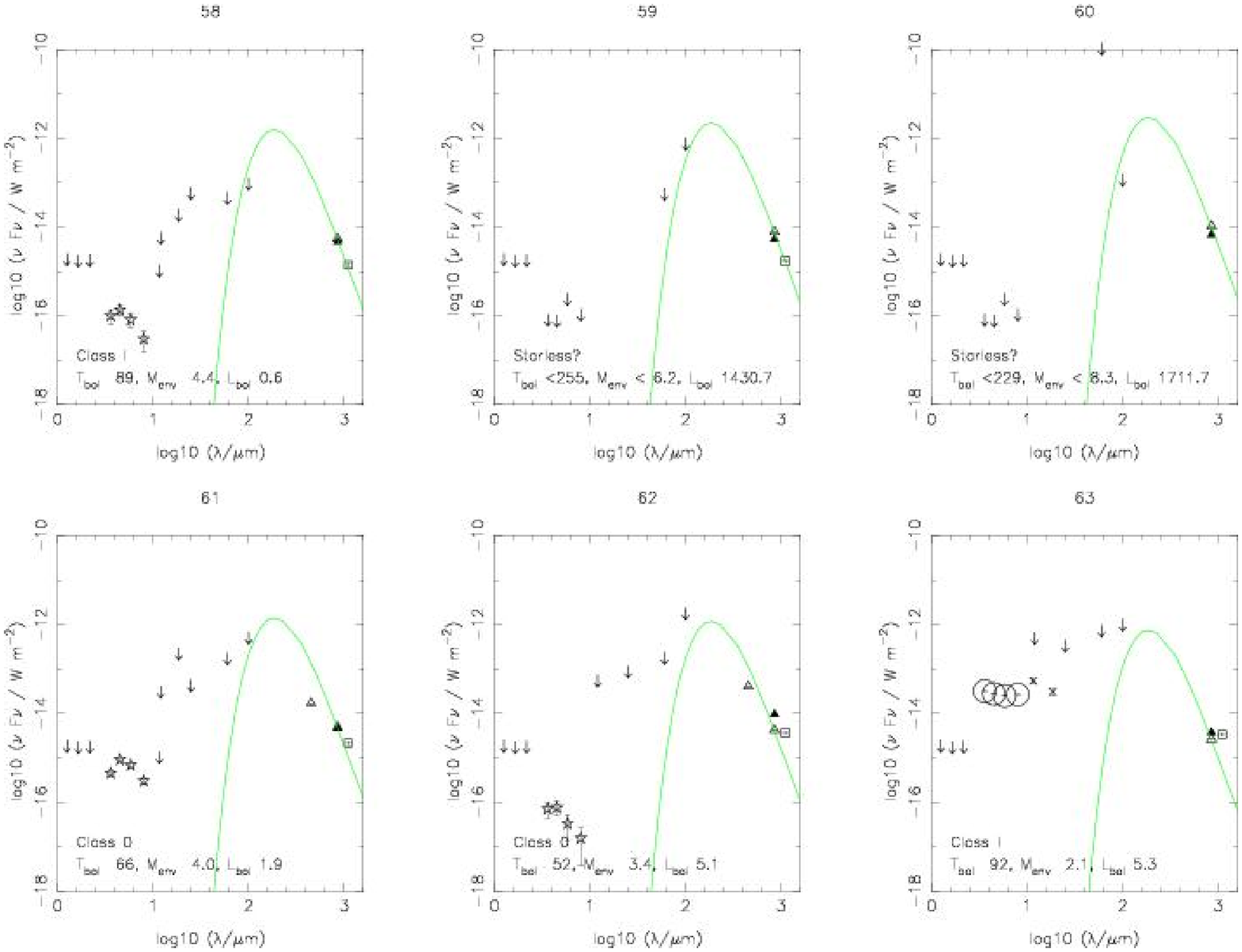}\\
\includegraphics[scale=0.60,angle=-90]{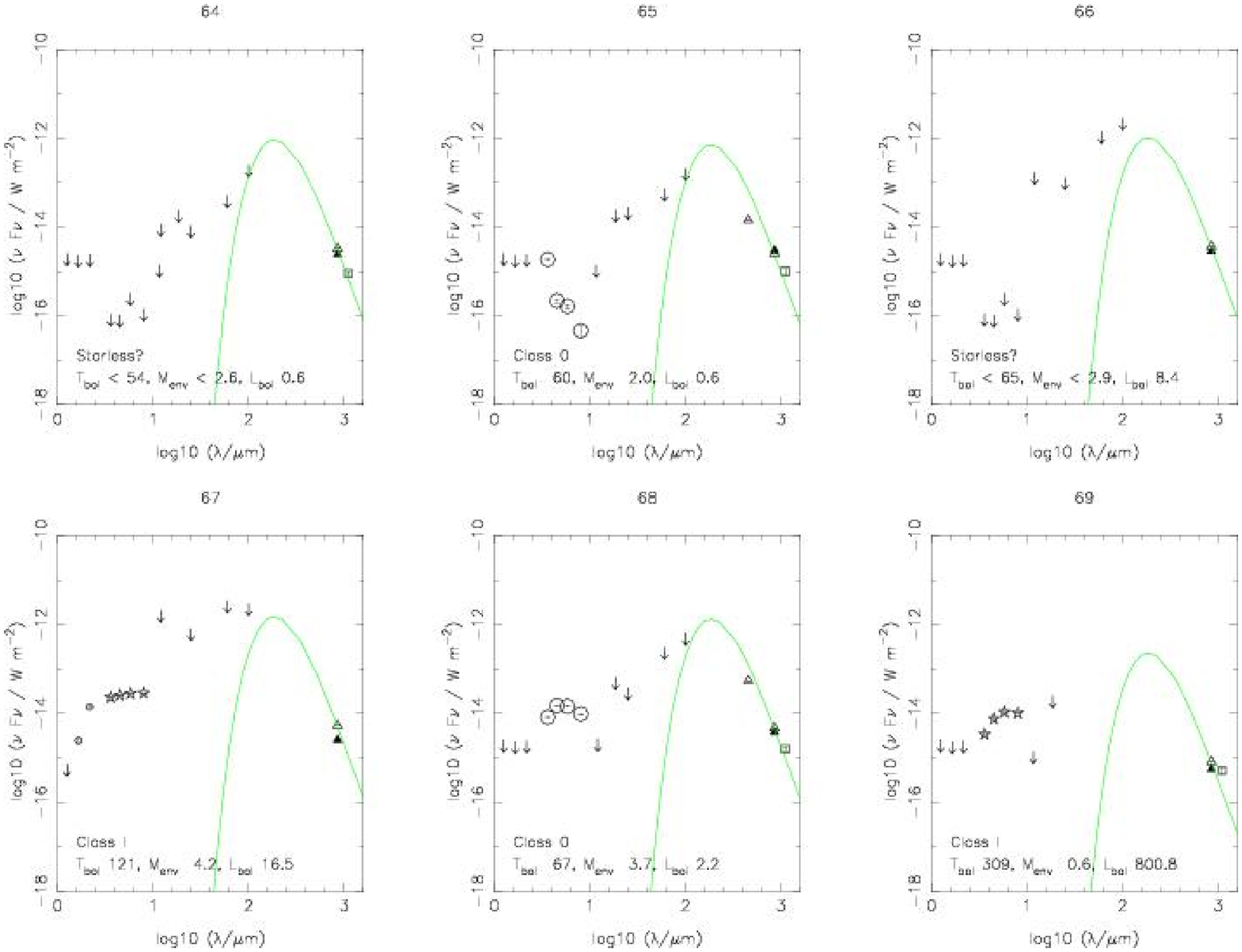}\\
\caption{Continued.}
\end{figure*}
\setcounter{figure}{6}
\begin{figure*}
\includegraphics[scale=0.60,angle=-90]{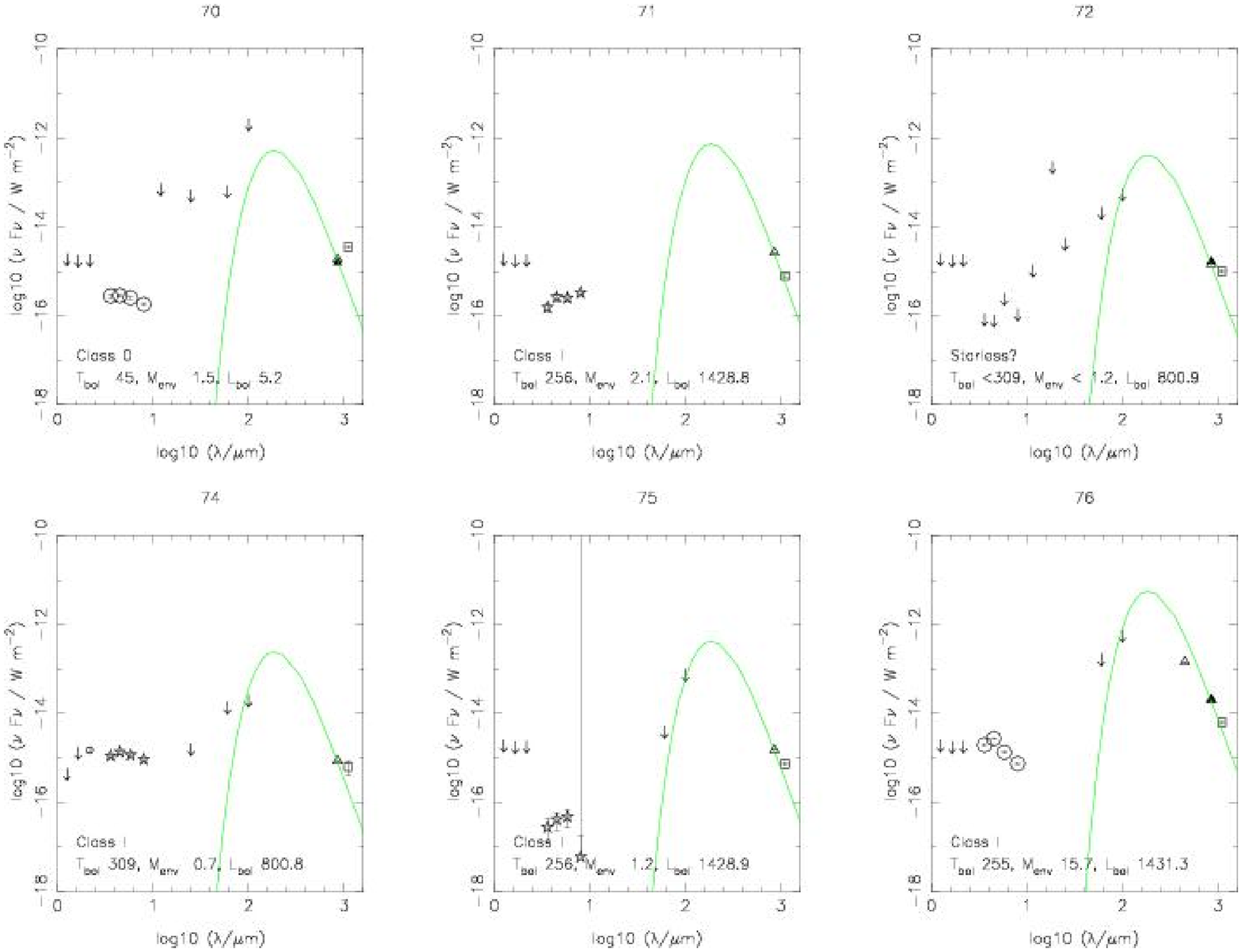}\\
\includegraphics[scale=0.60,angle=-90]{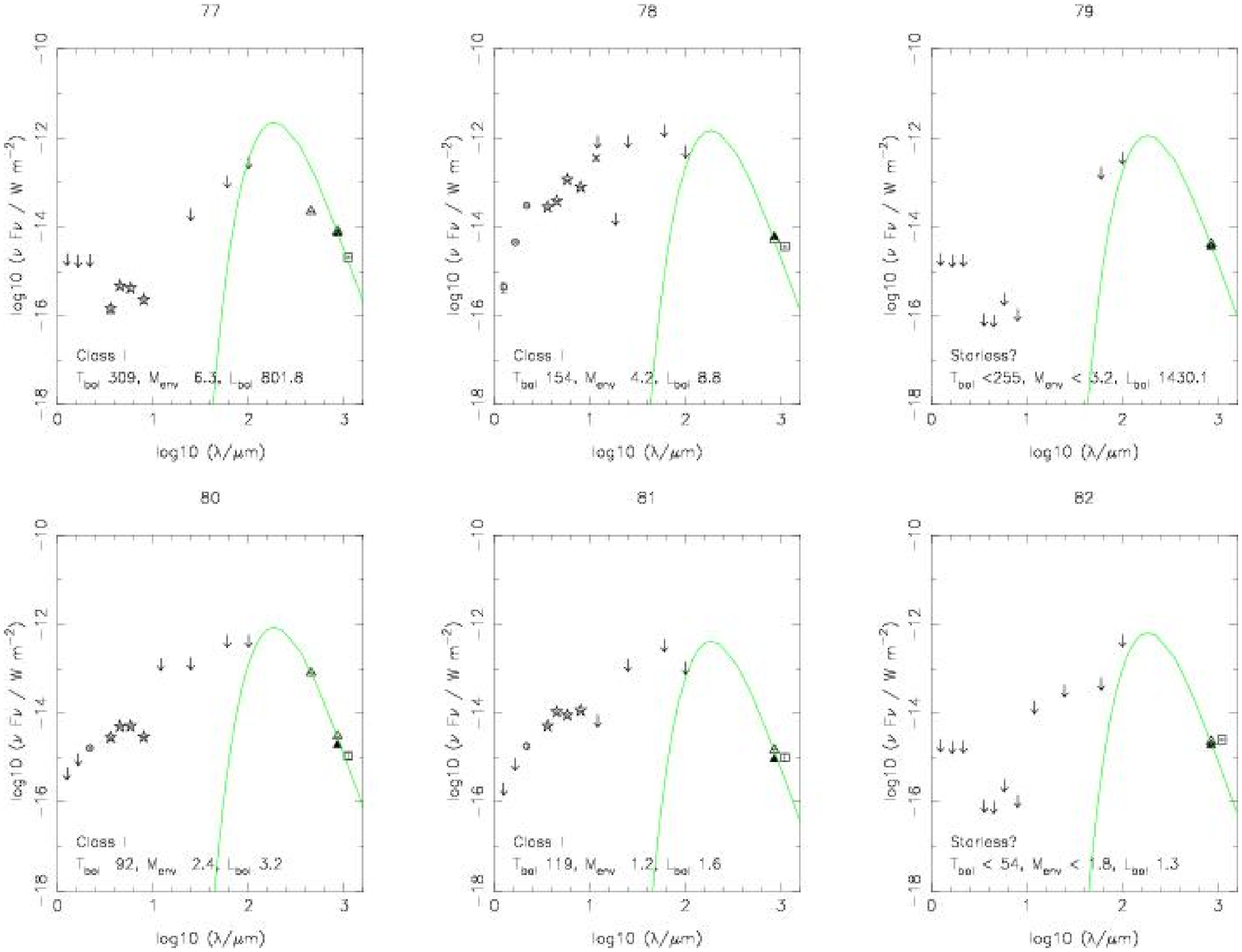}\\
\caption{Continued.}
\end{figure*}
\setcounter{figure}{6}
\begin{figure*}
\includegraphics[scale=0.60,angle=-90]{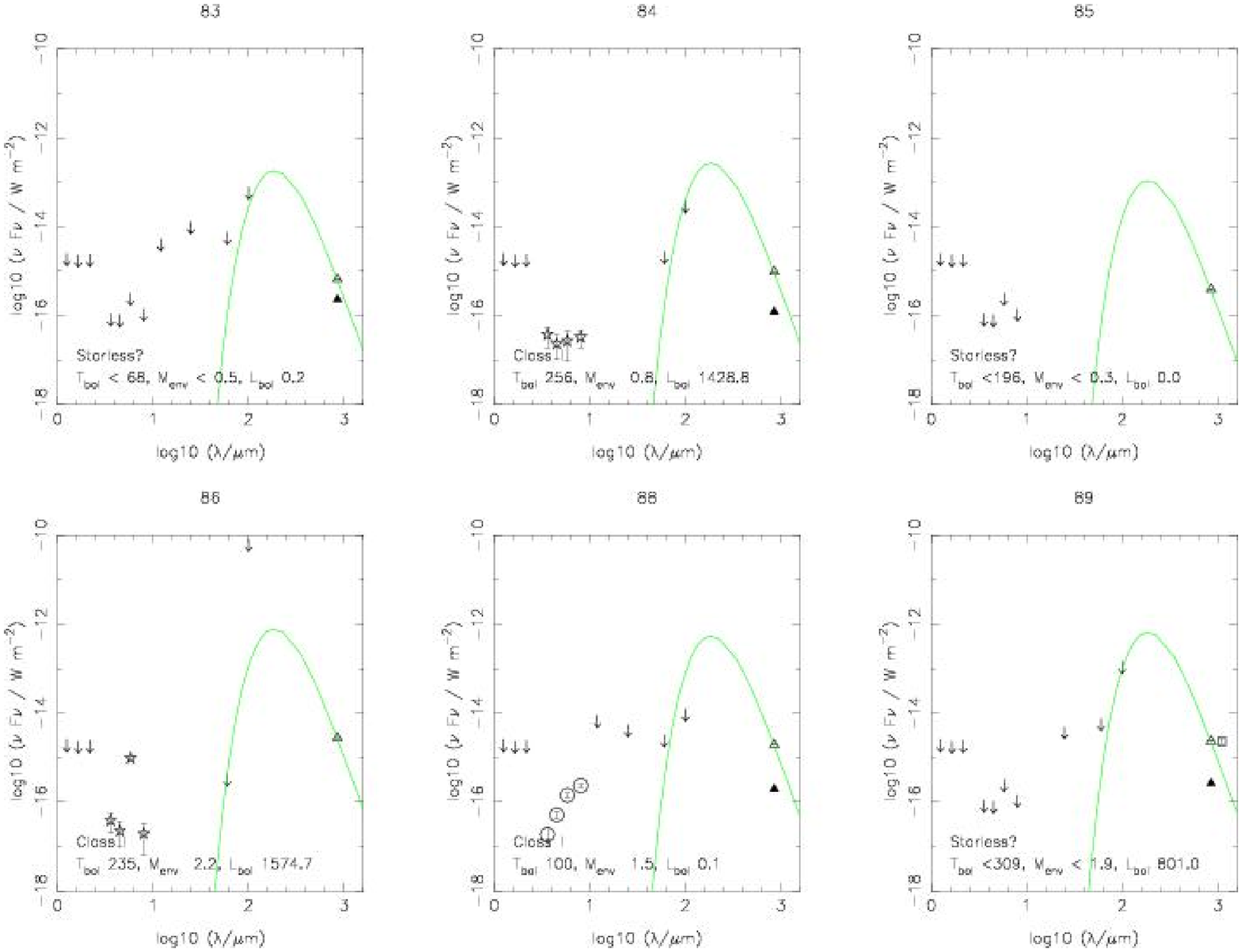}\\
\includegraphics[scale=0.60,angle=-90]{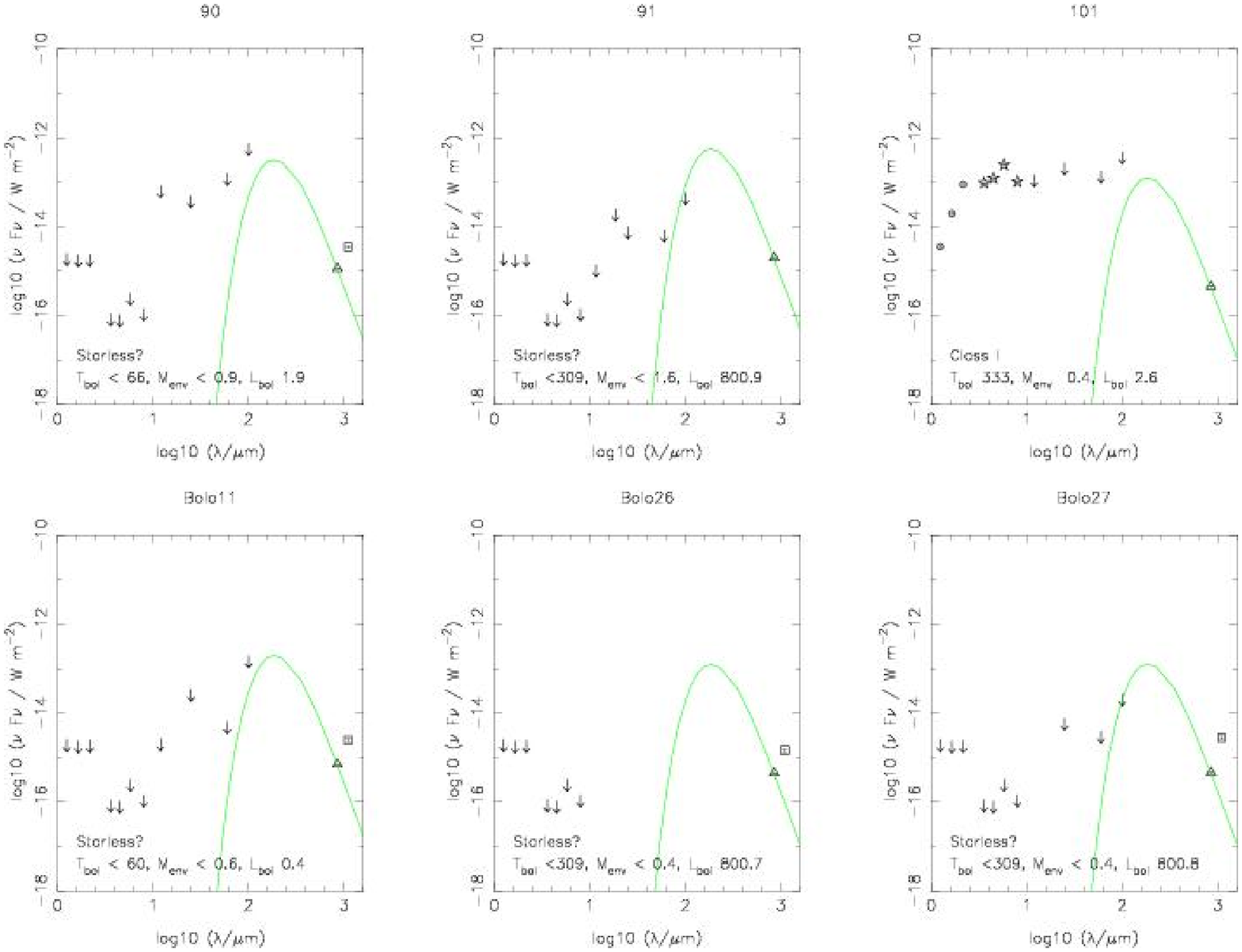}\\
\caption{Continued.}
\end{figure*}
\setcounter{figure}{6}
\begin{figure*}
\includegraphics[scale=0.60,angle=-90]{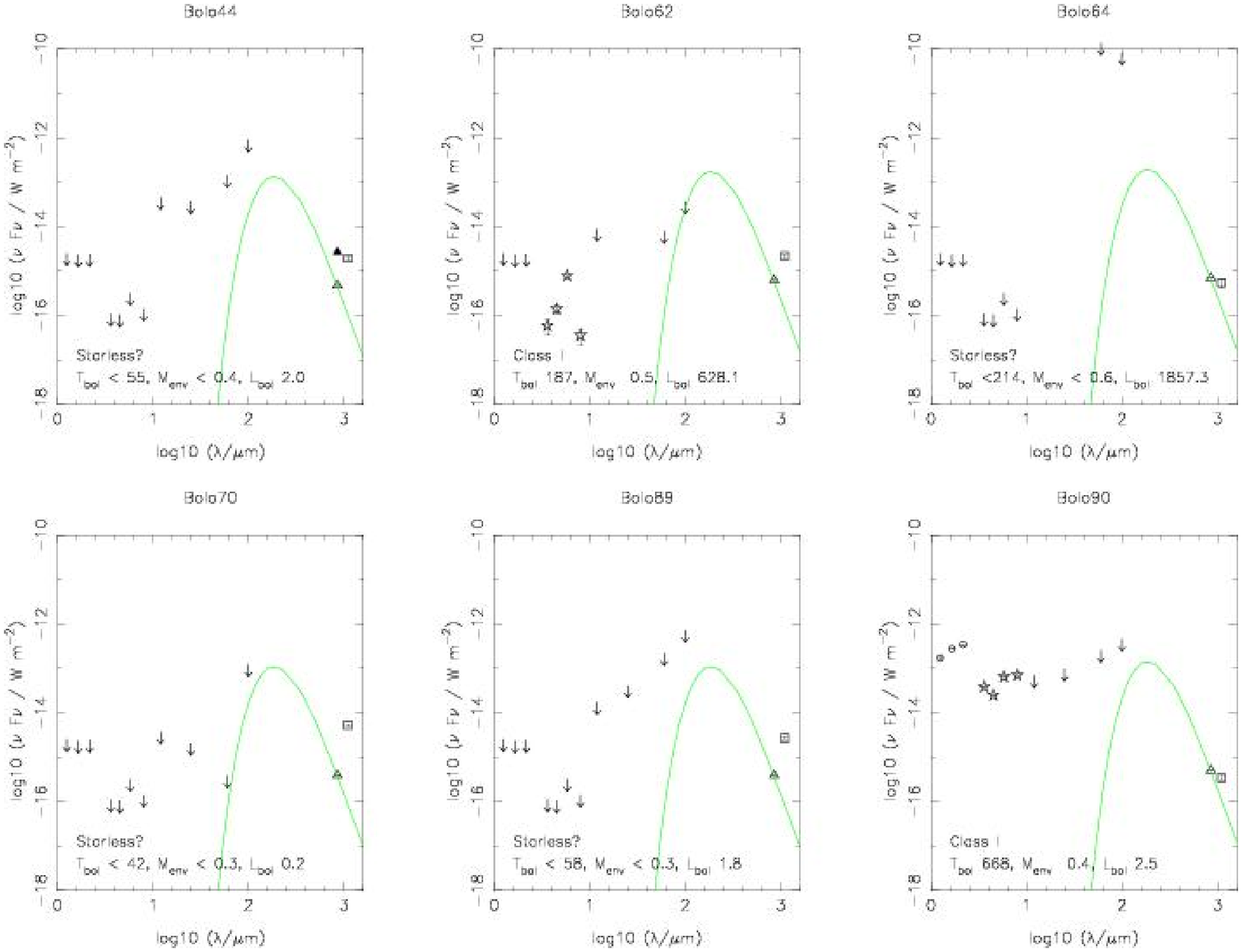}\\
\includegraphics[scale=0.60,angle=-90]{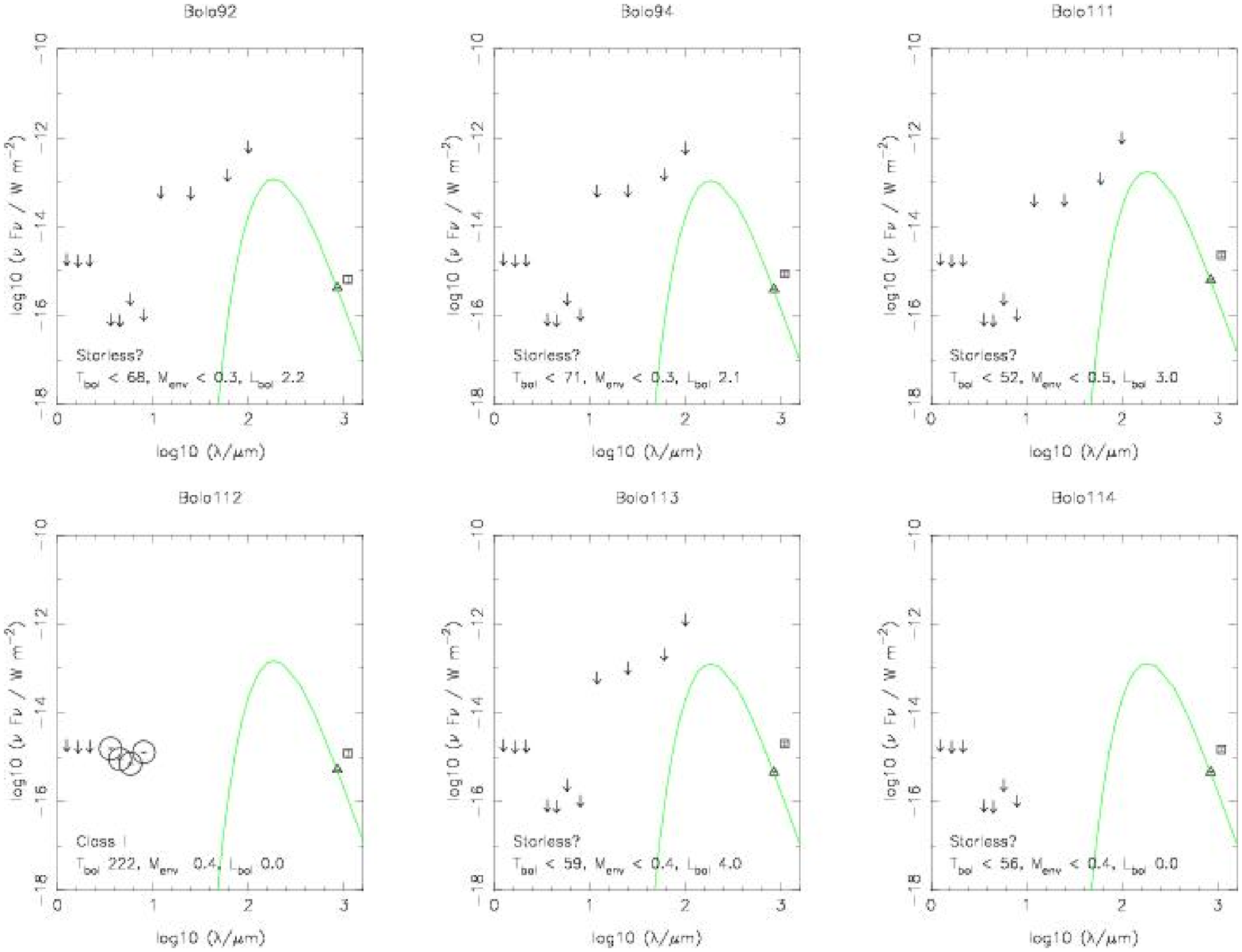}\\
\caption{Continued.}
\end{figure*}
\setcounter{figure}{6}
\begin{figure*}
\includegraphics[scale=0.60,angle=-90]{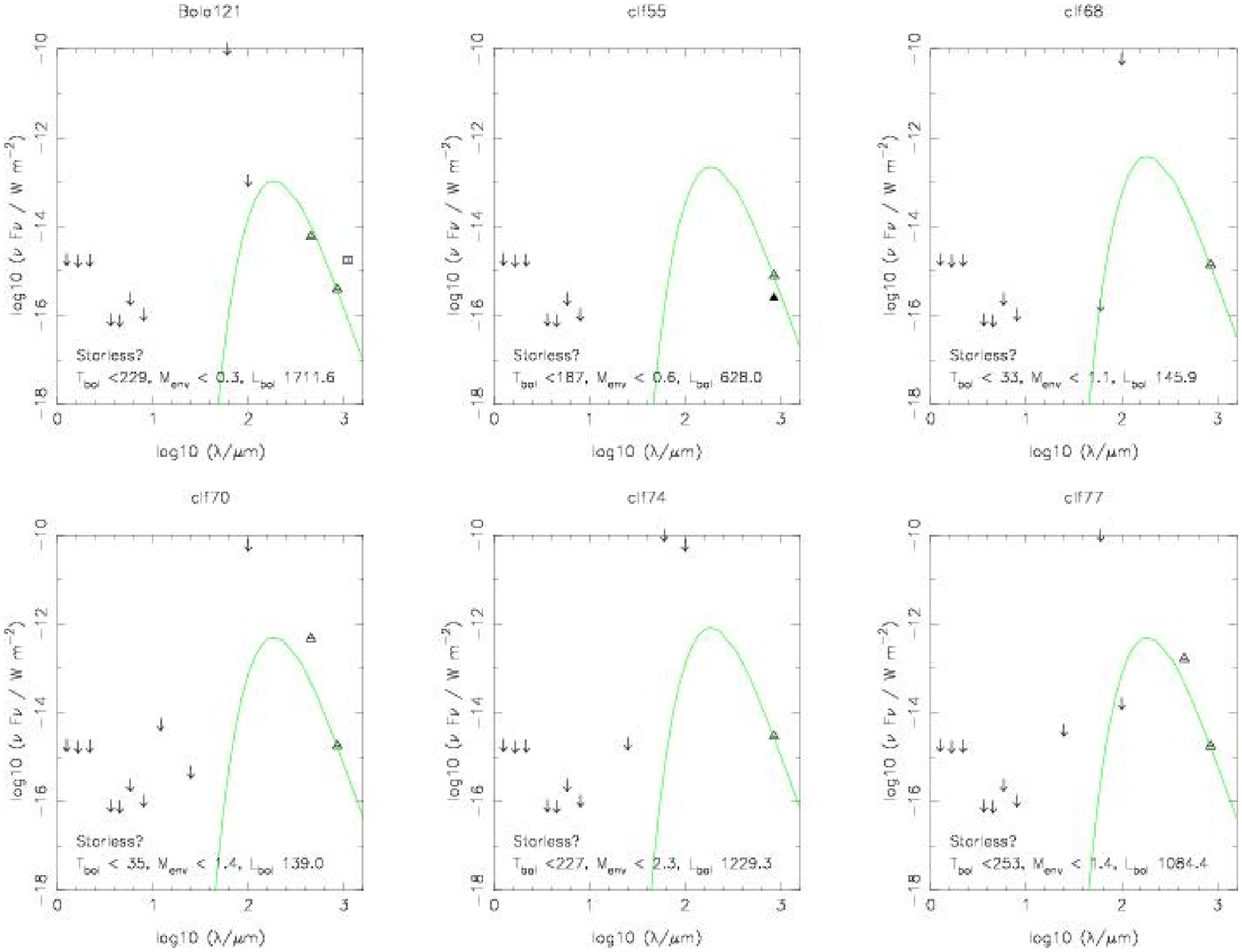}\\
\includegraphics[scale=0.60,angle=-90]{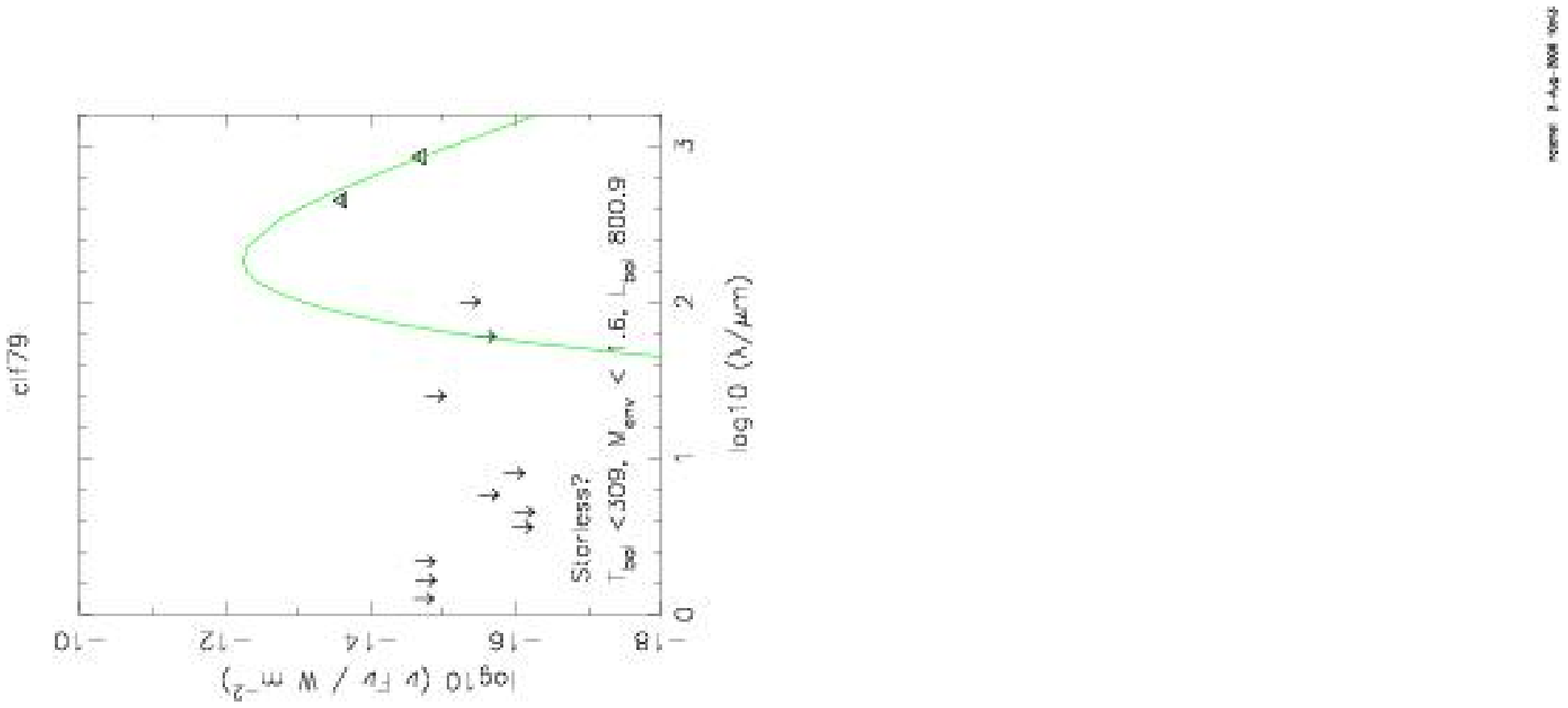}\\
\caption{Continued.}
\end{figure*}

\onecolumn
\include{class_properties_table}

\include{class_crossref_table}

%\include{testlong}
% \caption{Source properties derived from SEDs.  Integrated 850\micron\ flux, integrated 450\micron\ flux, Bolometric temperature, Bolometric luminosity, and envelope mass calculated based on bolometric temperature.}
% \label{tbl:sourcelist}
%\end{centering}

\begin{tjh}
%\section*{Appendix}
\appendix
\label{tjh_appendix}

\section{Class~0 models}

Below we provide details of the Class~0 structure used in
the radiative transfer modelling.

\subsection{Density structure}

We describe the geometry of the Class~0 sources using a similar
density structure to that of \cite{whitney03}. In summary, the
geometry consists of a central source of mass $M_*$, radius $R_*$, and
temperature $T_{\rm eff}=4000$\,K surrounded by a canonical flared disc \citep{SS1973}:
\begin{equation}
\rho(r, z) = \rho_0 \left( \frac{R_*}{r} \right)^\alpha  \exp \left(-\frac{1}{2}
  \left( \frac{z}{h} \right) ^2 \right)
\end{equation}
where $r$ is the radial distance in the midplane and $z$ is the
distance perpendicular to the midplane. The flare is imposed
by a power-law dependence of the scale-height $h$ on $r$:
$h = h_0 (r / R_*)^\beta$. The disc mass $M_{\rm disc}$ is then
used to define $\rho_0$ via
\begin{equation}
\rho_ 0 = \frac{M_{\rm disc} (\beta-\alpha+2)} {  (2 \pi)^{3/2}  h_0  R_*^{\alpha-\beta} 
         (R_{\rm outer}-R_{\rm inner})^{\beta-\alpha+2}}
\end{equation}
For all the models described below we adopt the following disc
parameters: $\beta=1.25$, $\alpha=2.25$, $h=0.01 R_*$, $M_{\rm
disc}=0.01$\,M$_\odot$, $R_{\rm inner}=7.5$\,R$_\odot$, and $R_{\rm
outer}=10$\,AU \citep{whitney03}.

The disc/star system is enshrouded by an infalling envelope of mass
$M_{\rm env}$, which has a density structure described by a
rotationally flattened infalling envelope \citep{Ulrich1976}:
\begin{equation}
  \rho = \frac{\dot{M}_{\rm env}} {4 \pi} \left( \frac{G M_*}{r^3}
       \right)^{-1/2} 
       \left( 1 + \frac{\mu}{\mu_0} \right)^{-1/2}
       \left( \frac{\mu}{\mu_0} + \frac{2 \mu_0^2  R_c}{r} \right)^{-1}
\end{equation}
where $\dot{M}_{\rm env}$ is the mass infall rate, $R_c$ is the
centrifugal radius, $\mu = \cos \theta = z/\sqrt{(z^2 + r^2)}$ and
$\mu_0$ is the cosine of the polar angle of a streamline of infalling
material as $r \rightarrow \infty$, and is found by solving:
\begin{equation}
\mu_0^3 + \mu_0 ( r/R_c -1) - \mu (r/R_c) = 0
\end{equation}
We fix the inner radius of the envelope at $R_{\rm inner}$ and set the
outer radius to 5000\,AU. We include narrow, fixed opening angle
($5^\circ$) opening angle, low density ($10^{-24}$ g\,cm$^{-3}$)
cavities in the polar directions. The density structure of the inner
region of this Class~0 model can be seen in
Figure~\ref{fig:class0_density}.

\begin{figure}
\begin{center}
\includegraphics[clip,  scale=0.45]{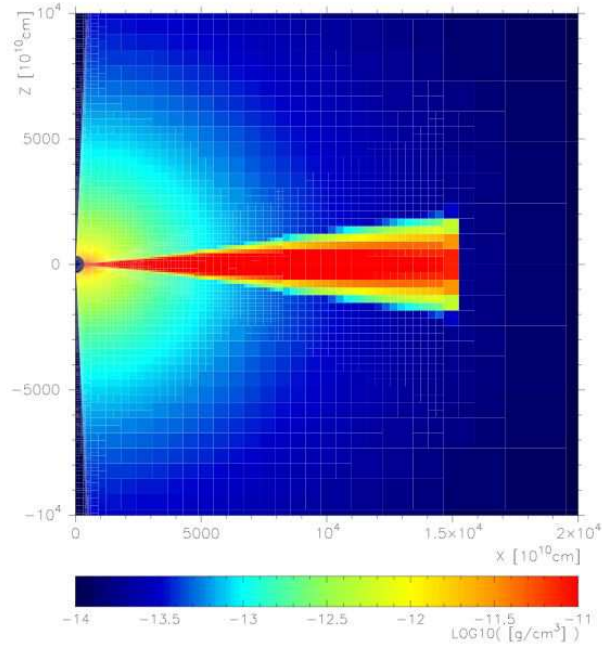}
\end{center}
\caption{The density structure of the inner region of the Class~0
  source used in the radiative-transfer modelling.}
\label{fig:class0_density}
\end{figure}

\begin{figure}
\begin{center}
\includegraphics[clip,  width=60mm,angle=-90]{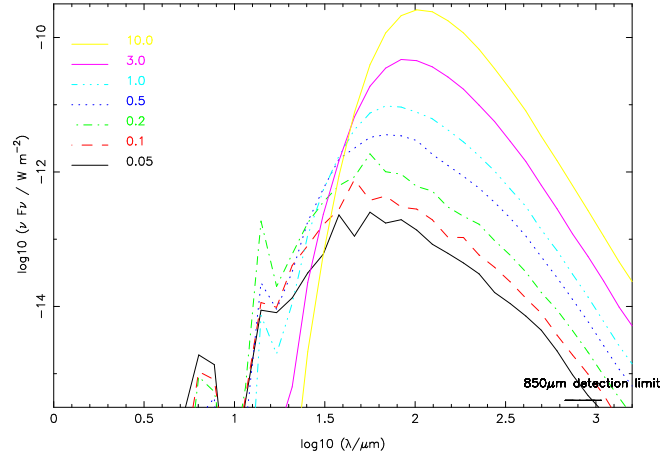}
\end{center}
\caption{Model SEDs for Class 0 objects. The flux of the object, in
  $\lambda F_\lambda$ (W\,m$^{-3}$) is plotted against wavelength in microns.}
\label{fig:model_seds}
\end{figure}

\begin{table}
\caption{Model parameters. The envelope mass ($M_{\rm env}$), stellar mass ($M_*$), and
stellar luminosity ($L_*$) are listed in solar units.}
\begin{center}
\begin{tabular}{ccc}
\hline
$M_{\rm env}$ ($M_\odot$) & $M_*$ ($M_\odot$) & $L_*$ ($L_\odot$) \\
\hline
10.00 &3.33 & 75.9 \\
3.00 & 1.00 & 14.9 \\
1.00 & 0.33 & 3.4   \\
0.50 & 0.17 & 1.3 \\
0.20 & 0.07  & 0.38 \\
0.10 & 0.03  & 0.15 \\
0.05 & 0.02  & 0.06 \\
\hline
\end{tabular}
\end{center}
\end{table}
\label{tab:model_params}

\subsection{Dust properties}
\begin{tjh}
We adopt a standard MRN \citep{MRN1977} grain-size distribution
\begin{equation}
n(a)\,da \propto a^{-3.5}\, da
\end{equation}
where $n(a)$ is the number density of grains with diameter $a$. We use
minimum and maximum grain sizes of 0.005\,$\mu$m and 1\,$\mu$m
respectively.
\end{tjh}
We assume that the dust is composed of 62.5\% `astronomical' silicate and 37.5\%
graphite grains, with the usual 2:1 ratio of ortho- and para-
graphite \citep{Draine1993}. The cross-sections and Mie phase matrices were
computed directly from the refractive indices tabulated by
\cite{DL1984}. A dust:gas ratio of 1:100 was used for all
models.
\end{tjh}

\end{document}

%% file: class_properties_table.tex
%From class_table.tex autogenerated by maketable.py plus class_sources.tex to form landscape table
% in aa longtable format
\begin{longtable}{c|r r|r r|r r r|l| p{4cm}}

\caption{ \label{tbl:sourceproperties}
Source properties: integrated 850\micron\ flux $F_{850}$; integrated 450\micron\ flux $F_{450}$;  bolometric luminosity \Lbol; envelope mass \Menv\ calculated at 10~K; the three evolutionary indicators -- bolometric temperature \Tbol; \Lsmm/Lbol; and $F_{3.6}/F_{850}$ (the number in brackets is the power of ten); and classification.  Sources are classed as starless (S) if there is no infrared source or outflow, otherwise the classification is based on the three evolutionary indicators discussed in Sect.\ref{sect:classification}, which, giving the condition for a Class~I are: $T_\mathrm{bol} > 70$~K; $L_\mathrm{bol}/L_\mathrm{smm} > 3000$; and $F_{3.6}/F_{850} > 0.003$.  The final classification is based on the majority of the indicators with results of individual indicators given in brackets in the order listed here. The commonly used source name is given in the last column; see \ref{tbl:crossrefs} for alternative names and references.}\\
\hline\hline
Source &$F_{850}$ &$F_{450}$ &\Lbol &\Menv &\Tbol &\Lbol/\Lsmm &$F_{3.6}/F_{850}$ &Class &Name\\
       &Jy  &Jy &\Lsun &\Msun\ &K &$\times 1/3000$&&\\
\hline
\endfirsthead
\caption{continued.}\\
\hline\hline
Source &$F_{850}$ &$F_{450}$ &\Lbol &\Menv &\Tbol &\Lbol/\Lsmm &$F_{3.6}/F_{850}$ &Class &Name$^1$\\
       &Jy  &Jy &\Lsun &\Msun\ &K &$\times 1/3000$&&\\
\hline
\endhead
\hline
\endfoot
     1  &$6.31$ &      37.8 &$ 3.7$ &$17.7$  &$  53$     &$0.5 $    &$2.4(-5  )$ &0  (000)  &b1-c\\
     2  &$9.31$ &      52.2 &$<2.5$ &$26.1$  &$ <25$     &$<0.2$    &$ <1.0(-5)$ &0  (000)  &b1-bS\\
     4  &$5.37$ &      50.4 &$ 2.6$ &$15.0$  &$  32$     &$0.4 $    &$5.1(-5  )$ &0  (000)  &b1-d\\
     5  &$3.67$ &         - &$<0.3$ &$10.3$  &$ <33$     &$<0.1$    &$ <2.6(-5)$ &S         &\\                                                                                                                           
     7  &$1.00$ &         - &$ 1.3$ &$ 2.8$  &$ 158$     &$1.2 $    &$1.4(-2  )$ &I  (III)  &IRAS 03301+3057, B1 SMM6\\
    10  &$0.32$ &         - &$ 0.4$ &$ 0.9$  &$ 117$     &$1.1 $    &$7.0(-3  )$ &I  (III)  &B1 SMM11\\
    12  &$8.19$ &      83.3 &$ 4.5$ &$22.9$  &$  31$     &$0.5 $    &$1.8(-4  )$ &0  (000)  &HH211\\
    13  &$3.59$ &      17.6 &$ 1.9$ &$10.1$  &$  59$     &$0.5 $    &$4.6(-4  )$ &0  (000)  &IC348 MMS\\
    14  &$1.98$ &      15.9 &$ 5.6$ &$ 5.5$  &$ 180$     &$2.6 $    &$5.8(-2  )$ &I  (III)  &\\                                                                                                  
    15  &$1.57$ &       2.0 &$ 1.9$ &$ 4.4$  &$  67$     &$1.1 $    &$9.1(-4  )$ &0  (0I0)  &\\
    16  &$1.43$ &       7.2 &$<1.6$ &$ 4.0$  &$<103$     &$<1.0$    &$ <6.7(-5)$ &S         &\\
    17  &$1.18$ &         - &$<2.0$ &$ 3.3$  &$ <78$     &$<1.6$    &$ <8.1(-5)$ &S         &\\
    18  &$0.61$ &         - &$<0.6$ &$ 1.7$  &$ <86$     &$<0.8$    &$ <1.6(-4)$ &S         &\\
    19  &$0.68$ &         - &$<1.2$ &$ 1.9$  &$ <61$     &$<1.6$    &$ <1.4(-4)$ &S         &\\
    20  &$1.44$ &         - &$<0.9$ &$ 4.0$  &$ <70$     &$<0.5$    &$ <6.7(-5)$ &S         &\\
    21  &$0.62$ &         - &$<1.2$ &$ 1.7$  &$ <81$     &$<1.8$    &$ <1.6(-4)$ &S         &\\
    23  &$0.32$ &         - &$<1.2$ &$ 0.9$  &$ <89$     &$<3.3$    &$ <3.0(-4)$ &S         &\\
    24  &$0.56$ &         - &$<0.9$ &$ 1.6$  &$ <93$     &$<1.4$    &$ <1.7(-4)$ &S         &\\
    25  &$0.39$ &         - &$<1.0$ &$ 1.1$  &$ <85$     &$<2.4$    &$ <2.5(-4)$ &S         &\\
    26  &$152.43$ &         &$<0.8$ &$2.7$ - &$ <84$    &$<0.7$    &$ <1.0(-4)$ &S          &\\
    27  &$11.03$ &         - &$ 5.4$ &$30.9$  &$  53$    &$0.5 $    &$1.1(-4  )$  &0  (000) &L1448 NW \\
    28  &$5.83$ &     196.9 &$17.0$ &$16.3$  &$  53$     &$2.7 $    &$3.3(-4  )$ &0  (0I0)  &L1448 N A/B\\
    29  &$6.51$ &      68.4 &$ 6.3$ &$18.2$  &$  49$     &$0.9 $    &$1.1(-3  )$ &0  (000)  &L1448C\\
    30  &$5.85$ &      51.1 &$ 4.1$ &$16.4$  &$  50$     &$0.7 $    &$1.2(-4  )$ &0  (000)  &L1448~IRS2\\
    31  &$4.26$ &      26.5 &$ 1.9$ &$11.9$  &$  38$     &$0.4 $    &$1.7(-5  )$ &0  (000)  &\\
    32  &$4.36$ &       2.9 &$<0.2$ &$12.2$  &$ <20$     &$<0.0$    &$ <2.2(-5)$ &S         &\\
    35  &$1.89$ &      18.4 &$ 7.0$ &$ 5.3$  &$ 105$     &$3.4 $    &$5.3(-4  )$ &I  (II0)  &L1455 FIR4\\
    36  &$1.91$ &       9.8 &$ 3.1$ &$ 5.3$  &$  59$     &$1.5 $    &$5.6(-4  )$ &0  (0I0)  &\\
    37  &$2.52$ &      20.5 &$ 3.3$ &$ 7.1$  &$ 672$     &$1.2 $    &$5.6(-2  )$ &I  (III)  &L1455 PP9\\
    39  &$0.72$ &       1.6 &$ 0.8$ &$ 2.0$  &$ 394$     &$1.0 $    &$7.3(-2  )$ &I  (I0I)  &L1455 FIR1/2 \\
    40  &$0.89$ &         - &$<0.6$ &$ 2.5$  &$ <63$     &$<0.6$    &$ <1.1(-4)$ &S         &\\
    41  &$18.61$ &     131.0 &$ 8.5$ &$52.1$  &$  35$    &$0.4 $    &$9.1(-6  )$  &0  (000) &NGC1333 IRAS 4A \\
    42  &$9.00$ &      94.6 &$ 7.6$ &$25.2$  &$  39$     &$0.8 $    &$8.9(-5  )$ &0  (000)  &NGC1333 IRAS 4B \\
    43  &$12.70$ &     159.9 &$29.5$ &$35.6$  &$ 180$    &$2.1 $    &$1.1(-2  )$  &I  (III) &NGC1333 SVS13 \\
    44  &$9.73$ &      95.6 &$22.7$ &$27.2$  &$  58$     &$2.2 $    &$2.9(-4  )$ &0  (0I0)  &NGC1333 IRAS 2A \\
    45  &$8.82$ &      91.3 &$10.2$ &$24.7$  &$ 164$     &$1.1 $    &$1.6(-2  )$ &I  (III)  &NGC1333 ASR 114\\
    46  &$6.10$ &      32.8 &$ 6.1$ &$17.1$  &$  67$     &$0.9 $    &$6.4(-4  )$ &0  (000)  &NGC1333 ASR 32/33\\
    47  &$4.33$ &      26.3 &$<1.8$ &$12.1$  &$ <54$     &$<0.4$    &$ <2.2(-5)$ &0  (000)  &NGC1333 SK31\\
    48  &$2.92$ &      20.9 &$ 3.2$ &$ 8.2$  &$  49$     &$1.0 $    &$2.0(-4  )$ &0  (000)  &NGC1333 IRAS 4C\\
    49  &$2.59$ &      21.1 &$ 6.8$ &$ 7.3$  &$ 140$     &$2.4 $    &$1.2(-2  )$ &I  (III)  &NGC1333 SK6\\
    50  &$2.58$ &       7.1 &$ 5.6$ &$ 7.2$  &$  88$     &$2.0 $    &$2.7(-4  )$ &I  (II0)  &NGC1333 HH 7-11 MMS~4\\
    51  &$3.01$ &      10.1 &$<1.5$ &$ 8.4$  &$ <37$     &$<0.5$    &$ <3.2(-5)$ &S         &NGC1333 SK16\\
    52  &$2.17$ &         - &$ 4.4$ &$ 6.1$  &$  57$     &$1.9 $    &$3.5(-4  )$ &0  (0I0)  &NGC1333 HH7--11 MMS~6\\
    53  &$2.14$ &      29.7 &$<3.7$ &$ 6.0$  &$ <54$     &$<1.6$    &$ <4.5(-5)$ &S         &NGC1333 SK26\\
    54  &$3.01$ &      78.8 &$106.9$ &$ 8.4$ &$ 138$     &$32.7$    &$ 1.6(-2 )$  &I  (III) &NGC1333 SK28 \\
    55  &$2.96$ &      14.4 &$ 1.0$ &$ 8.3$  &$  50$     &$0.3 $    &$5.8(-6  )$ &0  (000)  &\\
    56  &$1.96$ &         - &$19.1$ &$ 5.5$  &$ 124$     &$9.0 $    &$7.1(-2  )$ &I  (III)  &NGC1333 SK29\\
    57  &$2.10$ &      15.1 &$<8.0$ &$ 5.9$  &$ <81$     &$<3.5$    &$ <4.6(-5)$ &S         &NGC1333 SK33\\
    58  &$1.56$ &         - &$ 0.3$ &$ 4.4$  &$  82$     &$0.2 $    &$7.8(-5  )$ &0  (I00)  &\\
    59  &$2.21$ &         - &$<1.3$ &$ 6.2$  &$ <83$     &$<0.5$    &$ <4.3(-5)$ &S         &\\
    60  &$2.95$ &         - &$<1.2$ &$ 8.3$  &$<215$     &$<0.4$    &$ <3.3(-5)$ &S         &\\
    61  &$1.42$ &       2.5 &$ 1.0$ &$ 4.0$  &$  64$     &$0.7 $    &$3.8(-4  )$ &0  (000)  &\\
    62  &$1.21$ &       6.1 &$ 2.7$ &$ 3.4$  &$  50$     &$2.0 $    &$7.1(-5  )$ &0  (0I0)  &NGC1333 SK18\\
    63  &$0.76$ &         - &$ 2.8$ &$ 2.1$  &$ 103$     &$3.4 $    &$5.1(-2  )$ &I  (III)  &NGC1333 SK32\\
    64  &$0.92$ &         - &$<0.3$ &$ 2.6$  &$ <52$     &$<0.3$    &$ <1.0(-4)$ &S         &NGC1333 Per 4A3/4D\\
    65  &$0.72$ &       2.1 &$ 0.4$ &$ 2.0$  &$  56$     &$0.5 $    &$3.1(-3  )$ &0  (00I)  &NGC1333 IRAS 4B1\\
    66  &$1.03$ &         - &$<4.2$ &$ 2.9$  &$ <65$     &$<3.8$    &$ <9.3(-5)$ &S         &NGC1333 SK30\\
    67  &$1.51$ &         - &$ 8.4$ &$ 4.2$  &$ 127$     &$5.1 $    &$1.8(-2  )$ &I  (III)  &\\
    68  &$1.34$ &       8.2 &$ 1.3$ &$ 3.7$  &$  69$     &$0.9 $    &$7.4(-3  )$ &0  (00I)  &\\
    69  &$0.23$ &         - &$0.04$ &$ 0.6$  &$ 436$     &$0.2 $    &$1.8(-2  )$ &I  (I0I)  &\\
    70  &$0.53$ &         - &$ 2.6$ &$ 1.5$  &$  45$     &$4.5 $    &$6.3(-4  )$ &0  (0I0)  &NGC1333 SK22\\
    71  &$0.75$ &         - &$0.03$ &$ 2.1$  &$  37$     &$0.0 $    &$2.5(-4  )$ &0  (000)  &NGC1333 HH 340B\\
    72  &$0.42$ &         - &$<0.1$ &$ 1.2$  &$ <64$     &$<0.3$    &$ <2.3(-4)$ &S         &\\                                                                                                                                    
    74  &$0.24$ &         - &$0.06$ &$ 0.7$  &$ 144$     &$0.2 $    &$5.4(-3  )$ &I  (I0I)  &\\
    75  &$0.43$ &         - &$ 0.1$ &$ 1.2$  &$  67$     &$0.3 $    &$7.7(-5  )$ &0  (000)  &\\
    76  &$5.60$ &      21.5 &$ 2.4$ &$15.7$  &$ 109$     &$0.4 $    &$4.2(-4  )$ &0  (I00)  &B1 SMM5\\
    77  &$2.26$ &       3.3 &$ 0.7$ &$ 6.3$  &$  60$     &$0.3 $    &$7.7(-5  )$ &0  (000)  &B1 SMM7\\
    78  &$1.49$ &         - &$ 4.7$ &$ 4.2$  &$ 181$     &$2.9 $    &$2.3(-2  )$ &I  (III)  &B5 IRS1\\
    79  &$1.16$ &         - &$<1.4$ &$ 3.2$  &$<175$     &$<1.1$    &$ <8.3(-5)$ &S         &\\
    80  &$0.86$ &      12.1 &$ 1.9$ &$ 2.4$  &$  86$     &$2.0 $    &$3.9(-3  )$ &I  (III)  &IRAS 03235+3004\\                                                                                                         
    81  &$0.42$ &         - &$ 0.8$ &$ 1.2$  &$ 133$     &$1.8 $    &$1.5(-2  )$ &I  (III)  &IRAS 03271+3013\\                                                                                                         
    82  &$0.65$ &         - &$<0.7$ &$ 1.8$  &$ <53$     &$<0.9$    &$ <1.5(-4)$ &S         &B1 SMM1\\
    83  &$0.19$ &         - &$<0.10$ &$ 0.5$ &$ <68$     &$<0.6$    &$ <6.8(-4)$  &S        &\\
    84  &$0.28$ &         - &$0.06$ &$ 0.8$  &$  74$     &$0.2 $    &$1.6(-4  )$ &0  (I00)  &\\
    85  &$0.11$ &         - &$<0.00$ &$ 0.3$ &$<196$     &$<0.1$    &$ <3.9(-3)$  &S        &IRAS 03254+3050\\                                                                                                       
    86  &$0.79$ &         - &$0.06$ &$ 2.2$  &$  49$     &$0.1 $    &$5.8(-5  )$ &0  (000)  &\\
    88  &$0.55$ &         - &$0.05$ &$ 1.5$  &$  83$     &$0.1 $    &$4.0(-5  )$ &0  (I00)  &\\
    89  &$0.66$ &         - &$<0.2$ &$ 1.9$  &$ <45$     &$<0.2$    &$ <1.4(-4)$ &S         &B1 SMM9\\
    90  &$0.33$ &         - &$<0.9$ &$ 0.9$  &$ <66$     &$<2.7$    &$ <2.9(-4)$ &S         &Per 7\\
    91  &$0.58$ &         - &$<0.1$ &$ 1.6$  &$ <74$     &$<0.2$    &$ <1.7(-4)$ &S         &Per 4A \\
   101  &$0.13$ &         - &$ 1.6$ &$ 0.4$  &$ 463$    &$35.2$    &$ 2.7(0)$    &I  (III)  &IRAS 03410+3152\\
Bolo11  &$0.20$ &         - &$<0.2$ &$ 0.6$  &$ <60$     &$<1.0$    &$ <4.8(-4)$ &S         &\\
Bolo26  &$0.13$ &         - &$<0.01$ &$ 0.4$ &$ <64$     &$<0.0$    &$ <8.7(-4)$  &S        &\\
Bolo27  &$0.13$ &         - &$<0.05$ &$ 0.4$ &$<103$     &$<0.4$    &$ <9.4(-4)$  &S        &\\
Bolo44  &$0.14$ &         - &$<1.0$ &$ 0.4$  &$ <55$     &$<10.6$   &$ <1.1(-3)$ &S         &\\
Bolo62  &$0.18$ &         - &$0.07$ &$ 0.5$  &$ 103$     &$0.4 $    &$4.0(-4  )$ &0  (I00)  &\\
Bolo64  &$0.20$ &         - &$<0.1$ &$ 0.6$  &$<237$     &$<0.5$    &$ <4.8(-4)$ &S         &\\
Bolo70  &$0.11$ &         - &$<0.1$ &$ 0.3$  &$ <42$     &$<1.7$    &$ <1.5(-3)$ &S         &\\
Bolo89  &$0.11$ &         - &$<0.9$ &$ 0.3$ &$ <58$      &$<13.2$   &$ <1.5(-3)$ &S         &\\
Bolo90  &$0.14$ &         - &$ 1.7$ &$ 0.4$  &$ 943$     &$24.9$    &$ 7.3(-1 )$ &I  (III)  &IRAS 03380+3135\\                                                                                                     
Bolo92  &$0.12$ &         - &$<1.1$ &$ 0.3$ &$ <68$      &$<19.2$   &$ <1.9(-3)$ &S         &\\
Bolo94  &$0.11$ &         - &$<1.1$ &$ 0.3$ &$ <71$      &$<28.4$   &$ <2.8(-3)$ &S         &\\
Bolo111  &$0.18$ &         - &$<1.5$ &$ 0.5$  &$ <52$    &$<7.6$    &$ <5.3(-4)$  &S        &\\
Bolo112  &$0.15$ &         - &$0.01$ &$ 0.4$  &$ 222$    &$0.1 $    &$1.3(-2  )$  &I  (I0I) &\\
Bolo113  &$0.13$ &         - &$<2.0$ &$ 0.4$ &$ <59$     &$<14.2$   &$ <7.4(-4)$  &S        &\\
Bolo114  &$0.13$ &         - &$<0.01$ &$ 0.4$ &$ <56$    &$<0.0$    &$ <7.2(-4)$  &S       &\\
Bolo121  &$0.11$ &       0.9 &$<1.1$ &$ 0.3$  &$<231$    &$<34.2$   &$ <3.3(-3)$  &S        &\\
 clf55  &$0.23$ &         - &$<0.01$ &$ 0.6$ &$ <67$     &$<0.0$    &$ <7.8(-4)$  &S        &\\
 clf68  &$0.39$ &         - &$<0.03$ &$ 1.1$ &$ <42$     &$<0.1$    &$ <2.5(-4)$  &S        &\\
 clf70  &$0.51$ &      70.2 &$<5.1$ &$ 1.4$  &$ <34$     &$<9.3$    &$ <1.9(-4)$ &S         &\\
 clf74  &$0.83$ &         - &$<0.2$ &$ 2.3$  &$ <84$     &$<0.2$    &$ <1.2(-4)$ &S         &\\
 clf77  &$0.50$ &      24.5 &$<1.1$ &$ 1.4$  &$ <26$     &$<2.0$    &$ <1.9(-4)$ &S         &\\
 clf79  &$0.58$ &       3.7 &$<0.2$ &$ 1.6$  &$ <27$   &$<0.3$    &$ <1.7(-4)$   &S         &\\
%&&&&&&&&&&&\\
\end{longtable}

%% file: class_crossref_table.tex
%From class_table_landscape.tex
% in aa longtable format
\begin{longtable}{c|l l|c |l| p{2.7cm}| p{7cm}}
\caption{ \label{tbl:crossrefs}
Source peak positions and cross-identification with \citet{paperI}, Bolocam \citep{enoch06}, 2MASS, and other names commonly associated with the source. }\\
\hline\hline
Source &RA(J2000) &Dec(J2000) &Paper~I &Bolocam &2MASS &Other associated source names\\
       &h:m:s   &d:$'$:$''$    &&&&\\
\hline
\endfirsthead
\caption{continued.}\\
\hline\hline
Source &RA(J2000) &Dec(J2000) &Paper~I &Bolocam &2MASS &Other associated source names\\
       &h:m:s   &d:$'$:$''$    &&&&\\
\hline
\endhead
\hline
\endfoot
     1 &03:33:17.9 &31:09:33.4  &1 &Bolo80  & &b1-c \citep{hkm99}, B1 SMM2 \citep{walawender05}\\                                                                                         
     2 &03:33:21.4 &31:07:30.7  &2,3 &Bolo81 &  &b1-bS; b1-bN \citep{hkm99}, B1 SMM1 \citep{walawender05}\\                                                                             
     4 &03:33:16.3 &31:06:54.3  &4 & ---       & &b1-d \citep{hkm99}, B1 SMM3 \citep{walawender05}\\                                                                                      
     5 &03:33:01.9 &31:04:23.2  &5,6,8,9,11&Bolo74 &  &none\\                                                                                                                           
     7 &03:33:16.5 &31:07:51.3  &7 & ---       &03331667+3107548 &IRAS 03301+3057, B1 SMM6 \citep{walawender05}\\                                                                         
    10 &03:33:27.3 &31:07:10.1  &10 & ---       & &B1 SMM11 \citep{walawender05}\\                                                                                                        
    12 &03:43:56.5 &32:00:49.9  &12 &Bolo103 & &HH211 \citep{mccaughrean94,chandlerricher00}; HH211 VLA2 \citep{arc01}\\                                                                  
    13 &03:43:56.9 &32:03:04.8  &13 &Bolo104 & &IC348 MMS \citep{eisloeffel03}\\                                                                                                          
    14 &03:44:43.9 &32:01:32.0  &14 &Bolo116 &03444389+3201373, 03444330+3201315 &none\\                                                                                                  
    15 &03:43:50.8 &32:03:24.2  &15 &Bolo102 &03435056+3203180 &none\\                                                                                                                    
    16 &03:44:01.0 &32:01:54.8  &16 &Bolo106 & &none\\                                                                                                                                    
    17 &03:43:57.9 &32:04:01.5  &17 &Bolo105 & &none\\                                                                                                                                    
    18 &03:44:03.0 &32:02:24.3  &18 &Bolo107 & &CXOPZ 15 \citep{preibisch01}\\                                                                                                            
    19 &03:44:36.8 &31:58:48.9  &19 &Bolo115 & &none\\                                                                                                                                    
    20 &03:44:05.5 &32:01:56.7  &20,22 &Bolo110 & &none\\                                                                                                                               
    21 &03:44:02.3 &32:02:48.5  &21 &Bolo107 & &none\\                                                                                                                                    
    23 &03:43:37.8 &32:03:06.2  &23 &Bolo99  & &none\\                                                                                                                                    
    24 &03:43:42.3 &32:03:23.2  &24 &---        & &none\\                                                                                                                                 
    25 &03:44:48.5 &32:00:30.8  &25 &Bolo117 & &none\\                                                                                                                                    
    26 &03:43:44.4 &32:02:55.7  &26 &Bolo100 &&none\\                                                                                                                                    
    27 &03:25:35.9 &30:45:30.0  &27,34 &Bolo8   &03253643+3045252 &L1448 NW \citep{barsony98,bachiller86nh3,cr90,looney00}\\                                                            
    28 &03:25:36.4 &30:45:15.0  &28,34 &Bolo8   &03253652+3045070, 03253643+3045252 &L1448 N A/B \citep{barsony98,cr90,looney00}, L1448 IRS 3A/3B \citep{bachiller86nh3}, IRS2 VLA6 \citep{anglada02} \\
    29 &03:25:38.8 &30:44:03.6  &29 &Bolo10  & &L1448C \citep{barsony98,bachiller91,cr90}\\                                                                                               
    30 &03:25:22.4 &30:45:10.7  &30 &Bolo5   & &IRAS 03222+3034, L1448~IRS2 \citep{olinger99}, L1448 IRS2 VLA4 \citep{anglada02}\\                                                        
    31 &03:25:25.9 &30:45:02.7  &31 &---        & &none\\                                                                                                                                 
    32 &03:25:49.0 &30:42:24.6  &32,33 &Bolo13  &&none\\                                                                                                                                
    35 &03:27:39.1 &30:13:00.6  &35 &Bolo22  & &IRAS 03245+3002, L1455 FIR4 \citep{tpb97}, RNO15 FIR \citep{rengel02}\\                                                                   
    36 &03:27:42.9 &30:12:28.5  &36 &Bolo23  & &none\\                                                                                                                                    
    37 &03:27:48.4 &30:12:08.8  &37,38 &Bolo24  &03274767+3012043 &L1455 PP9 \citep{tpb97}, GN 03.24.7 \citep{magakian03}\\                                                             
    39 &03:27:38.1 &30:13:57.3  &39 &Bolo21  &03273825+3013585 &L1455 FIR1/2 \citep{tpb97}\\                                                                                              
    40 &03:27:39.9 &30:12:09.8  &40 &---        & &none\\                                                                                                                                 
    41 &03:29:10.4 &31:13:30.0  &41 &Bolo48 & &NGC1333 IRAS 4A \citep{jcc87}, VLA25 \citep{rac97}, SK4 \citep{sandellknee01}\\                                                            
    42 &03:29:12.0 &31:13:10.0  &42 &Bolo48  & &NGC1333 IRAS 4B \citep{jcc87}, SK3 \citep{sandellknee01}\\                                                                                
    43 &03:29:03.2 &31:15:59.0  &43 &Bolo43  &03290375+3116039 &NGC1333 SVS13 \citep{svs76}, ASR 2 \citep{asr94}, IRAS 03259+3105,HH7-11 MMS1 \citep{chini97}, VLA 4a/b \citep{rac97}\\   
    44 &03:28:55.3 &31:14:36.4  &44 &Bolo38  & &NGC1333 IRAS 2A \citep{jcc87}\\                                                                                                           
    45 &03:29:01.4 &31:20:28.6  &45 &Bolo42  &03290116+3120244, 03290149+3120208, 03290051+3120284 &NGC1333 ASR 114 \citep{asr94}, VLA~42/43 \citep{rac97}, SK24 \citep{sandellknee01}\\  
    46 &03:29:11.0 &31:18:27.4  &46 &Bolo49  & &NGC1333 ASR 32/33 \citep{asr94}, IRAS 7 SM1/2 \citep{sandellknee01},VLA 2 \citep{sb86}\\                                                  
    47 &03:28:59.7 &31:21:34.2  &47 &Bolo40  & &NGC1333 SK31 \citep{sandellknee01}\\                                                                                                      
    48 &03:29:13.6 &31:13:55.0  &48 &---        & &NGC1333 IRAS 4C \citep{jcc87,looney00}, SK5 \citep{sandellknee01}, VLA29 \citep{rac99}\\                                               
    49 &03:28:36.7 &31:13:29.6  &49 &Bolo29  &03283609+3113346 &NGC1333 SK6 \citep{sandellknee01}, IRAS 03255+3103\\                                                                      
    50 &03:29:06.5 &31:15:38.6  &50 &---        &03290642+3115348 &NGC1333 HH 7-11 MMS~4 \citep{crs97} ASR6 \citep{asr94}, SK15 \citep{sandellknee01}, HH8 \citep{hh88}\\                 
    51 &03:29:08.8 &31:15:18.1  &51 &Bolo46  & &NGC1333 SK16 \citep{sandellknee01}\\                                                                                                      
    52 &03:29:03.7 &31:14:53.1  &52 &---        & &NGC1333 SK14 \citep{sandellknee01}, HH7--11 MMS~6 \citep{crs97}\\                                                                      
    53 &03:29:04.5 &31:20:59.1  &53 &---        & &NGC1333 SK26 \citep{sandellknee01}\\                                                                                                   
    54 &03:29:10.7 &31:21:45.3  &54 &---        &03291063+3121469 &NGC1333 SK28 \citep{sandellknee01}\\                                                                                   
    55 &03:28:40.4 &31:17:51.3  &55 &Bolo31  & &none\\                                                                                                                                    
    56 &03:29:07.7 &31:21:56.8  &56 &---        &03290773+3121575, 03290704+3121578 &NGC1333 SK29 \citep{sandellknee01}\\                                                                 
    57 &03:29:18.2 &31:25:10.8  &57 &Bolo53  & &NGC1333 SK33 \citep{sandellknee01}\\                                                                                                      
    58 &03:29:24.0 &31:33:20.8  &58 &Bolo57  & &none\\                                                                                                                                    
    59 &03:29:16.5 &31:12:34.6  &59 &Bolo51  & &none\\                                                                                                                                    
    60 &03:28:39.4 &31:18:27.1  &60,73 &---   & &none\\                                                                                                                                 
    61 &03:29:17.3 &31:27:49.6  &61 &Bolo52  & &none\\                                                                                                                                    
    62 &03:29:07.1 &31:17:23.7  &62 &Bolo45  & &NGC1333 SK18 \citep{sandellknee01}\\                                                                                                      
    63 &03:29:18.8 &31:23:16.9  &63 &Bolo54  & &NGC1333 SK32 \citep{sandellknee01}\\                                                                                                      
    64 &03:29:25.5 &31:28:18.1  &64 &Bolo58  & &NGC1333 Per 4A3/4D \citep{ladd94}\\                                                                                                       
    65 &03:29:00.4 &31:12:01.5  &65 &Bolo41  & &NGC1333 IRAS 4B1 \citep{jcc87}, SK1 \citep{sandellknee01}\\                                                                               
    66 &03:29:05.3 &31:22:11.3  &66 &---        & &NGC1333 SK30 \citep{sandellknee01}\\                                                                                                   
    67 &03:29:19.7 &31:23:56.0  &67 &---        &03292003+3124076 &none\\                                                                                                                 
    68 &03:28:56.2 &31:19:12.5  &68 &Bolo39  & &NGC1333 MBO146 \citep{wilking04}\\                                                                                                        
    69 &03:28:34.4 &31:06:59.2  &69 &Bolo28  & &none\\                                                                                                                                    
    70 &03:29:15.3 &31:20:31.2  &70 &Bolo50  & &NGC1333 SK22 \citep{sandellknee01}\\                                                                                                      
    71 &03:28:38.7 &31:05:57.1  &71 &Bolo30  & &HH 340B \citep{hh88}\\                                                                                                                    
    72 &03:29:19.1 &31:11:38.1  &72 &Bolo55  & &none\\                                                                                                                                    
    74 &03:28:32.5 &31:11:07.7  &74 &Bolo25  &03283258+3111040 &NGC1333 38 \citep{lal96}\\                                                                                                
    75 &03:28:42.6 &31:06:10.0  &75 &Bolo33  & &none\\                                                                                                                                    
    76 &03:32:17.8 &30:49:46.3  &76 &Bolo66  & &IRAS 03292+3039, B1 SMM5 \citep{walawender05}\\                                                                                           
    77 &03:31:21.0 &30:45:27.8  &77 &Bolo65  & &IRAS 03282+3035, B1 SMM7 \citep{walawender05}\\                                                                                           
    78 &03:47:41.6 &32:51:44.0  &78 &Bolo122 &03474160+3251437 &IRAS03445+3242, B5 IRS1 \citep{motteandre01}, HH 366 VLA 1 \citep{rr98}\\                                                 
    79 &03:47:39.1 &32:52:17.9  &79 &---        & &none\\                                                                                                                                 
    80 &03:26:37.6 &30:15:24.2  &80 &Bolo18  &03263742+3015283 &IRAS 03235+3004\\                                                                                                         
    81 &03:30:15.5 &30:23:42.8  &81 &Bolo60  &03301515+3023493 &IRAS 03271+3013\\                                                                                                         
    82 &03:33:13.1 &31:19:51.0  &82 &Bolo78  & &B1 SMM1 \citep{walawender05}\\                                                                                                            
    83 &03:32:48.9 &31:09:40.1  &83 &---        & &none\\                                                                                                                                 
    84 &03:32:21.9 &31:04:55.6  &84,87 &---   & &none\\                                                                                                                                 
    85 &03:28:32.5 &31:00:53.0  &85 &---        & &IRAS 03254+3050 \citep{dent98}\\                                                                                                       
    86 &03:26:30.9 &30:32:27.6  &86 &---        & &none\\                                                                                                                                 
    88 &03:31:31.6 &30:43:32.2  &88 &---        & &none\\                                                                                                                                 
    89 &03:32:25.9 &30:59:05.0  &89 &Bolo67  & &B1 SMM9 \citep{walawender05}\\                                                                                                            
    90 &03:45:16.5 &32:04:47.1  &90 &Bolo119 & &Per 7 \citep{bvp04}\\                                                                                                                     
    91 &03:29:23.3 &31:36:08.6  &91 &---        & &Per 4A \citep{ladd94}\\                                                                                                                
   101 &03:44:12.8 &32:01:33.9  &101 &---       &03441297+3201354 &IRAS 03410+3152\\
Bolo11 &03:25:46.5 &30:44:17.8  &--- &Bolo11  & &none\\                                                                                                                                
Bolo26 &03:28:32.7 &31:04:55.9  &--- &Bolo26  & &none\\                                                                                                                                
Bolo27 &03:28:32.7 &30:19:51.1  &--- &Bolo27  & &none\\                                                                                                                                
Bolo44 &03:29:04.9 &31:18:41.2  &--- &Bolo44  & &none\\                                                                                                                                
Bolo62 &03:30:32.3 &30:26:27.4  &--- &Bolo62  & &none\\                                                                                                                                
Bolo64 &03:30:51.3 &30:49:14.7  &--- &Bolo64  & &none\\                                                                                                                                
Bolo70 &03:32:44.3 &31:00:09.7  &--- &Bolo70  & &none\\                                                                                                                                
Bolo89 &03:40:49.3 &31:48:50.1  &--- &Bolo89  & &none\\                                                                                                                                
Bolo90 &03:41:09.3 &31:44:38.1  &--- &Bolo90  &03410913+3144378 &IRAS 03380+3135\\                                                                                                     
Bolo92 &03:41:40.7 &31:57:59.8  &--- &Bolo92  & &none\\                                                                                                                                
Bolo94 &03:41:46.1 &31:57:22.9  &--- &Bolo94  & &none\\                                                                                                                                
Bolo111 &03:44:14.4 &31:57:57.6 &--- &Bolo111 & &none\\                                                                                                                               
Bolo112 &03:44:15.6 &32:09:15.0 &--- &Bolo112 & &CL* IC 348 LRL 329 \citep{lrl98}\\                                                                                                   
Bolo113 &03:44:23.9 &31:59:25.2 &--- &Bolo113 & &none\\                                                                                                                               
Bolo114 &03:44:23.2 &32:10:10.1&--- &Bolo114 & &none\\                                                                                                                               
Bolo121 &03:47:32.7 &32:50:50.2 &--- &Bolo121 & &none\\                                                                                                                               
 clf55 &03:32:14.7 &31:23:36.8  &--- &---        & &none\\                                                                                                                              
 clf68 &03:28:51.4 &30:13:40.2  &--- &---        & &none\\                                                                                                                              
 clf70 &03:28:21.5 &30:20:30.7  &--- &---        & &none\\                                                                                                                              
 clf74 &03:28:24.5 &30:20:28.3  &--- &---        & &none\\                                                                                                                              
 clf77 &03:28:41.1 &30:25:31.6  &--- &---        & &none\\                                                                                                                              
 clf79 &03:28:24.8 &30:20:49.4   &--- &---        & &none\\                                                                                                                              
%&&&&&&&&&&&\\

\end{longtable}